\RequirePackage{fix-cm}
\documentclass[smallextended]{svjour3}       
\smartqed  
\makeatletter
\def\@seccntformat#1{\@ifundefined{#1@cntformat}%
   {\csname the#1\endcsname\quad}  
   {\csname #1@cntformat\endcsname}
}
\let\oldappendix\appendix 
\renewcommand\appendix{%
    \oldappendix
    \newcommand{\section@cntformat}{\appendixname~\thesection\quad}
}
\makeatother

\usepackage{graphicx}
\usepackage[breaklinks=true]{hyperref}
\usepackage{natbib}
\usepackage{xspace}
\usepackage{mathptmx}
\newcommand{\degr}{$^{o}$\xspace}
\journalname{Celestial Mechanics \& Dynamical Astronomy}
\begin{document}

\title{Recent arrivals to the main asteroid belt
}


\author{Carlos~de~la~Fuente~Marcos \and
        Ra\'ul~de~la~Fuente~Marcos 
}


\institute{C.~de~la~Fuente~Marcos (0000-0003-3894-8609) \at
              Universidad Complutense de Madrid, 
              Ciudad Universitaria, E-28040 Madrid, Spain \\
              \email{nbplanet@ucm.es}           
           \and
           R.~de~la~Fuente~Marcos (0000-0002-5319-5716) \at
              AEGORA Research Group,
              Facultad de Ciencias Matem\'aticas,
              Universidad Complutense de Madrid,
              Ciudad Universitaria, E-28040 Madrid, Spain
}

\date{Received: November 15, 2021 / Accepted: July 12, 2022}

\maketitle

\begin{abstract}
The region where the main asteroid belt is now located may have started
empty, to become populated early in the history of the Solar system with
material scattered outward by the terrestrial planets and inward by 
the giant planets. These dynamical pathways toward the main belt may 
still be active today. Here, we present results from a data mining 
experiment aimed at singling out present-day members of the main 
asteroid belt that may have reached the belt during the last few hundred 
years. Probable newcomers include 2003~BM$_{1}$, 2007~RS$_{62}$, 457175 
(2008~GO$_{98}$), 2010~BG$_{18}$, 2010~JC$_{58}$, 2010~JV$_{52}$, 
2010~KS$_{6}$, 2010~LD$_{74}$, 2010~OX$_{38}$, 2011~QQ$_{99}$, 
2013~HT$_{149}$, 2015~BH$_{103}$, 2015~BU$_{525}$, 2015~RO$_{127}$, 
2015~RS$_{139}$, 2016~PC$_{41}$, 2016~UU$_{231}$, 2020~SA$_{75}$, 
2020~UO$_{43}$, and 2021~UJ$_{5}$, all of them in the outer belt. Some 
of these candidates may have been inserted in their current orbits after 
experiencing relatively recent close encounters with Jupiter. We also 
investigated the likely source regions of such new arrivals. Asteroid 
2020~UO$_{43}$, if real, has a non-negligible probability of having an 
origin in the Oort cloud or even interstellar space. Asteroid 
2003~BM$_{1}$ may have come from the neighborhood of Uranus. However, 
most newcomers ---including 457175, 2011~QQ$_{99}$, and 2021~UJ$_{5}$--- 
might have had an origin in Centaur orbital space. The reliability of 
these findings is assessed within the context of the uncertainties of 
the available orbit determinations.
\end{abstract}

\section{Introduction}
\label{intro}
Between the orbits of Mars and Jupiter, there is a diverse population of small bodies known collectively as the main 
asteroid belt (see for example the review by \citealt{2020arXiv201207932R}). Although there is general agreement that 
the main belt formed early in the history of the Solar system, the exact details of its origins remain uncertain. 
Shortly after the discovery of the first members of the main asteroid belt, H.~W.~M.~Olbers and others proposed that 
these objects could be debris from a destroyed planet (see for example the reviews and notes by 
\citealt{1931PASP...43..324B,1950AJ.....55..164K,1972Natur.239..508O,1973Natur.242..250N}) or leftover material that 
never made into a planet due to, namely, collisional processes (see for example \citealt{1950AJ.....55..164K,
1964Icar....3...52A,1974MNRAS.166..469N,2001Icar..153..338P}). However, \citet{2017SciA....3E1138R} pointed out that 
dark, water-rich carbonaceous C-type asteroids dominate the outer belt and drier, siliceous S-type asteroids are more 
common in the inner belt. Such a trend led these authors to propose that the region where the main asteroid belt is now 
located started empty and that the currently observed asteroids were inserted there from the inner Solar system 
(S-types) and from the region of the giant planets (C-types). In this scenario, material spread outward by the 
terrestrial planets contributed to the S-type asteroid population currently inhabiting the main belt; debris scattered 
inward mainly by Jupiter and Saturn ended up contributing to the C-type component of the belt. The presence in the main 
asteroid belt of material formed well beyond Jupiter and Saturn has been dramatically confirmed by 
\citet{2021ApJ...916L...6H}, who found two extremely red main-belt asteroids, 203~Pompeja and 269~Justitia. Such red 
bodies may share an origin with trans-Neptunian objects (TNOs) and Centaurs that have surfaces covered with complex 
organics (see for example \citealt{2008ssbn.book..143B}). However, red carbonaceous asteroids may turn less red over 
time because of space weathering (see for example \citealt{2022ApJ...924L...9H}). 

As the current orbital architecture of the coupled subsystems made of the inner and outer planets may have remained in 
its present form for Gyr (see for example \citealt{1998AJ....116.2055I,1999Icar..139..336I,2002MNRAS.336..483I,
2007PASJ...59..989T,2022CeMDA.134...20M}) it is reasonable to assume that the dynamical pathways that might have 
populated the main asteroid belt in the past could still be open today. In order to investigate such a possibility, 
numerical integrations may be used to identify present-day members of the main asteroid belt that may have followed 
significantly different orbits in the relatively recent past. There are, however, over 10$^{6}$ known main-belt members 
and a blind search would require a very substantial amount of computer power in order to complete the task. In sharp 
contrast, a data mining exploration of a reliable database may lead to robust results using just modest resources. Here, 
we present results from a data mining experiment aimed at singling out present-day members of the main asteroid belt 
that may have reached the belt during the last few hundred years. In a way, our investigation is the exact opposite of 
the one discussed by \citet{2017A&A...598A..52G} that focused on the escape of asteroids from the main belt. This paper 
is organized as follows. In Sect.~\ref{sec:2}, we present the data and methods used in our analyses, which are shown in 
Sect.~\ref{sec:3}. In Sect.~\ref{sec:4}, we apply $N$-body simulations to further study some objects of interest singled 
out by our statistical analyses. In Sect.~\ref{sec:5}, we discuss our results. Our conclusions are summarized in 
Sect.~\ref{sec:6}.

\section{Data and methods}
\label{sec:2}
As of April 22, 2022, the Minor Planet Center (MPC)\footnote{\url{https://minorplanetcenter.net/}} had 344.0 million of 
observations of 1,194,113 asteroids and 4405 comets (see for example \citealt{2015IAUGA..2253004R,2016IAUS..318..265R,
2019AAS...23324503H}). The vast majority of the data corresponded to members of the main asteroid belt. The data are 
available from the MPC itself and, in somewhat processed form, from several other online resources that include the Jet 
Propulsion Laboratory (JPL)\footnote{\url{https://www.jpl.nasa.gov}}-hosted ensemble of small-bodies websites and the 
European Asteroids Dynamic Site (ASTDyS)\footnote{\url{https://newton.spacedys.com/astdys/index.php?pc=0}} online 
information service (see for example \citealt{2012IAUJD...7P..18K,2015IAUGA..2235184B}). These databases can be queried 
using tools available from the websites themselves, but also through several packages written in the {\tt Python} 
language \citep{van1995python,10.5555/1593511}.

\subsection{Data}
\label{sec:2.1}
Here, we work with publicly available data (orbit determinations, input Cartesian vectors, ephemerides) from JPL's 
Small-Body Database (SBDB)\footnote{\url{https://ssd.jpl.nasa.gov/tools/sbdb_lookup.html\#/}} and {\tt Horizons} on-line 
solar system data and ephemeris computation service,\footnote{\url{https://ssd.jpl.nasa.gov/horizons/}} both provided by 
the Solar System Dynamics Group (SSDG,\footnote{\url{https://ssd.jpl.nasa.gov/}} \citealt{1996DPS....28.2504G,
1997DPS....29.2106C,1997BAAS...29.1099G,2011jsrs.conf...87G,2015IAUGA..2256293G}). The {\tt Horizons} ephemeris system 
has recently been updated, replacing the DE430/431 planetary ephemeris, used since 2013, with the new DE440/441 solution 
\citep{2021AJ....161..105P}. DE440 covers the years 1550--2650 while DE441 is tuned to cover a time range of $-$13,200 
to +17,191 years \citep{2021AJ....161..105P}. The new DE440/441 general-purpose planetary solution includes seven 
additional years of ground and space-based astrometric data, data calibrations, and dynamical model improvements, most 
significantly involving Jupiter, Saturn, Pluto, and the Kuiper Belt \citep{2021AJ....161..105P}. Most data were 
retrieved from JPL's SBDB and {\tt Horizons} using tools provided by the {\tt Python} package {\tt Astroquery} 
\citep{2019AJ....157...98G} and its {\tt HorizonsClass} 
class.\footnote{\url{https://astroquery.readthedocs.io/en/latest/jplhorizons/jplhorizons.html}} 

\subsection{Methods}
\label{sec:2.2}
The statistical analyses in Sect.~\ref{sec:3} have been carried out using tools from several {\tt Python} packages ({\tt 
Python~3.9} and the latest versions of the libraries were used). Some figures have been produced using the {\tt 
Matplotlib} library \citep{2007CSE.....9...90H} and statistical tools provided by {\tt NumPy} 
\citep{2011CSE....13b..22V,2020Natur.585..357H}. Sets of bins in frequency-based histograms were computed using {\tt 
NumPy} by applying the Freedman and Diaconis rule \citep{Freedman1981OnTH}.

The $N$-body simulations discussed in Sect.~\ref{sec:4} were carried out using a direct $N$-body code developed by 
\citet{2003gnbs.book.....A} that is publicly available from the website of the Institute of Astronomy of the University 
of Cambridge.\footnote{\url{http://www.ast.cam.ac.uk/~sverre/web/pages/nbody.htm}} This software uses the Hermite 
integration scheme implemented by \citet{1991ApJ...369..200M}. This scheme applies a predictor-corrector time 
integration method that uses an extrapolation of the equations of motion to predict positions and velocities from which 
the new accelerations are computed, then the predicted values are corrected using interpolation by applying finite 
differences techniques. The Hermite scheme allows efficient numerical integration of the entire Solar system thanks to 
the use of a block-step strategy \citep{2003gnbs.book.....A} in which suitably quantized time-steps allow the precise 
integration of the orbits of Mercury or planetary satellites and trans-Neptunian objects simultaneously; encounters at 
very close range can also be followed with sufficient precision (see for example the application to the Chelyabinsk 
event in \citealt{2015ApJ...812...26D}). Results from this code have been extensively discussed by 
\citet{2012MNRAS.427..728D} and compare well with those from \citet{2011A&A...532A..89L} among others. In our 
calculations, relative errors in the total energy are as low as 10$^{-16}$ to 10$^{-15}$. The relative error in the 
total angular momentum is several orders of magnitude smaller.

The physical model included gravitational perturbations from the eight major planets, the Moon, the barycentre of the 
Pluto-Charon system, and the three largest asteroids. When integrating the equations of motion, non-gravitational 
forces, relativistic or oblateness terms were not taken into account. Besides studying some representative orbits, we  
performed additional calculations that applied the Monte Carlo using the Covariance Matrix (MCCM) methodology described 
by \citet{2015MNRAS.453.1288D} in which a Monte Carlo process generates control or clone orbits based on the nominal 
orbit but adding random noise on each orbital element by making use of the covariance matrix (that was also retrieved 
from JPL's SSDG SBDB using the {\tt Python} package {\tt Astroquery} and its {\tt SBDBClass} 
class).\footnote{\url{https://astroquery.readthedocs.io/en/latest/jplsbdb/jplsbdb.html}} 

\section{Statistical analysis}
\label{sec:3}
In order to identify present-day members of the main belt that may have had very different orbits in the relatively 
recent past ($\sim$400~yr ago), we retrieved (from JPL's {\tt Horizons} via the {\tt astroquery.jplhorizons} package 
pointed out above) the heliocentric osculating orbital elements of each known member of the main belt for two epochs 
($t$ and $t_{0}$): 2459600.5 JD TDB (2022-Jan-21.0 00:00:00.0 TDB, Barycentric Dynamical Time, J2000.0 ecliptic and 
equinox) and 2305427.5 JD TDB (= A.D. 1599-Dec-12 00:00:00.0 TDB). Then, we computed the absolute relative variation of 
the value of the orbital elements semimajor axis, eccentricity, and inclination ($a$, $e$ and $i$, respectively),
$\Delta_{a}$, $\Delta_{e}$ and $\Delta_{i}$. For example, $\Delta_{a}=|a_{t} - a_{t_{0}}|/a_{t}$ gives the absolute 
relative variation in the value of the semimajor axis. By computing the distributions of $\Delta_{a}$, $\Delta_{e}$ and 
$\Delta_{i}$, we can single out known members of the main belt that had statistically significant different orbits in 
the recent past. The orbital evolution of such objects can be studied in further detail using $N$-body simulations to 
find out about their past dynamical history and most probable origin.

\subsection{Inner main belt}
\label{sec:3.1}
Following JPL's SBDB, inner main-belt asteroids have $a<2.0$~au and $q>1.666$~au. The current tally of this orbit class 
is 26,027 objects (as of April 22, 2022). Most objects in this region are Hungaria asteroids that populate the 5:1 
mean-motion resonance with Jupiter between 1.78~au and 2~au from the Sun (see for example 
\citealt{2010Icar..207..769M}). 

\begin{figure*}
  \centering
  \includegraphics[width=0.49\linewidth]{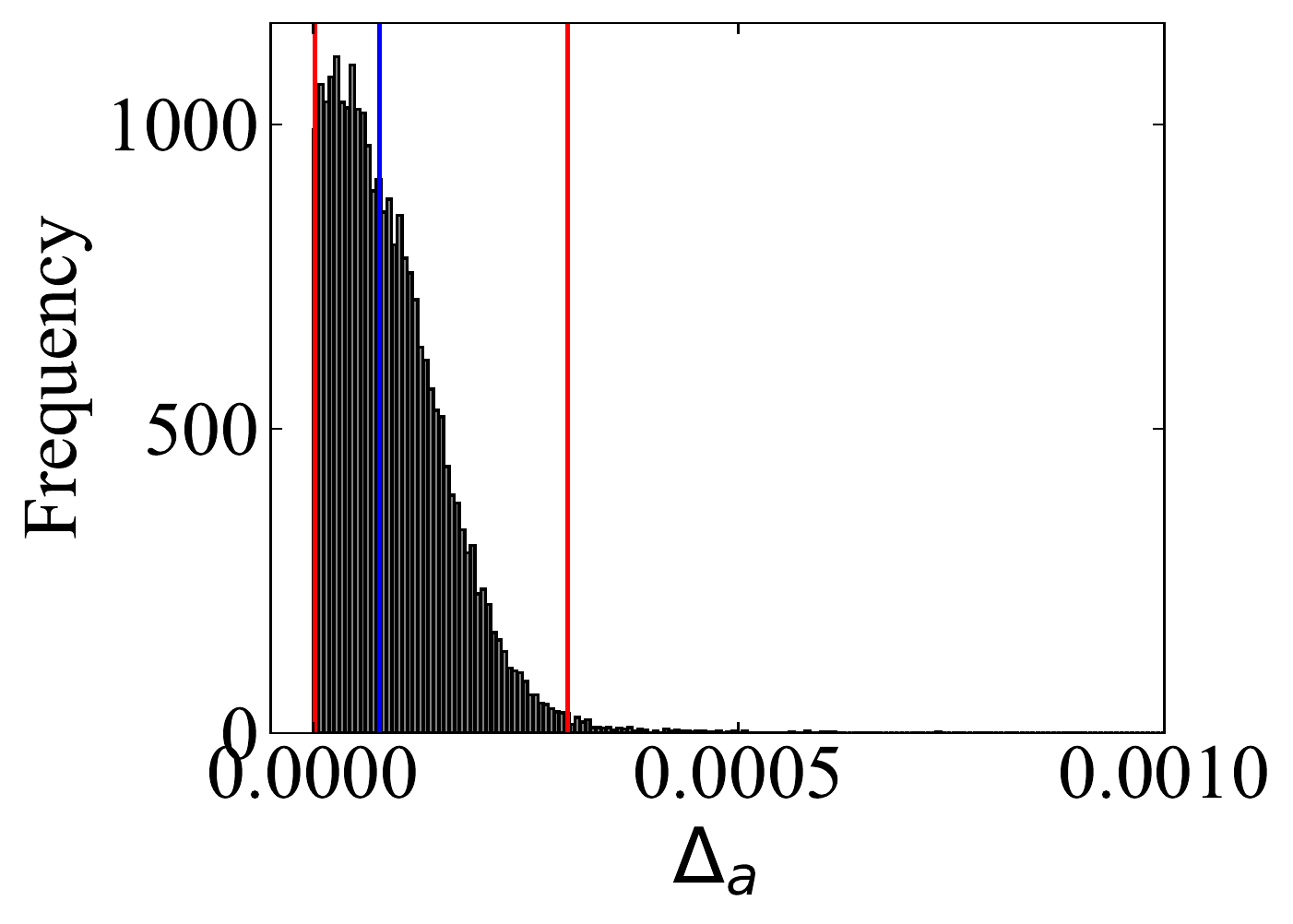}
  \includegraphics[width=0.49\linewidth]{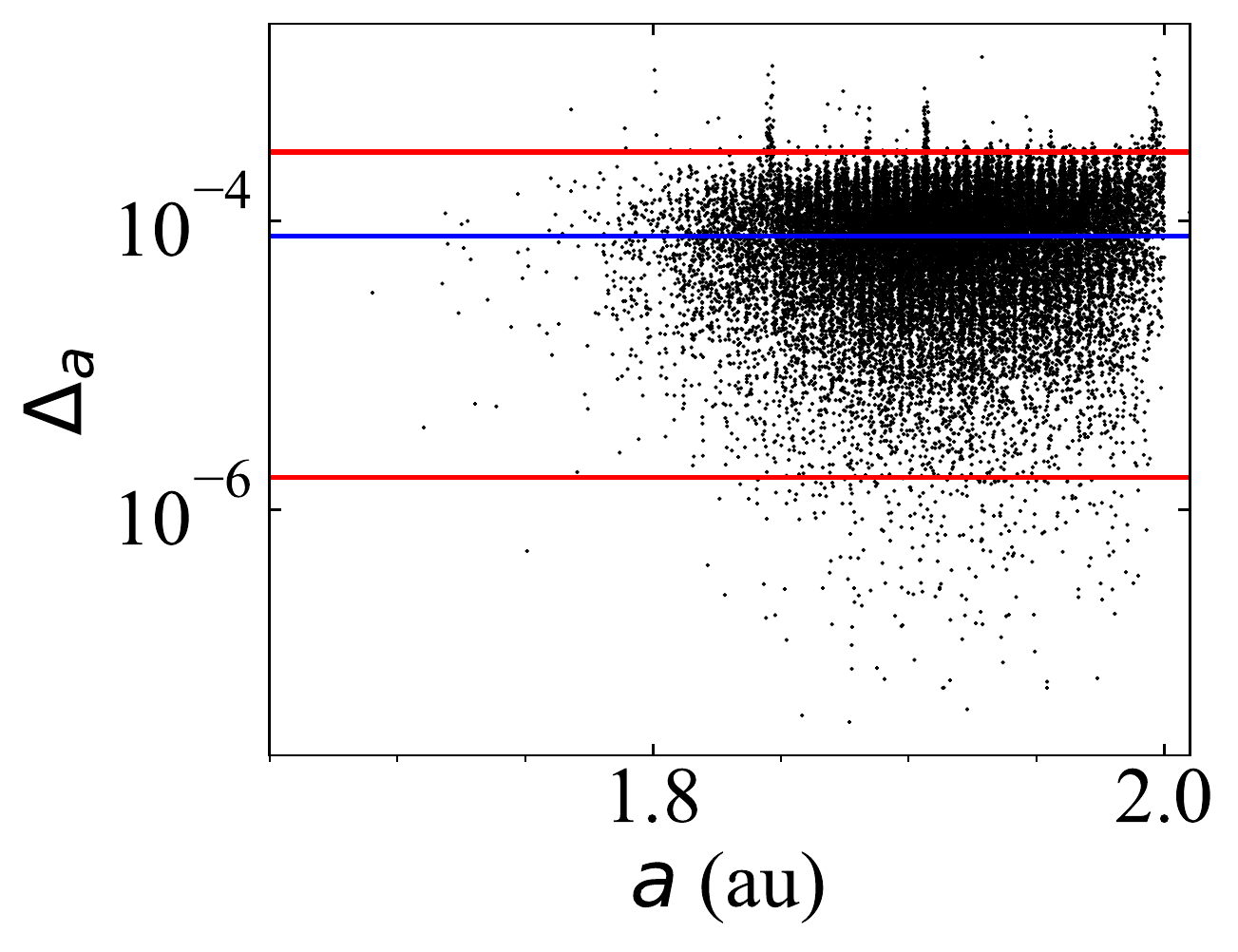}
  \includegraphics[width=0.49\linewidth]{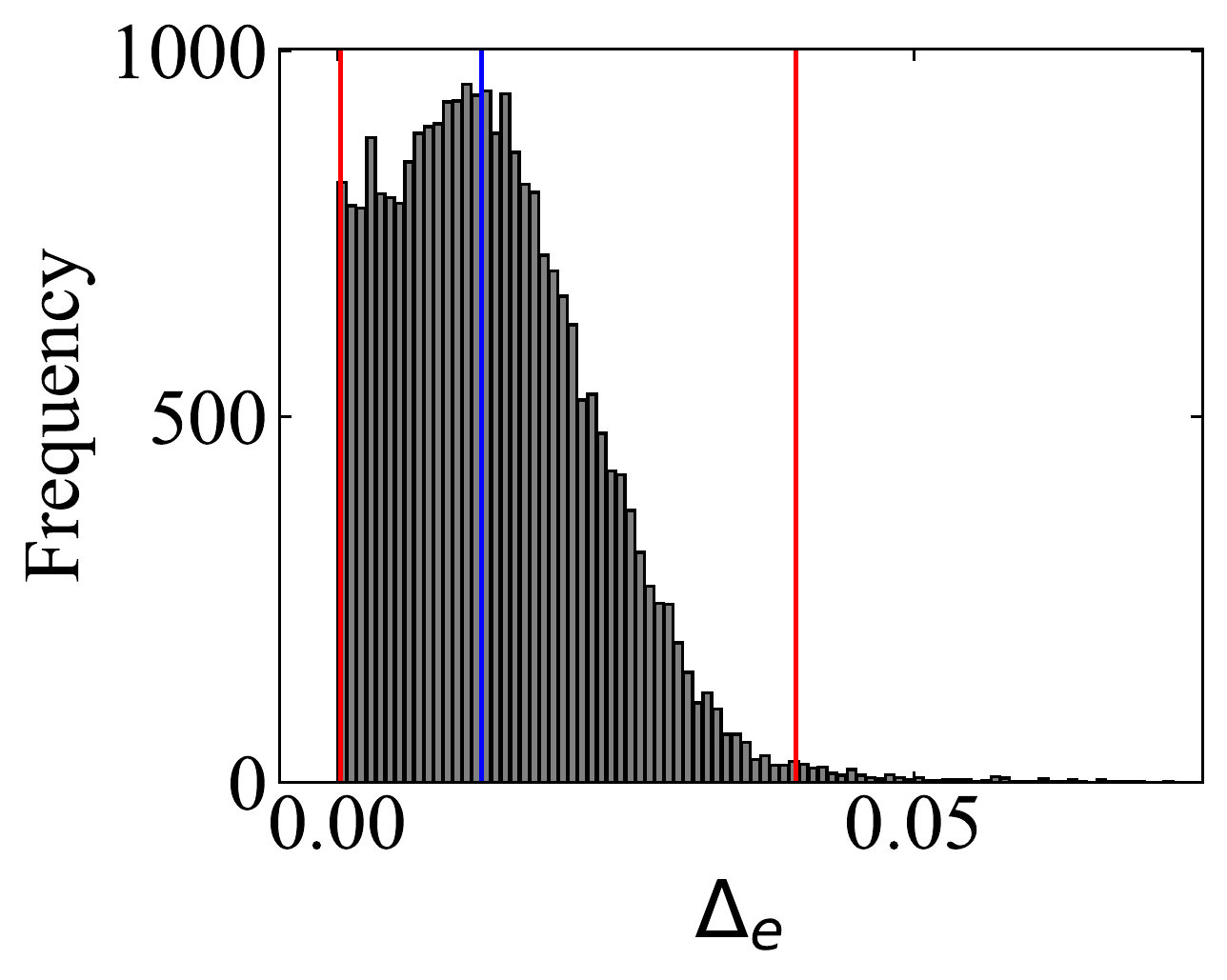}
  \includegraphics[width=0.49\linewidth]{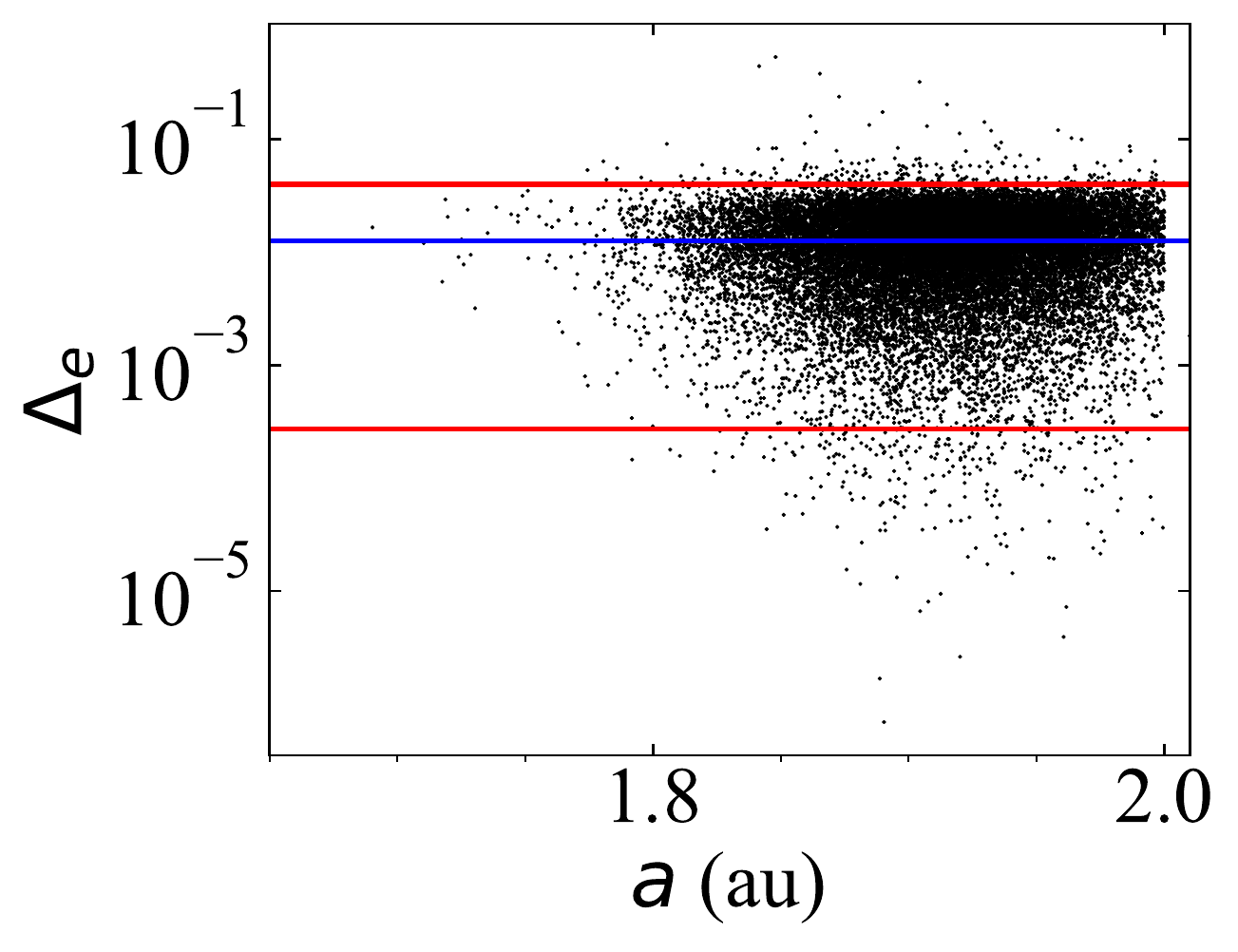}
  \includegraphics[width=0.49\linewidth]{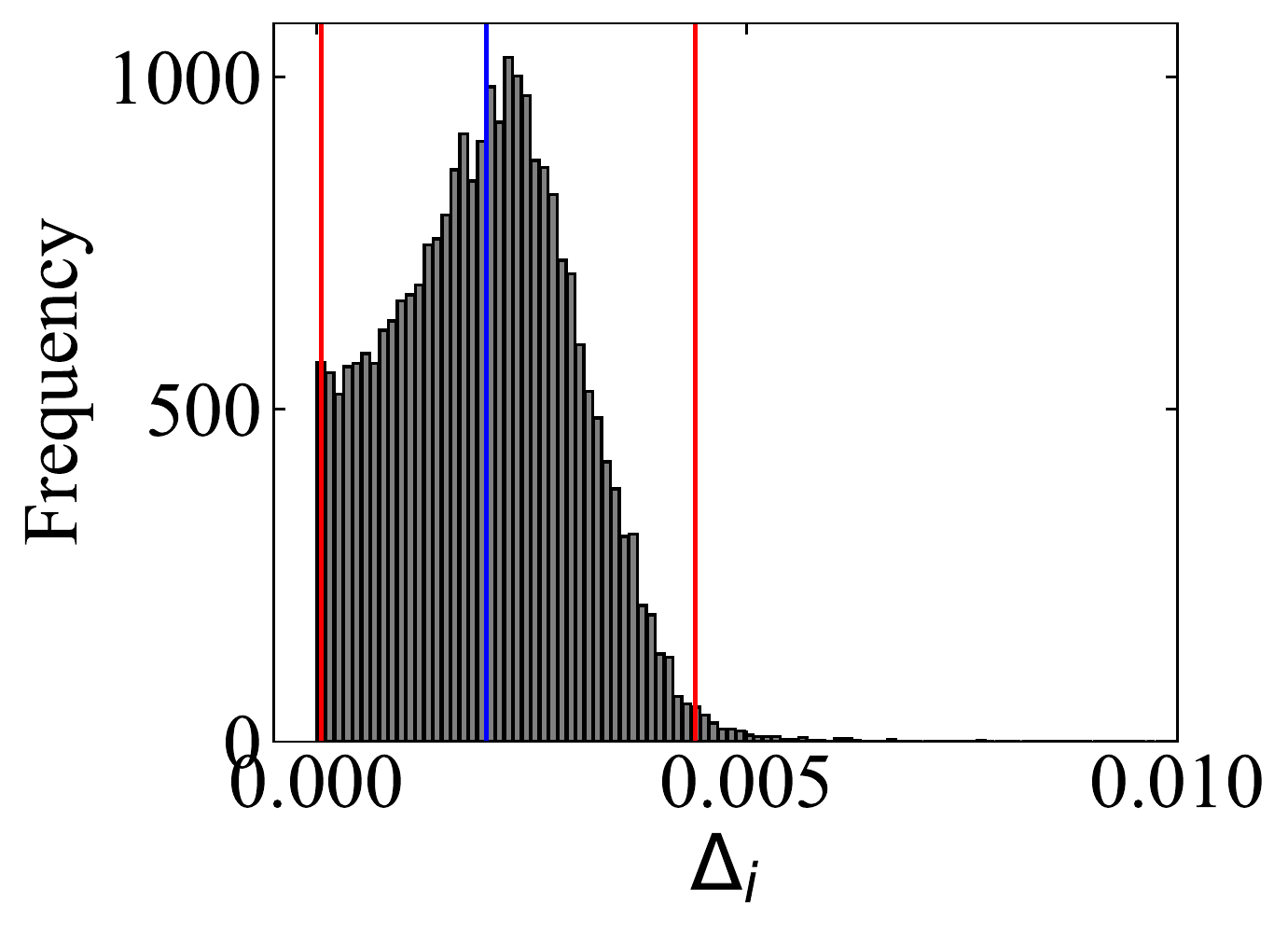}
  \includegraphics[width=0.49\linewidth]{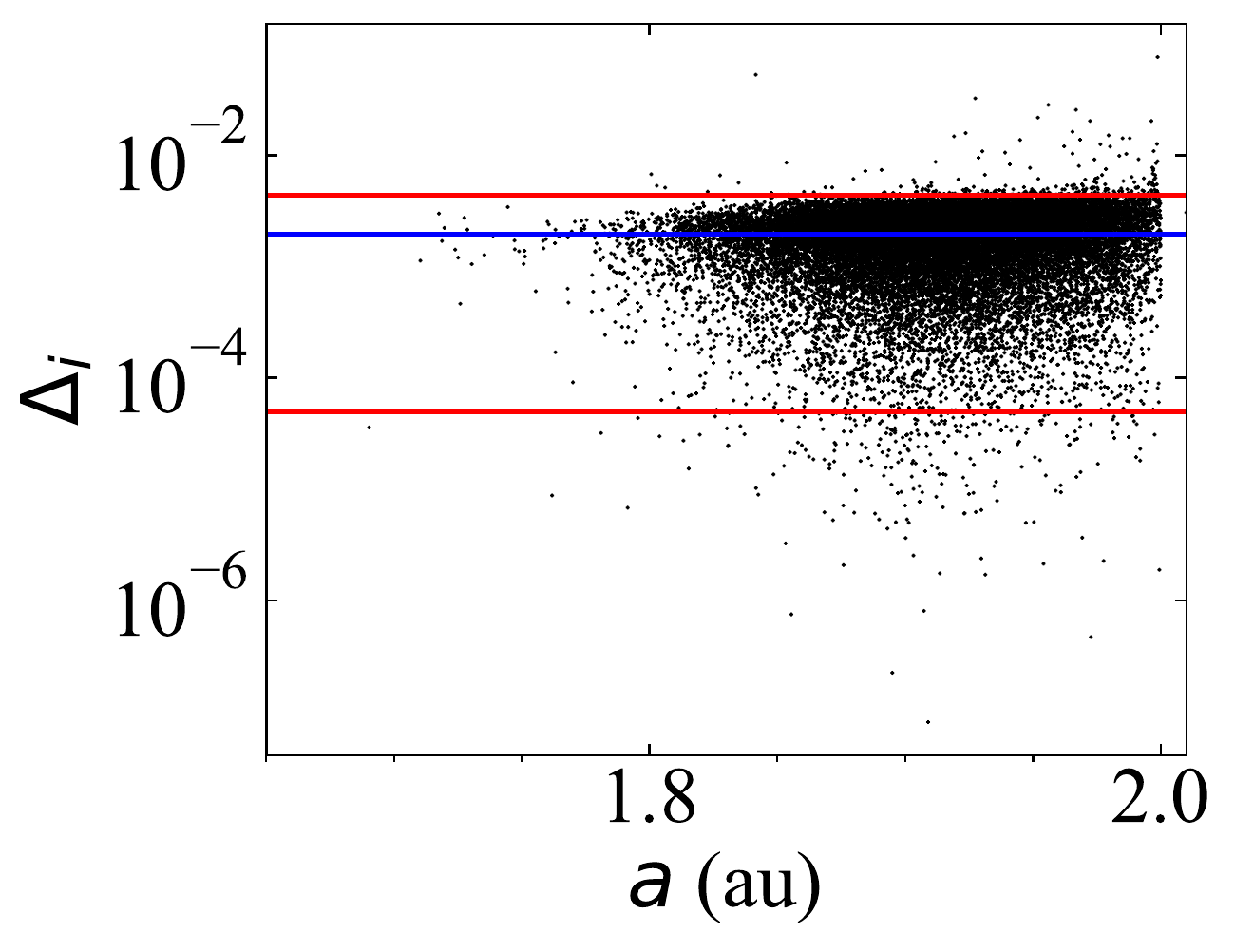}
\caption{Inner main belt (26,027 objects as of April 22, 2022). Absolute relative variation of the value of the orbital 
elements semimajor axis, eccentricity, and inclination: $\Delta_{a}$, $\Delta_{e}$ and $\Delta_{i}$. The left-hand side 
column of panels shows the distributions as frequency histograms; the right-hand side column of panels shows the 
distribution of absolute relative variations as a function of the semimajor axis in logarithmic scale. Median values are 
shown as vertical or horizontal blue lines and the 1st and 99th percentiles as red lines. Data source: JPL's 
{\tt Horizons}.}
\label{fig:inndist}       
\end{figure*}

Figure~\ref{fig:inndist} shows the resulting distributions of the absolute relative variation of the value of the 
orbital elements semimajor axis, eccentricity, and inclination: $\Delta_{a}$, $\Delta_{e}$ and $\Delta_{i}$. We use the 
1st and 99th percentiles of the distribution to identify severe outliers (see for example \citealt{2012psa..book.....W}). 
Objects in the 1st percentile of the distribution, particularly in the case of $\Delta_{a}$, can be considered as 
unusually dynamically stable; minor bodies in the 99th percentile are the least stable of the sample set and any 
newcomers to the main belt are expected to be above the 99th percentile of the distribution in $\Delta_{a}$. In the 
figure, median values are shown as vertical or horizontal blue lines and the 1st and 99th percentiles are displayed as 
red lines. For $\Delta_{a}$ the 1st, 16th, 50th, 84th, 99th percentiles are: 1.66$\times$10$^{-6}$, 
2.46$\times$10$^{-5}$, 7.83$\times$10$^{-5}$, 0.00016 and 0.00030. For $\Delta_{e}$, the same percentiles are: 0.00027, 
0.00424, 0.01253, 0.02226, and 0.03979. And for $\Delta_{i}$: 4.96$\times$10$^{-5}$, 0.00077, 0.00197, 0.00300, and 
0.00440. The distributions are not normal but in a normal distribution a value that is one standard deviation above the 
mean is equivalent to the 84th percentile and a value that is one standard deviation below the mean is equivalent to the 
16th percentile (see for example \citealt{2012psa..book.....W}). 
%
%
%
%

Our analysis shows that the most stable objects, those within the 1st percentile of the distributions, tend to occupy 
the central regions of the inner belt, with $a\sim1.9$~au. On the other hand, there are no objects with large values 
(for example $>1$) of the relative variations. In particular, the maximum value of $\Delta_{a}$ is of the order of 
0.0015 (see the right-hand side panels in Fig.~\ref{fig:inndist}), which means that the largest variation of the 
semimajor axis over the studied time interval is well under 1\%. The largest variations are observed in $\Delta_{e}$. 
Figure~\ref{fig:inndist}, right-hand side top panel, shows that the largest variations are linked to mean-motion 
resonances; the banded distribution observed is the result of the overlapping grid of mean-motion resonances with Earth 
and Mars discussed by \citet{2006Icar..184...29G, 2019Icar..317..121G}. The maxima in the distributions of $\Delta_{e}$ 
and $\Delta_{i}$ are linked to objects subjected to the von Zeipel-Lidov-Kozai mechanism \citep{1910AN....183..345V,
1962AJ.....67..591K,1962P&SS....9..719L,2019MEEP....7....1I} that drives anti-correlated eccentricity-inclination 
oscillations. The effects of the von Zeipel-Lidov-Kozai mechanism have been well documented not only within the inner 
main belt \citep{1996A&A...307..310M} but in other sections of the belt as well \citep{2017MNRAS.468.4719V}.

\subsection{Main belt}
\label{sec:3.2}
\begin{figure*}
  \centering
  \includegraphics[width=0.49\linewidth]{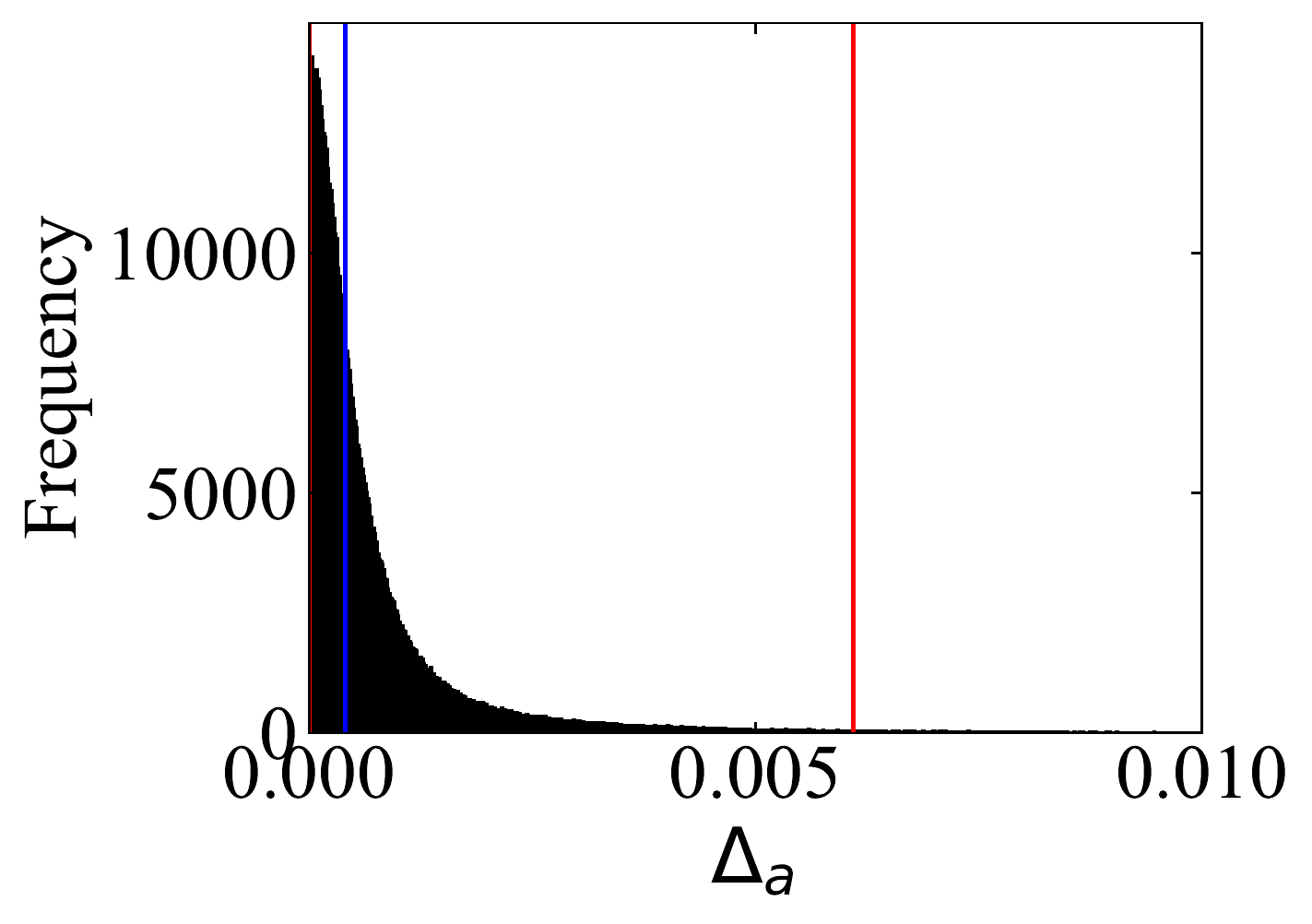}
  \includegraphics[width=0.49\linewidth]{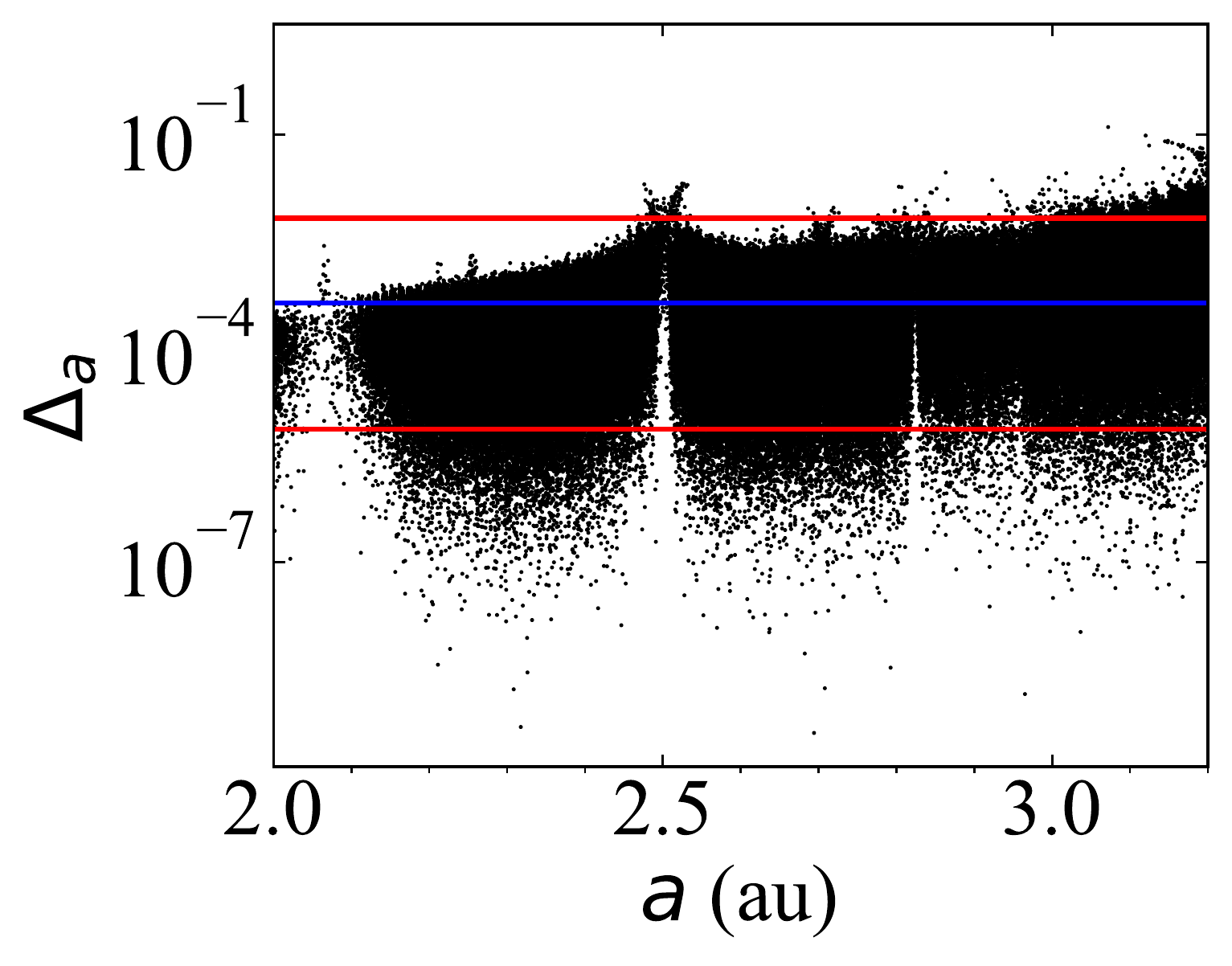}
  \includegraphics[width=0.49\linewidth]{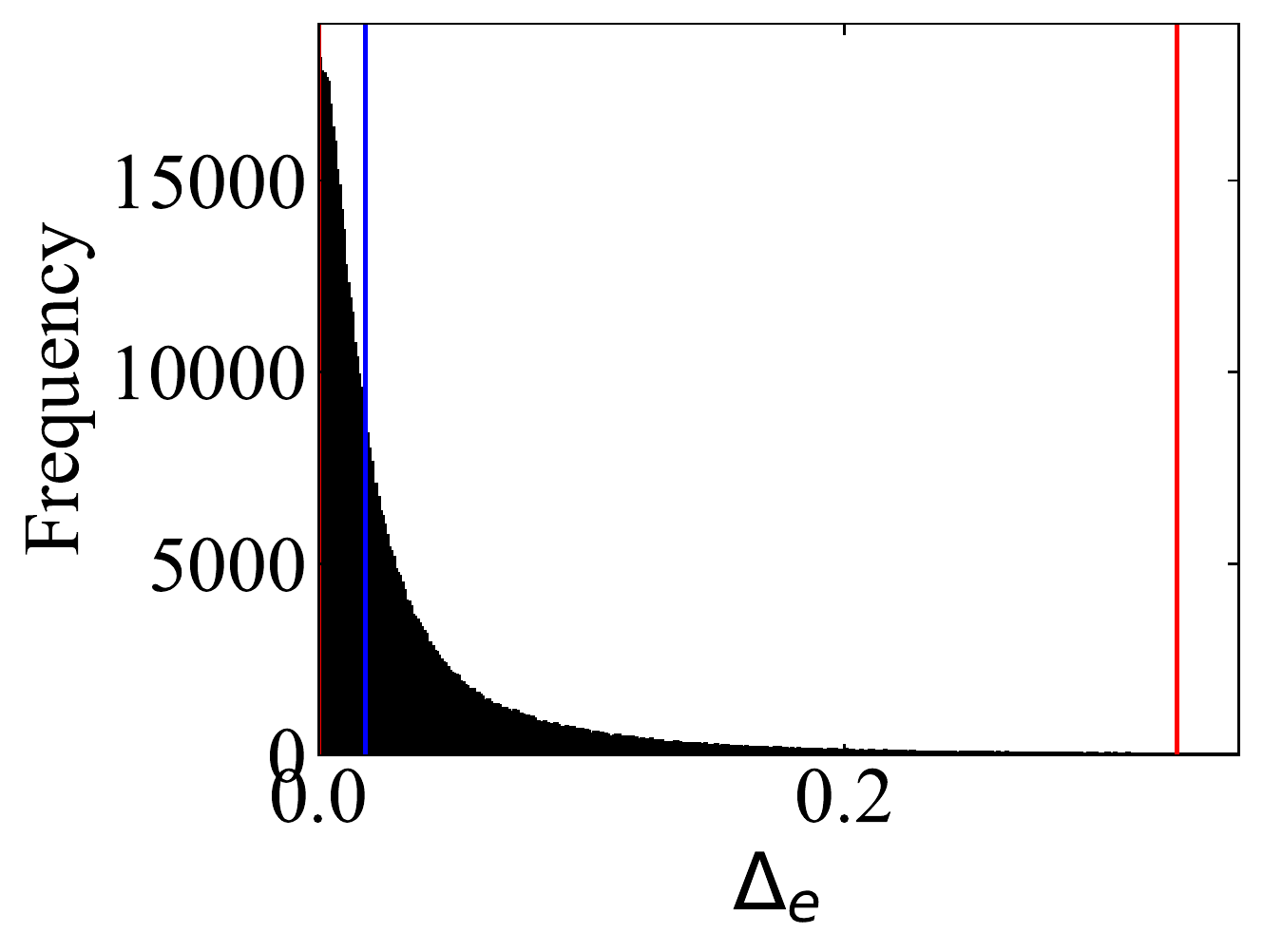}
  \includegraphics[width=0.49\linewidth]{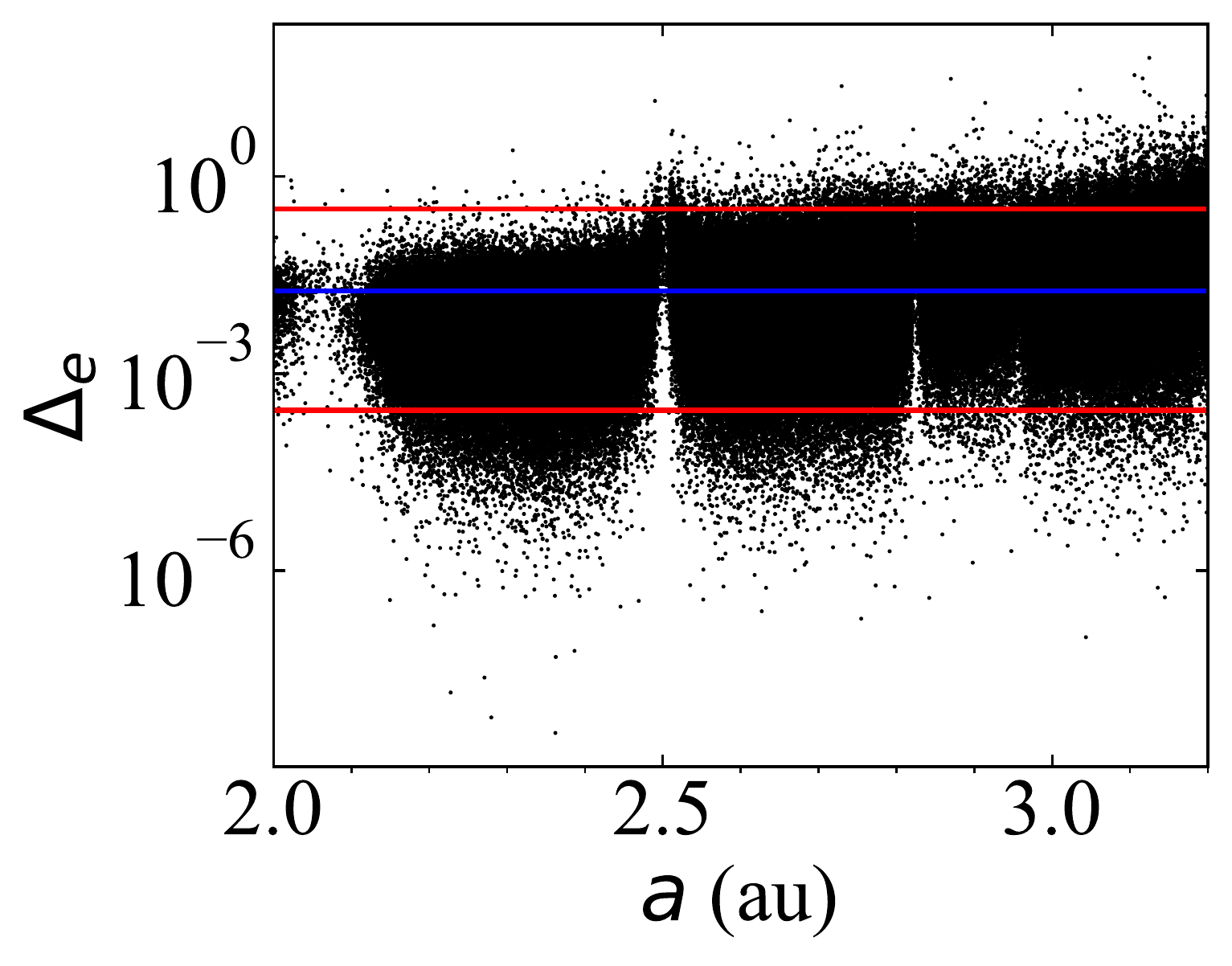}
  \includegraphics[width=0.49\linewidth]{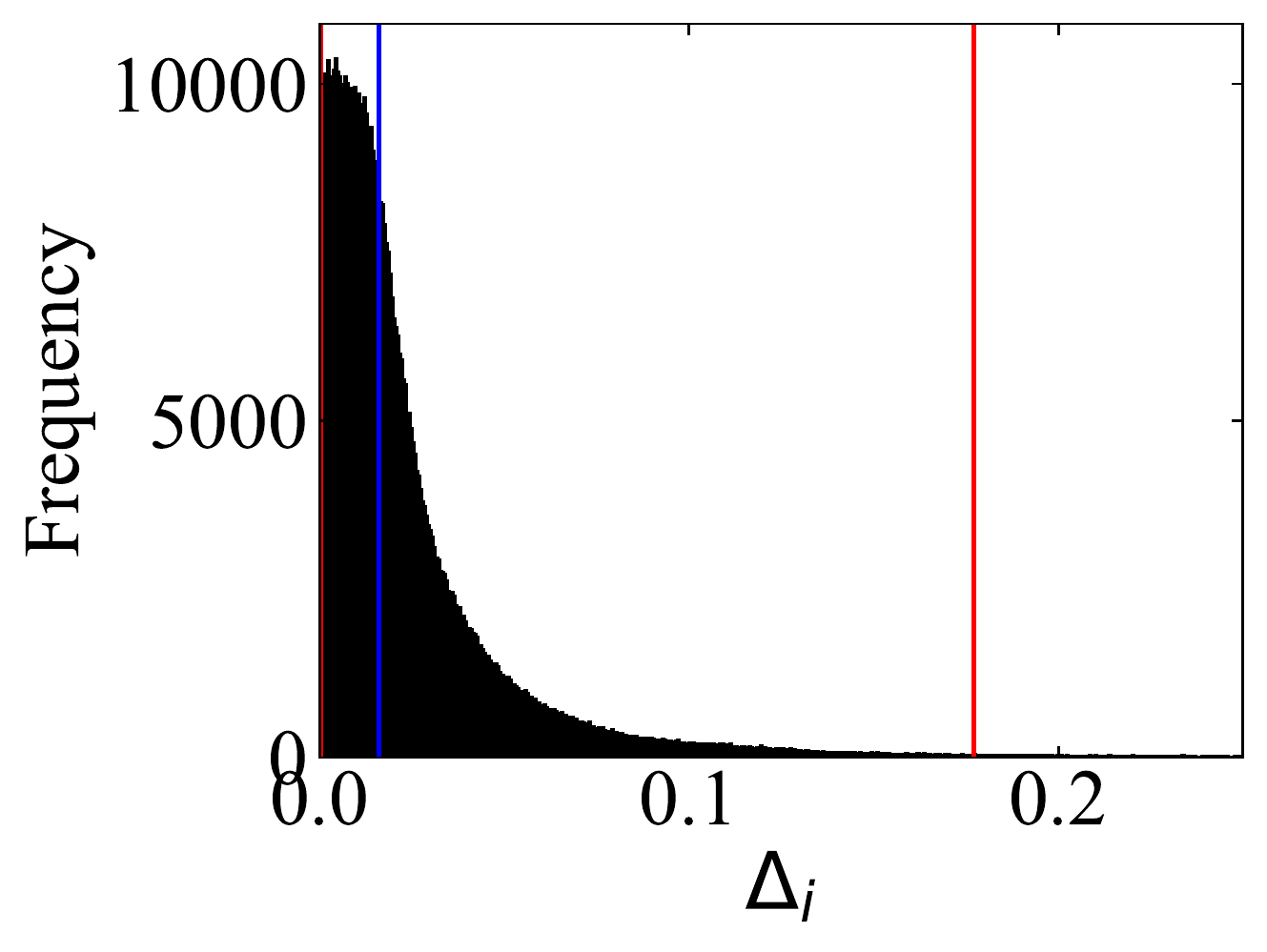}
  \includegraphics[width=0.49\linewidth]{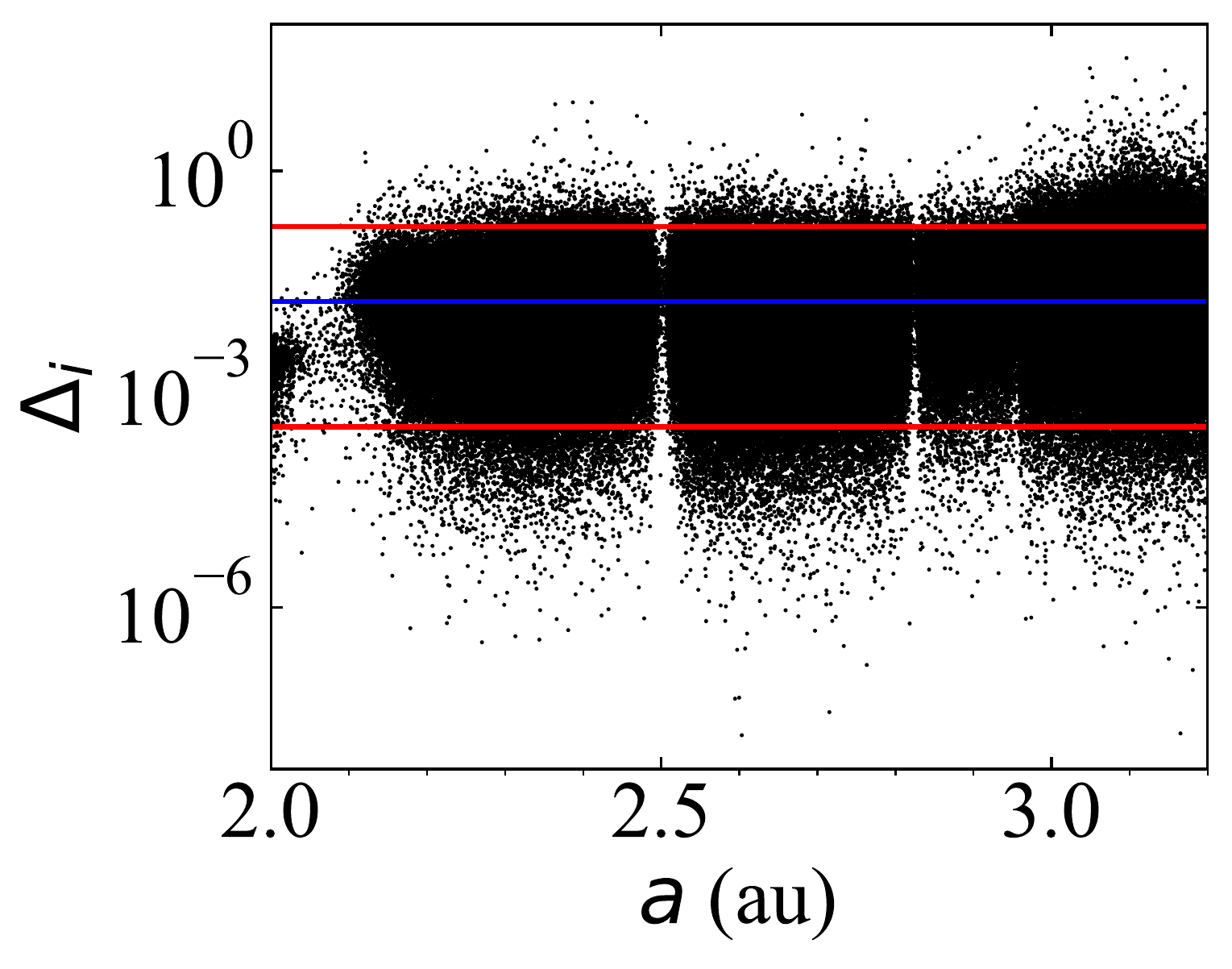}
\caption{Main belt's core (1,073,121 objects as of April 22, 2022). Absolute relative variation of the value of the 
orbital elements semimajor axis, eccentricity, and inclination: $\Delta_{a}$, $\Delta_{e}$ and $\Delta_{i}$. The 
left-hand side column of panels shows the distributions as frequency histograms; the right-hand side column of panels 
shows the distribution of absolute relative variations as a function of the semimajor axis in logarithmic scale. Median 
values are shown as vertical or horizontal blue lines and the 1st and 99th percentiles as red lines. Data source: JPL's 
{\tt Horizons}.}
\label{fig:maidist}       
\end{figure*}

Following JPL's SBDB, main-belt asteroids have 2.0~au $<a<$ 3.2~au and $q>1.666$~au and this orbit class includes 
1,073,121 objects. Figure~\ref{fig:maidist}, right-hand side top panel, shows the strong Kirkwood gap at 2.5~au due to 
the 3:1 mean-motion resonance with Jupiter, another one at 2.82~au caused by the 5:2 mean-motion resonance with Jupiter,
and a third one at 2.958~au linked to the 7:3 mean-motion resonance with Jupiter (see for example 
\citealt{1999ssd..book.....M}). 

The maximum values of $\Delta_{a}$, $\Delta_{e}$ and $\Delta_{i}$ are observed for objects with $a>3.0$~au. However, the 
maximum value of $\Delta_{a}$ is under 0.1 (with one exception); therefore, the largest variation over the studied time 
interval is still below 10\%, not high enough to signal objects that may have entered the main belt in the relatively 
recent past. Although not as clearly as in Figure~\ref{fig:inndist}, right-hand side top panel, Figure~\ref{fig:maidist}, 
right-hand side top panel, shows an overlapping grid of mean-motion resonances. For $\Delta_{a}$ the 1st, 16th, 50th, 
84th, 99th percentiles are: 7.34$\times$10$^{-6}$, 0.00012, 0.00044, 0.00134 and 0.00677. For $\Delta_{e}$, the same 
percentiles are: 0.00028, 0.00449, 0.01827, 0.06437, and 0.31803. And for $\Delta_{i}$: 0.00030, 0.00482, 0.01593, 
0.04076, and 0.17062. These values are higher than their equivalents for the inner main belt. Our analysis reveals that 
the most stable objects, those within the 1st percentile of the distributions, tend to occupy the regions with 2.0~au 
$<a<$ 2.5~au, and they could be as stable as their counterparts in the inner belt.
%
%
%
%

\subsection{Outer main belt}
\label{sec:3.3}
Following JPL's SBDB, outer main-belt asteroids have 3.2~au $<a<$ 4.6~au and the current membership of this orbit class 
includes 36,901 objects. This region hosts the Hildas (see for example \citealt{1998P&SS...46.1425F}), a concentration 
of asteroids trapped in the 3:2 mean-motion resonance with Jupiter (between 3.7~au and 4.2~au), the Cybele asteroids 
(see for example \citealt{2015MNRAS.451..244C}), in the 7:4 mean-motion resonance with Jupiter (between 3.27~au and 
3.7~au), and the Thule dynamical group (see for example \citealt{2008MNRAS.390..715B}) in the 4:3 mean-motion resonance 
with Jupiter (between 4.26~au and 4.3~au). At 3.27~au, we have the Hecuba gap (see for example 
\citealt{2002MNRAS.335..417R}). The dynamical space between the Cybele, Thule, and Hilda asteroidal populations is 
unstable because of the overlapping of numerous three-body mean-motion resonances, the so-called ``resonance sea" 
discussed by \citet{2006Icar..184...29G}.

\begin{figure*}
  \centering
  \includegraphics[width=0.49\linewidth]{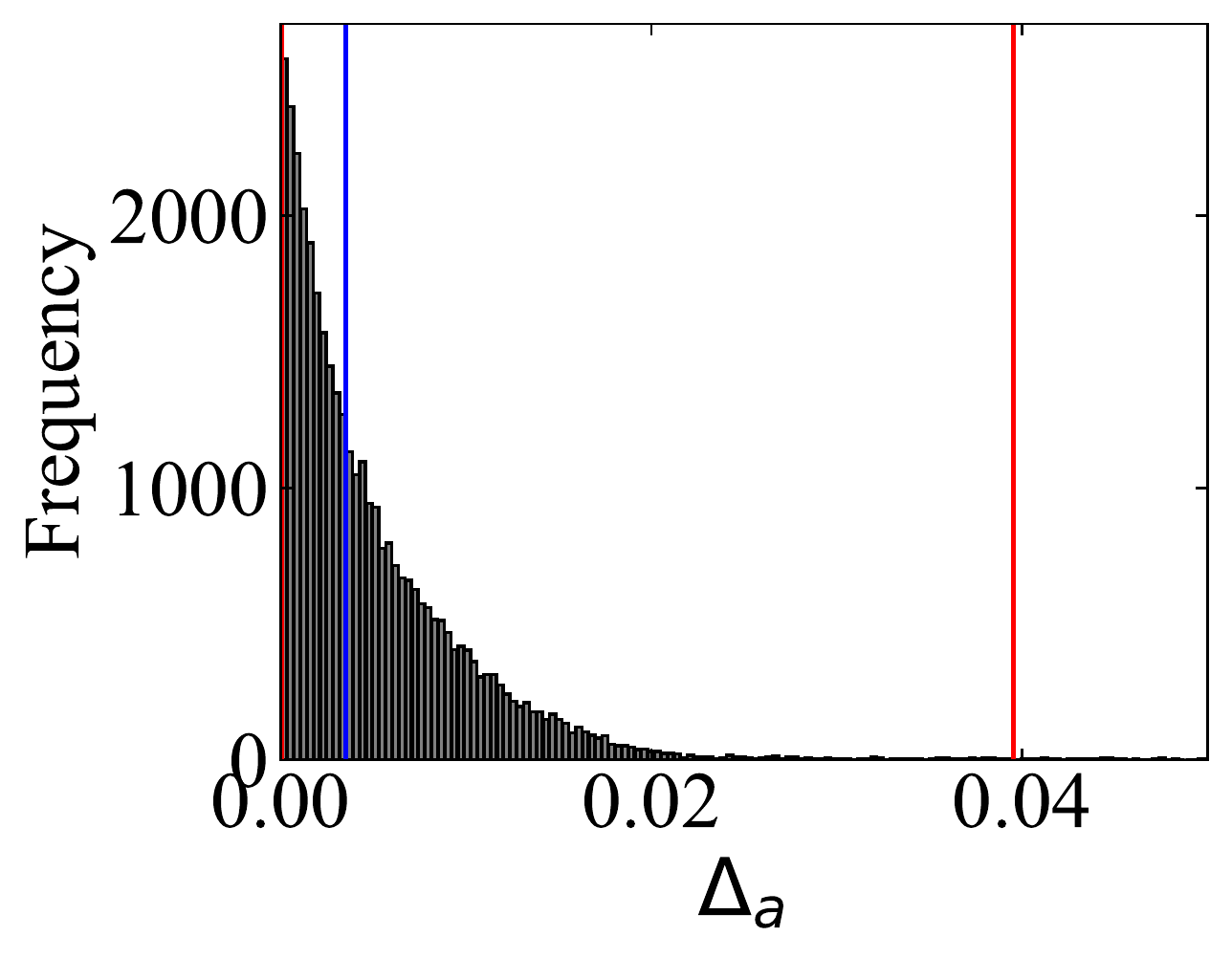}
  \includegraphics[width=0.49\linewidth]{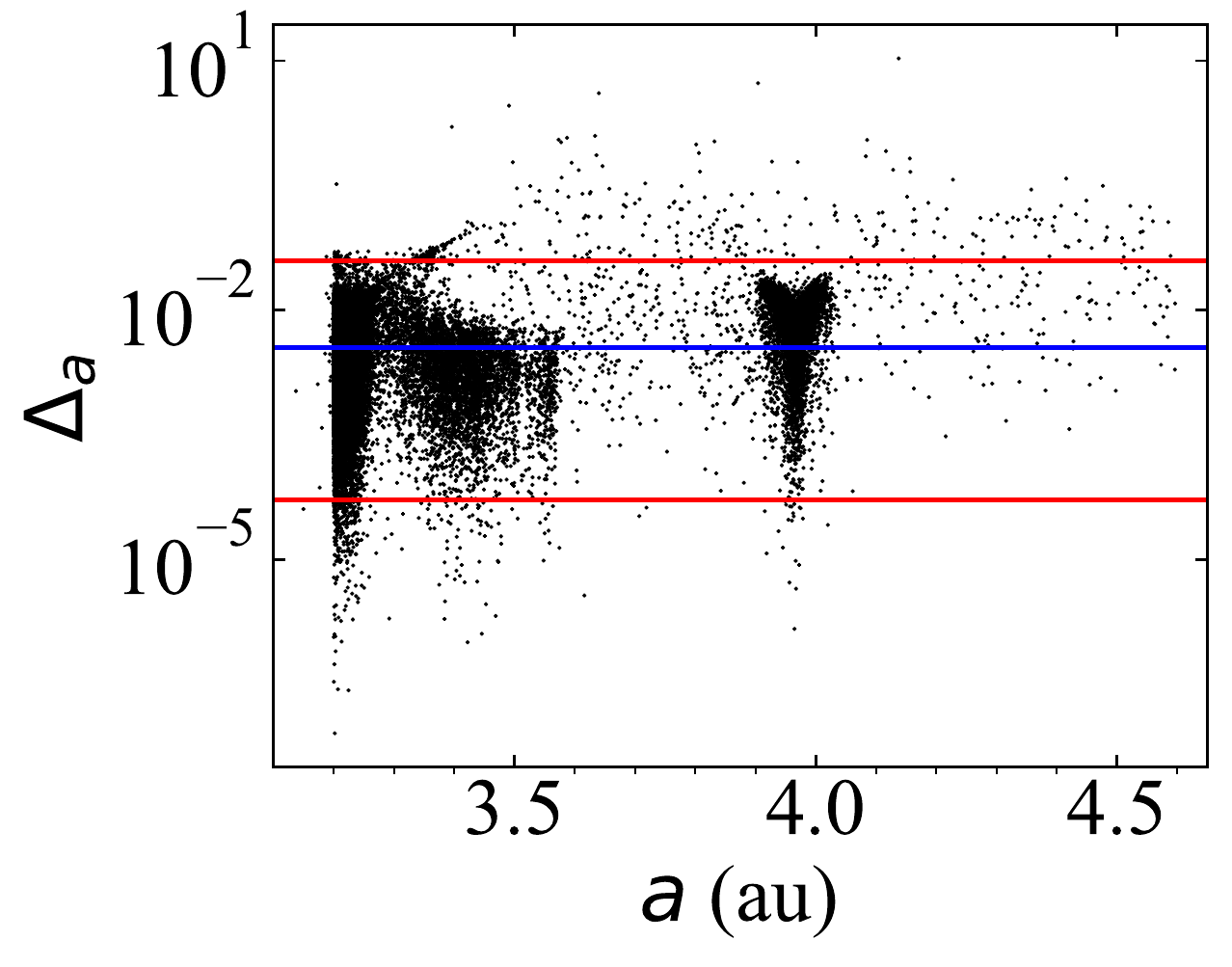}
  \includegraphics[width=0.49\linewidth]{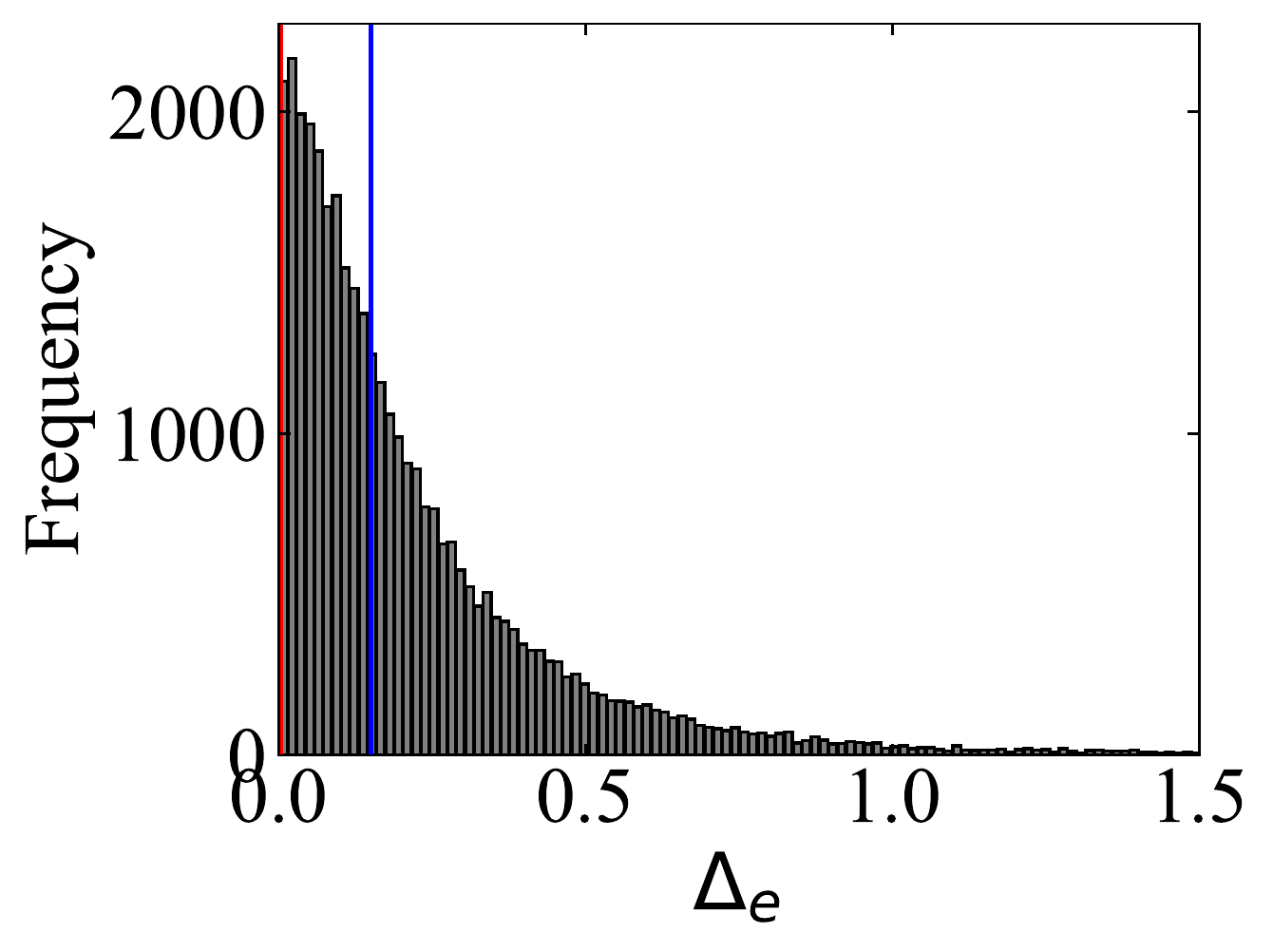}
  \includegraphics[width=0.49\linewidth]{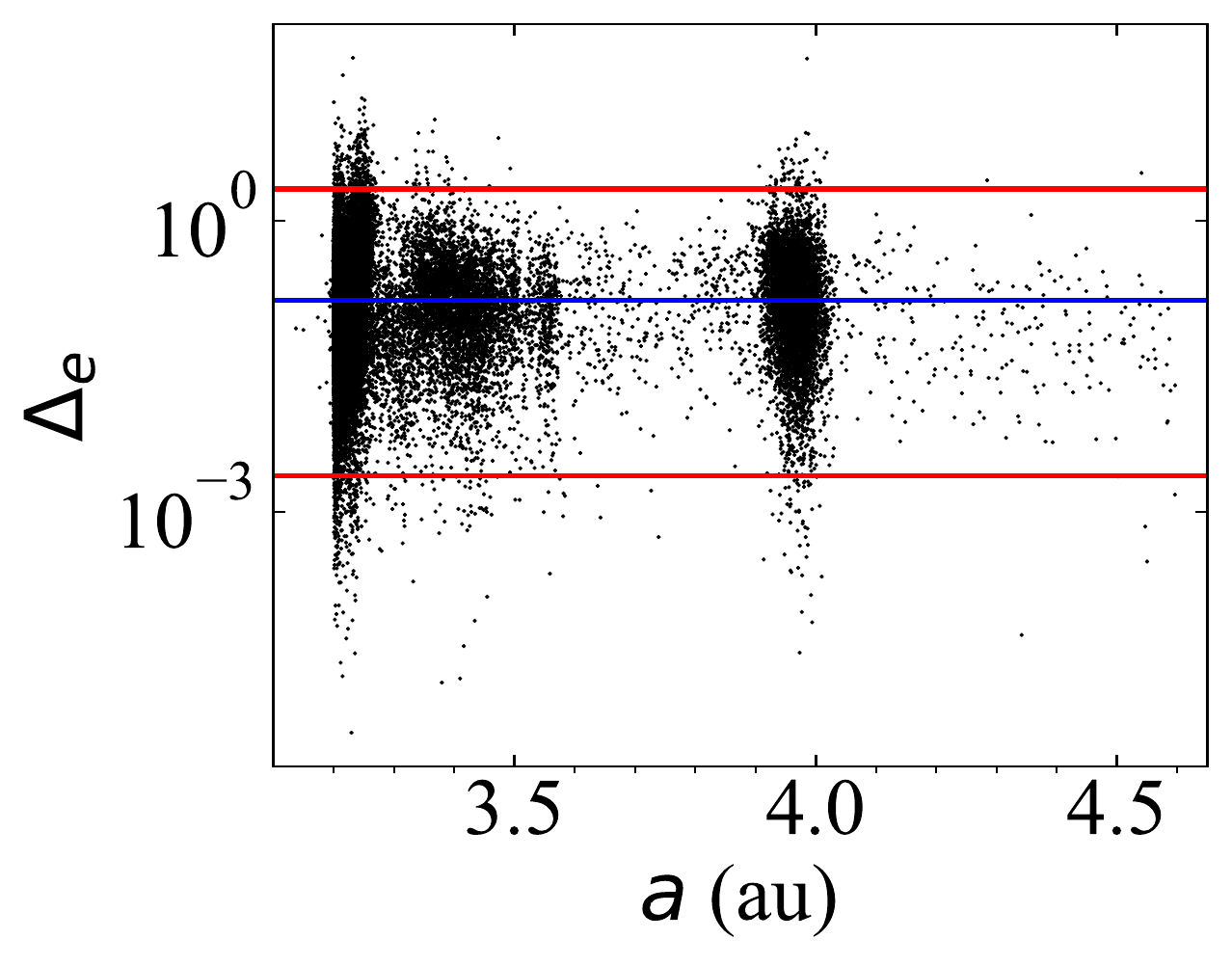}
  \includegraphics[width=0.49\linewidth]{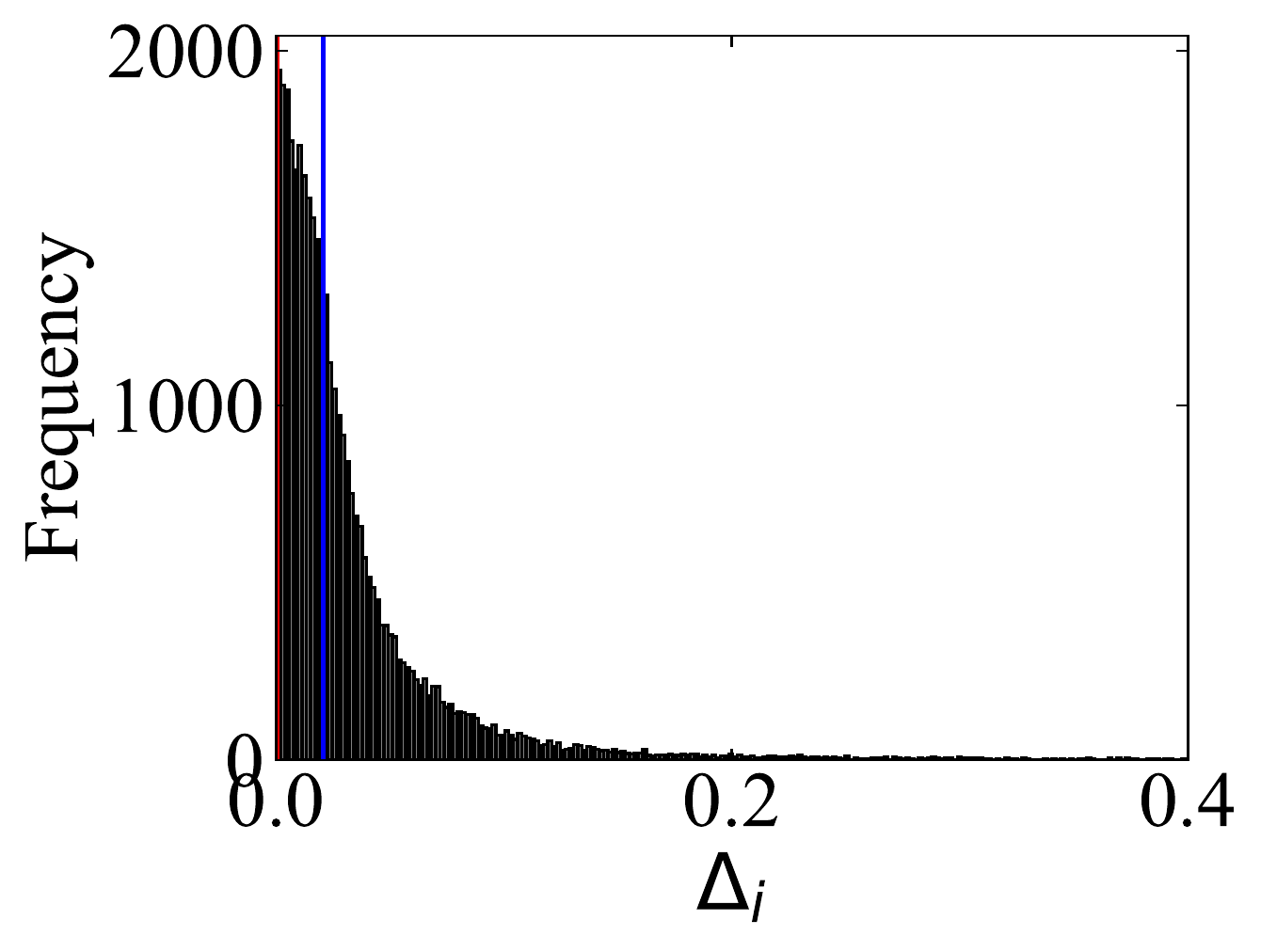}
  \includegraphics[width=0.49\linewidth]{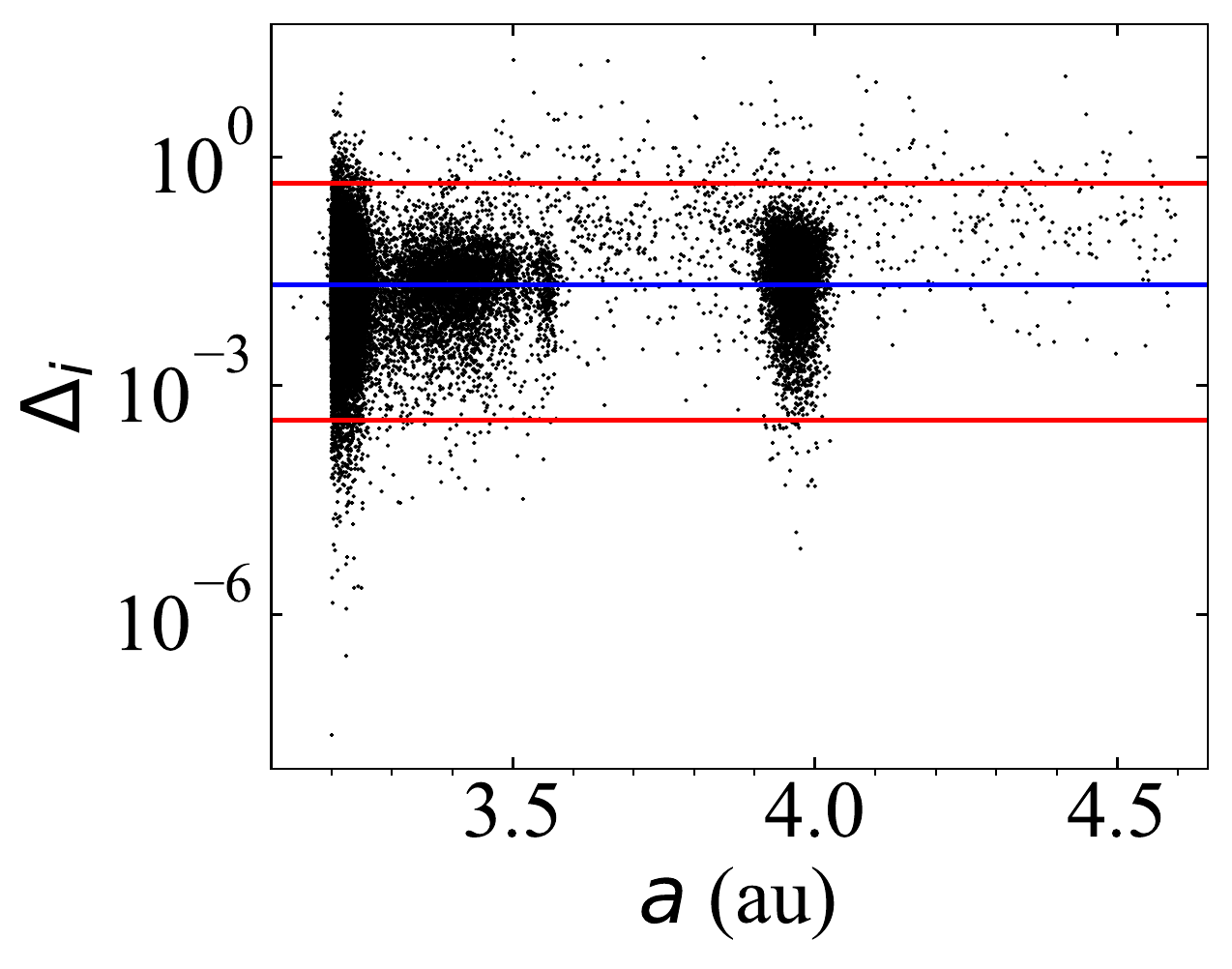}
\caption{Outer main belt (36,901 objects as of April 22, 2022). Absolute relative variation of the value of the orbital 
elements semimajor axis, eccentricity, and inclination: $\Delta_{a}$, $\Delta_{e}$ and $\Delta_{i}$. The left-hand side 
column of panels shows the distributions as frequency histograms; the right-hand side column of panels shows the 
distribution of absolute relative variations as a function of the semimajor axis in logarithmic scale. Median values are 
shown as vertical or horizontal blue lines and the 1st and 99th percentiles as red lines. Data source: JPL's 
{\tt Horizons}.}
\label{fig:outdist}       
\end{figure*}

Figure~\ref{fig:outdist} shows that this region is by far the most perturbed of the main asteroid belt. The values of 
the absolute relative variations of $a$, $e$, and $i$ are significantly higher than in other regions of the belt. For 
$\Delta_{a}$ the 1st, 16th, 50th, 84th, 99th percentiles are: 5.25$\times$10$^{-5}$, 0.00084, 0.00353, 0.00960 and 
0.03954. For $\Delta_{e}$, the same percentiles are: 0.00237, 0.04060, 0.15082, 0.42004, and 2.13023. And for 
$\Delta_{i}$: 0.00035, 0.00586, 0.02082, 0.05824, and 0.45447. Here, we find multiple objects with $\Delta_{a}$ well 
above the 99th percentile of the distribution. In the following section, we will focus on the subsample with 
$\Delta_{a}>0.6$ that we consider as highly likely to having arrived at the outer main belt in relatively recent times. 
It is however possible that some of such objects are trapped in secular resonances that may bring them back and forth 
into the belt from beyond Jupiter. If the orbit determinations are robust enough, numerical integrations should be able 
to confirm their true dynamical nature. 
%
%
%
%

\section{Recent arrivals}
\label{sec:4}
From the analyses in Sect.~\ref{sec:3}, we conclude that only the outer main belt hosts a population of minor bodies 
that may have entered the belt during the last few hundred years. Assuming a threshold of $\Delta_{a}>0.6$ as pointed 
out above, the list of recent arrivals in Table~\ref{tab:1} includes 2003~BM$_{1}$, 2007~RS$_{62}$, 457175 
(2008~GO$_{98}$), 2010~BG$_{18}$, 2010~JC$_{58}$, 2010~JV$_{52}$, 2010~KS$_{6}$, 2010~LD$_{74}$, 2010~OX$_{38}$, 
2011~QQ$_{99}$, 2013~HT$_{149}$, 2015~BH$_{103}$, 2015~BU$_{525}$, 2015~RO$_{127}$, 2015~RS$_{139}$, 2016~PC$_{41}$, 
2016~UU$_{231}$, 2020~SA$_{75}$, 2020~UO$_{43}$, and 2021~UJ$_{5}$.
%
%
\begin{table*}
\tabcolsep 0.14truecm
\caption{Recent arrivals to the outer main belt (absolute relative variation of the semimajor axis, $\Delta_{a}>0.6$). 
The data include number (if assigned by the MPC) and provisional designation, number of days spanned by the data arc, 
number of observations used to compute the orbit determination, semimajor axis, eccentricity, inclination (these 
three referred to epoch 2459600.5 JD TDB), absolute relative variation of $a$, $e$, and $i$. Data sources: JPL's SBDB 
and {\tt Horizons}.}
\label{tab:1}       
\begin{tabular}{rrrrrrrrr}
\hline\noalign{\smallskip}
object                   &  arc  & obs. & $a$     &  $e$    & $i$      & $\Delta_{a}$ & $\Delta_{e}$ & $\Delta_{i}$ \\
                         &  (d)  &      & (au)    &         & (\degr)  &             &             &                \\
\noalign{\smallskip}\hline\noalign{\smallskip}
        2003 BM$_{1}$    &  5140 &  90  & 3.64009 & 0.51436 & 11.34667 &  4.06475    & 0.43067     & 0.28590        \\
        2007 RS$_{62}$   &     1 &   9  & 3.49720 & 0.36435 &  0.61698 &  0.60015    & 0.97395     & 1.82065        \\
457175 (2008 GO$_{98}$)  &  6207 & 690  & 3.97017 & 0.27908 & 15.55621 &  0.60318    & 0.58341     & 0.03283        \\
        2010 BG$_{18}$   &     1 &  11  & 3.80613 & 0.25454 & 17.85287 &  0.77561    & 0.12023     & 0.35398        \\
        2010 JC$_{58}$   &     2 &  12  & 3.63599 & 0.37660 &  5.98705 &  0.73026    & 0.48396     & 0.46895        \\
        2010 JV$_{52}$   &     1 &   8  & 3.83195 & 0.21150 &  4.12106 &  1.07049    & 0.39983     & 0.74718        \\
        2010 KS$_{6}$    &     1 &  11  & 3.92736 & 0.24029 &  9.39648 &  0.61127    & 0.34709     & 0.29612        \\
        2010 LD$_{74}$   &     1 &  15  & 3.90426 & 0.32040 & 18.84908 &  5.35829    & 1.50536     & 0.66081        \\
        2010 OX$_{38}$   &     1 &  13  & 4.11663 & 0.26317 &  4.29817 &  0.81833    & 0.14387     & 0.40455        \\
        2011 QQ$_{99}$   &  8918 &  48  & 3.80163 & 0.42617 &  3.21247 &  0.98310    & 0.30898     & 3.38391        \\
        2013 HT$_{149}$  &     2 &  11  & 4.15614 & 0.18113 &  0.72244 &  0.66600    & 0.14155     & 5.98370        \\
        2015 BH$_{103}$  &     6 &  12  & 4.08323 & 0.18091 &  2.04186 &  0.70796    & 0.18405     & 1.07415        \\
        2015 BU$_{525}$  &     1 &  17  & 3.49100 & 0.51602 &  8.86846 &  2.87732    & 0.38818     & 0.07012        \\
        2015 RO$_{127}$  &     3 &  11  & 3.57322 & 0.36363 &  1.56916 &  1.12174    & 0.24970     & 0.37371        \\
        2015 RS$_{139}$  &     3 &   7  & 4.08529 & 0.32595 &  1.35669 &  1.11238    & 0.39090     & 7.32546        \\
        2016 PC$_{41}$   &     3 &  11  & 3.63385 & 0.39302 &  8.15787 &  1.24666    & 0.14065     & 0.74684        \\
        2016 UU$_{231}$  &     1 &  22  & 3.57738 & 0.34396 &  2.15470 &  1.04531    & 0.16766     & 0.53508        \\ 
        2020 SA$_{75}$   &    30 &  18  & 3.58730 & 0.53166 &  3.03217 &  1.18201    & 0.23644     & 3.66545        \\
        2020 UO$_{43}$   &   548 &  13  & 4.13775 & 0.64513 &  1.75827 & 10.65010    & 0.72835     & 1.23496        \\
        2021 UJ$_{5}$    &    68 &  51  & 3.39604 & 0.51475 &  9.38787 &  1.59925    & 0.23100     & 0.65926        \\
\noalign{\smallskip}\hline
\end{tabular}
\end{table*}
%
%

Unfortunately, most of the objects in Table~\ref{tab:1} have very short data arcs and, consistently, their orbit 
determinations are very uncertain. The objects with the highest probability of being newcomers are 2003~BM$_{1}$, 
2010~LD$_{74}$, 2015~BU$_{525}$ and 2020~UO$_{43}$. Asteroids 2003~BM$_{1}$ and 2020~UO$_{43}$ have robust orbit 
determinations (but see the comment on 2020~UO$_{43}$ later on), but 2010~LD$_{74}$ and 2015~BU$_{525}$ have very poor 
orbit determinations, based on observational arcs spanning just one day. Minor bodies 2003~BM$_{1}$, 2015~BU$_{525}$,
2020~SA$_{75}$, 2020~UO$_{43}$, and 2021~UJ$_{5}$ are included in the list of asteroids with comet-like orbits 
maintained by Y.~R. Fern\'andez.\footnote{\url{https://physics.ucf.edu/~yfernandez/lowtj.html}} None of these objects 
have been studied in detail yet and no cometary activity has ever been observed on any of them. 

In sharp contrast and among the rest of the sample, at least one object has been found to exhibit cometary activity,
457175 (2008 GO$_{98}$), and it has a dual designation as comet 362P \citep{2018P&SS..160...12G,2019Icar..317...44B,
2021AdSpR..67..639K}. From this group, only 457175, 2011~QQ$_{99}$, and 2021~UJ$_{5}$ (data arc of 68~d with 51 
observations) have robust orbit determinations. It is rather suspicious that out of 15 objects in Table~\ref{tab:1} with 
$0.6<\Delta_{a}<2$ just 20\% have sufficiently good orbit determinations, with 80\% having very short data arcs, mostly 
$\leq$3~d. The lack of recovery observations suggests that some of these objects may have been in outburst when 
discovered and their apparent magnitudes outside their active phases may be too low to enable casual recovery by 
non-targeted surveys such as those looking for TNOs or near-Earth objects (NEOs). A recent example of an object 
serendipitously discovered when in the active phase is Centaur 2020~MK$_{4}$ \citep{2021A&A...649A..85D}.   

\begin{figure}
  \centering
  \includegraphics[width=0.490\linewidth]{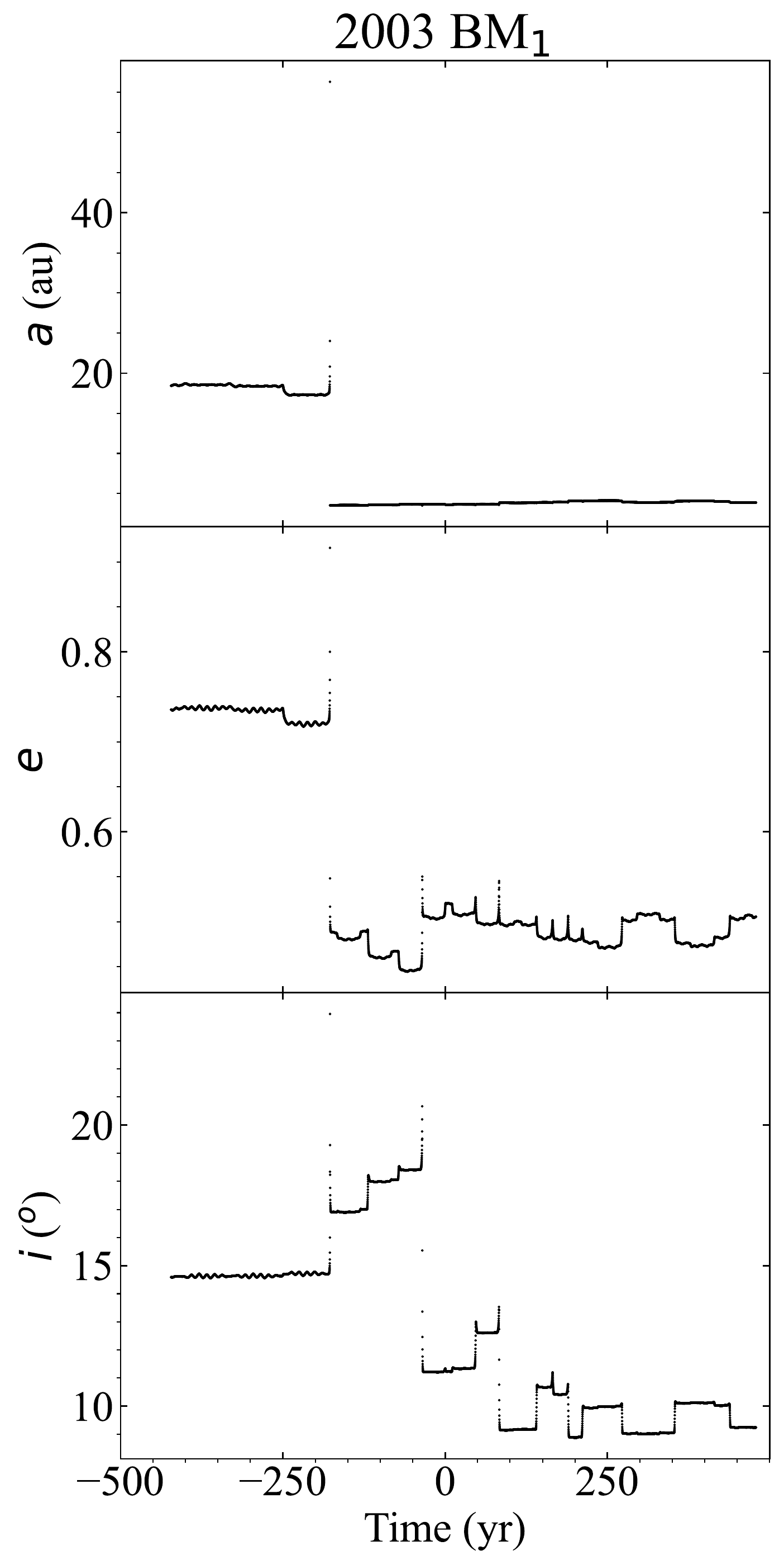}
  \includegraphics[width=0.500\linewidth]{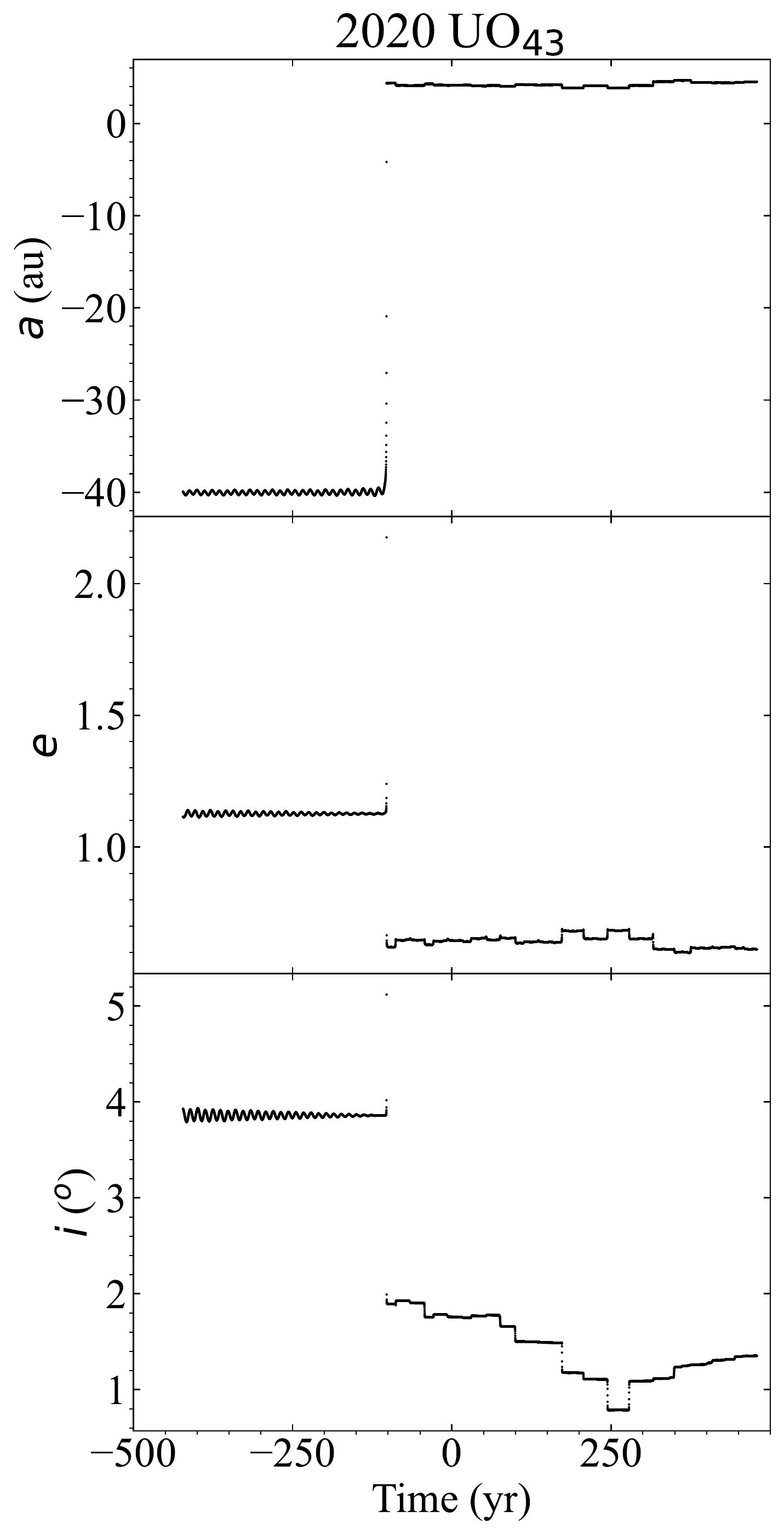}
  \caption{Evolution of the orbital elements semimajor axis (top panels), eccentricity (middle panels), and inclination 
           (bottom panels) for the nominal orbit of 2003~BM$_{1}$ (left-hand side panels) and 2020~UO$_{43}$ (right-hand 
           side panels). For 2020~UO$_{43}$, the semimajor axis and eccentricity are initially negative and larger than 
           1, respectively. These values are associated with unbound orbits. The origin of time is the epoch 2459600.5 
           JD Barycentric Dynamical Time (2022-Jan-21.0 00:00:00.0 TDB) and the output cadence is 30~d. The source of 
           the data is JPL's {\tt Horizons}. 
          }
  \label{fig:best2nom}
\end{figure}

Figure~\ref{fig:best2nom} shows the short-term past and future dynamical evolution of the nominal orbits of 
2003~BM$_{1}$ and 2020~UO$_{43}$ as computed by JPL's {\tt Horizons} (the ephemerides were retrieved using the 
{\tt Astroquery} package as pointed out above and plotted using {\tt Matplotlib}). The two objects exhibit a rather
chaotic evolution both in the recent past and into the immediate future. This is the result of planetary encounters at
close range (mainly with Jupiter and Saturn, see below). Figure~\ref{fig:best2nom}, left-hand side top panel, shows that 
2003~BM$_{1}$ came from the neighborhood of Uranus and arrived at its present location nearly 235~yr ago. However, the 
most striking case involves 2020~UO$_{43}$ that may have reached its current orbit coming from interstellar space. This 
object may have entered the Solar system nearly a century ago with an eccentricity of 1.1150, not too different from 
that of 1I/2017~U1~(`Oumuamua), 1.20113$\pm$0.00002, the first known interstellar body passing through the Solar system 
(see for example \citealt{2017RNAAS...1....9D,2017RNAAS...1....5D,2018Msngr.173...13H,2018Natur.559..223M,
2019NatAs...3..594O,2019ApJ...876L..26S}). 
%
%
%
%

\begin{figure}
  \centering
  \includegraphics[width=0.327\linewidth]{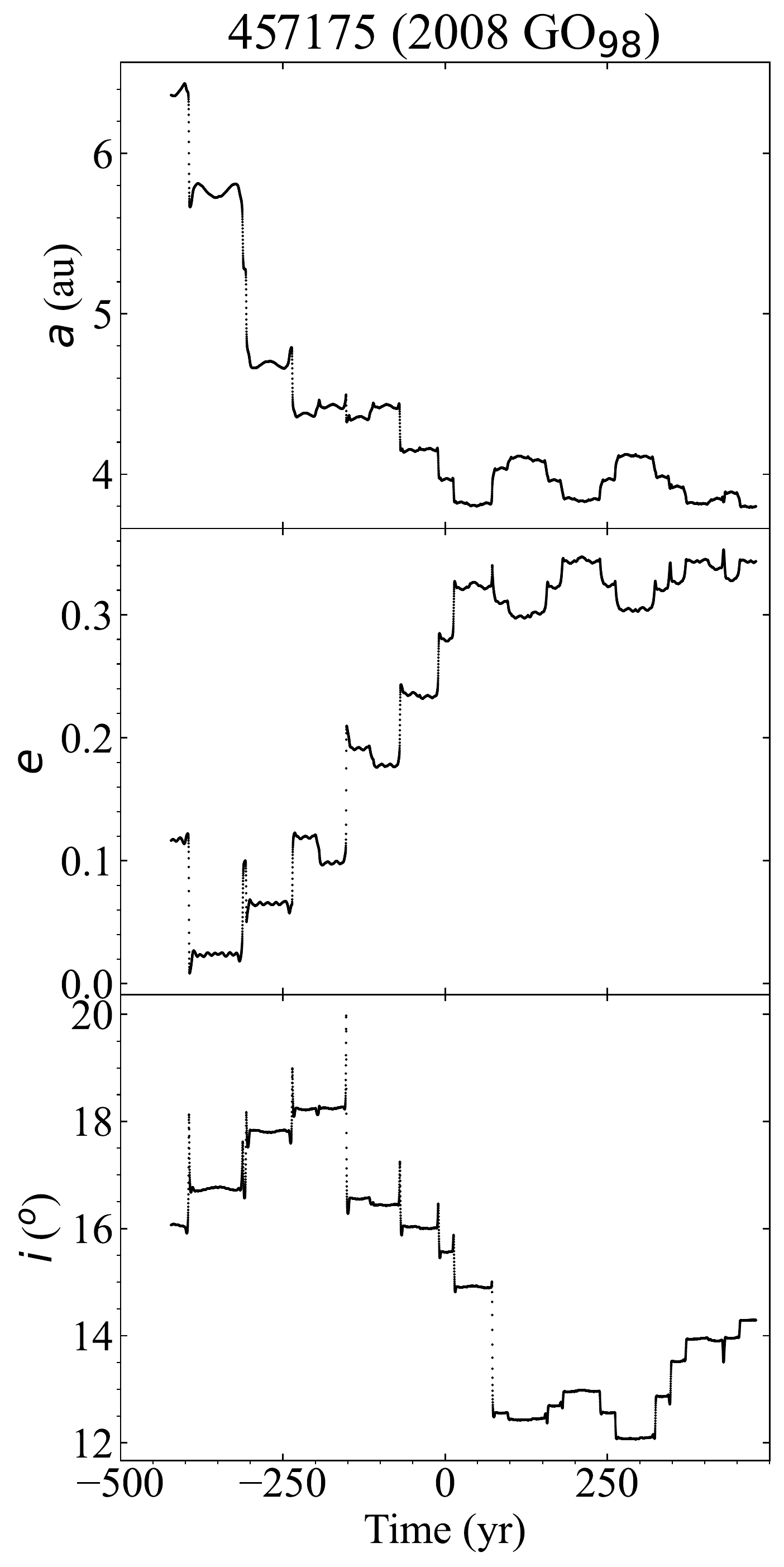}
  \includegraphics[width=0.327\linewidth]{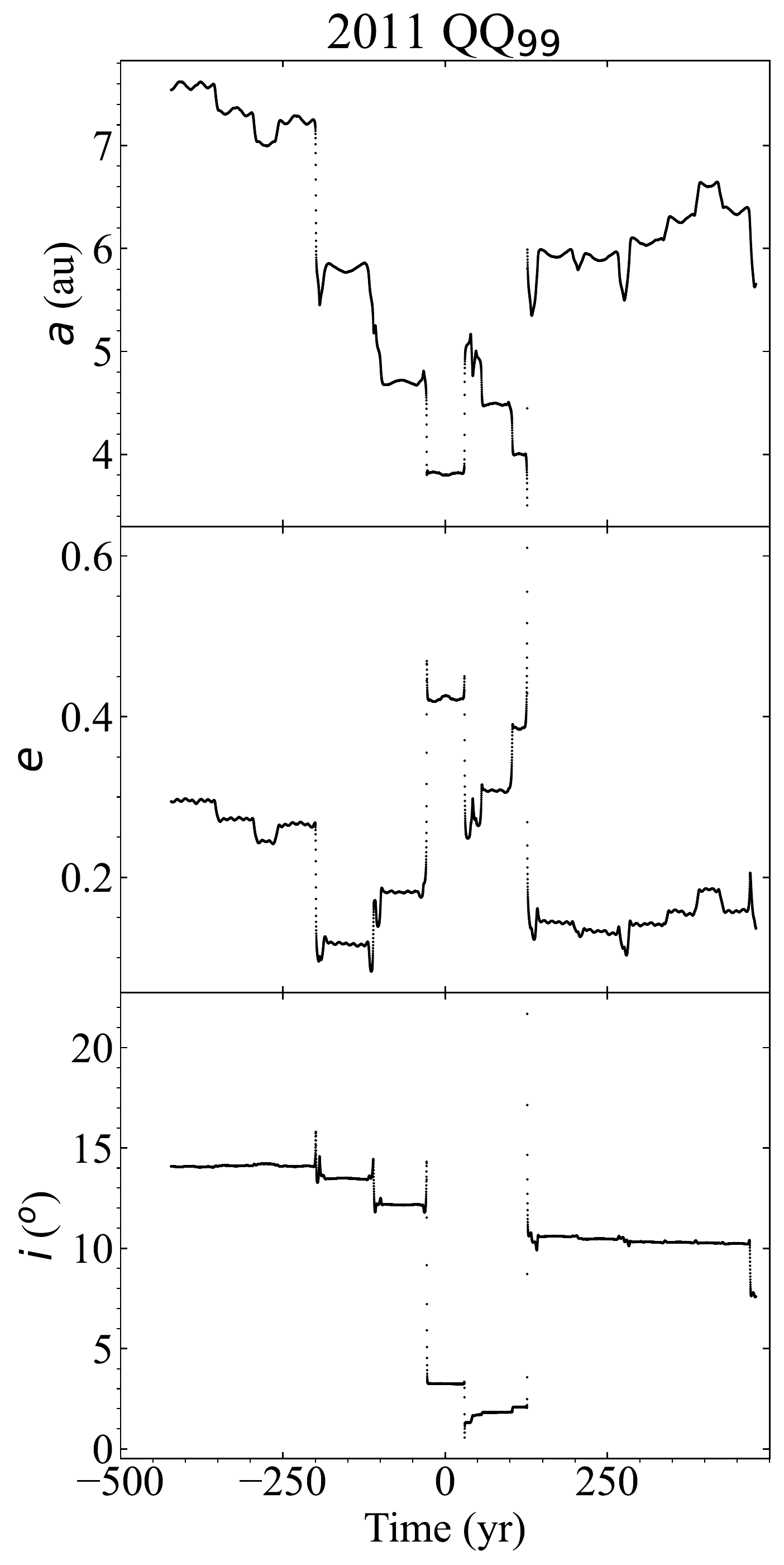}
  \includegraphics[width=0.327\linewidth]{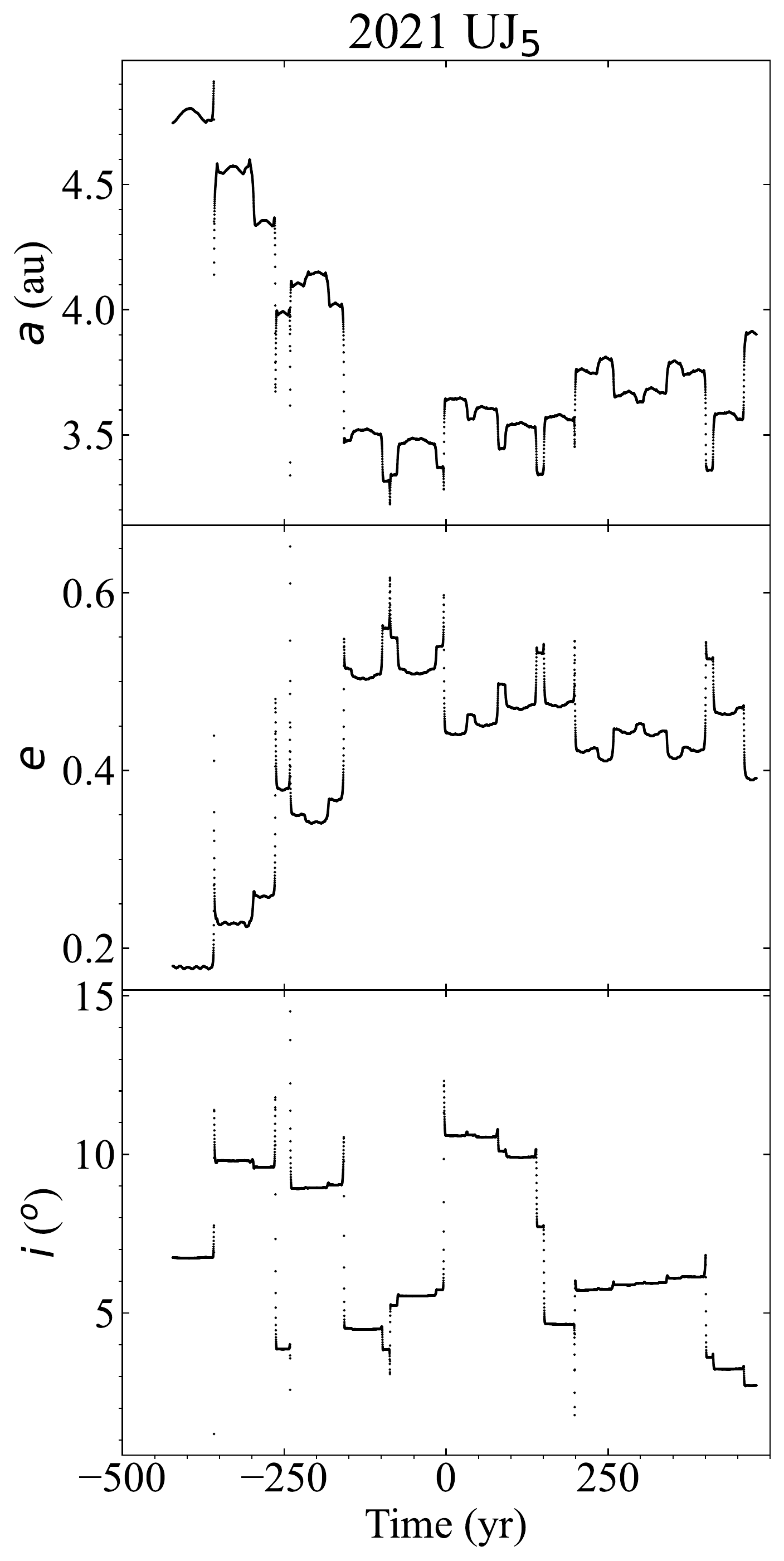}
  \caption{Evolution of the orbital elements semimajor axis (top panels), eccentricity (middle panels), and inclination 
           (bottom panels) for the nominal orbit of 457175 (2008~GO$_{98}$), left-hand side panels, 2011~QQ$_{99}$, 
           central panels, and 2021~UJ$_{5}$, right-hand side panels. The origin of time is the epoch 2459600.5 JD 
           Barycentric Dynamical Time (2022-Jan-21.0 00:00:00.0 TDB) and the output cadence is 30~d. The source of the 
           data is JPL's {\tt Horizons}. 
          }
  \label{fig:best3nom}
\end{figure}

The second group of newcomers, those with $\Delta_{a}<2$, may have come from Centaur orbital space (see for example 
\citealt{2007Icar..190..224D,2009Icar..203..155B,2009AJ....137.4296J,2019AJ....157..225W,2020CeMDA.132...36D,
2021Icar..35814201R}). Figure~\ref{fig:best3nom} shows the short-term past and future evolution of the three objects 
with the best orbit determinations of the sample: 457175, 2011~QQ$_{99}$, and 2021~UJ$_{5}$. The three objects have
chaotic dynamical histories, but 457175 appears to have followed a one-way path into the outer main belt; the other two 
objects seem to spend as much time inside the belt as outside of it and even beyond Jupiter's orbit (2011~QQ$_{99}$). 
The evolution of the relevant orbital elements of the remaining objects in Table~\ref{tab:1} is shown in 
Fig.~\ref{fig:others} of Appendix~\ref{app:3}. As in the previous case, the jumps in the values of the orbital 
parameters are the result of planetary encounters at close range (mainly with Jupiter and Saturn, see below).
%
%
%
%

In the following, we will study in more detail the past dynamical evolution of some of these objects. We will leave 
outside of our analyses those objects with short data arcs in Table~\ref{tab:1} and 457175 because its orbital evolution 
is affected by the non-gravitational force caused by outgassing. The orbit determination of 457175 available from JPL's 
SBDB includes the usual orbital elements ---$a$, $e$, $i$, longitude of the ascending node, $\Omega$, argument of 
perihelion, $\omega$, and time of perihelion passage, $\tau$--- and three non-gravitational parameters associated with 
the non-gravitational radial, transverse, and normal accelerations.

\subsection{2003~BM$_{1}$}
\label{sec:4.3}
This object was first observed on January 24, 2003 by the Near-Earth Asteroid Tracking (NEAT, 
\citealt{1997NYASA.822....6H,1999AJ....117.1616P}) program at Palomar Mountain \citep{2003MPEC....B...29L}. Its orbit 
determination is based on 90 observations with a data-arc span of 5140~d (see Table~\ref{elementsBM1} in 
Appendix~\ref{app:1}). Asteroid 2003~BM$_{1}$ follows a rather eccentric path that approaches Mars, but it does not 
cross its orbit, reaching aphelion beyond Jupiter, which is the only planet that can directly perturb its trajectory. 
The value of its semimajor axis, 3.64~au, places it in the orbital realm of the Cybele asteroids, but it is not one of 
them. Figure~\ref{fig:best2nom}, left-hand side panels, shows that the nominal orbit led 2003~BM$_{1}$ into the outer 
belt from the neighborhood of Uranus. \citet{2008A&A...487..363D} included this object in their study of asteroids in 
cometary orbits concluding that some of those objects are not dormant comet candidates from the Jupiter family ---those 
with Tisserand's parameter, $T_{\rm J}$ \citep{1999ssd..book.....M}, in the range 2--3 and orbital periods under 20~yr 
or 200~yr, depending on the authors--- but asteroids that reached their current orbits as a result of perturbations. The 
Tisserand parameter, which is a quasi-invariant, is given by the expression:
\begin{equation}
   T_{\rm J} = \frac{a_{\rm J}}{a} + 2 \ \cos{i} \ \sqrt{\frac{a}{a_{\rm J}} \ (1 - e^{2})} \,, \label{Tisserand}
\end{equation}
where $a$, $e$, and $i$ are the semimajor axis, eccentricity and inclination of the orbit of the small body under study,
and $a_{\rm J}$ is the semimajor axis of the orbit of Jupiter \citep{1999ssd..book.....M}.  

\begin{figure}
   \centering
   \includegraphics[width=\linewidth]{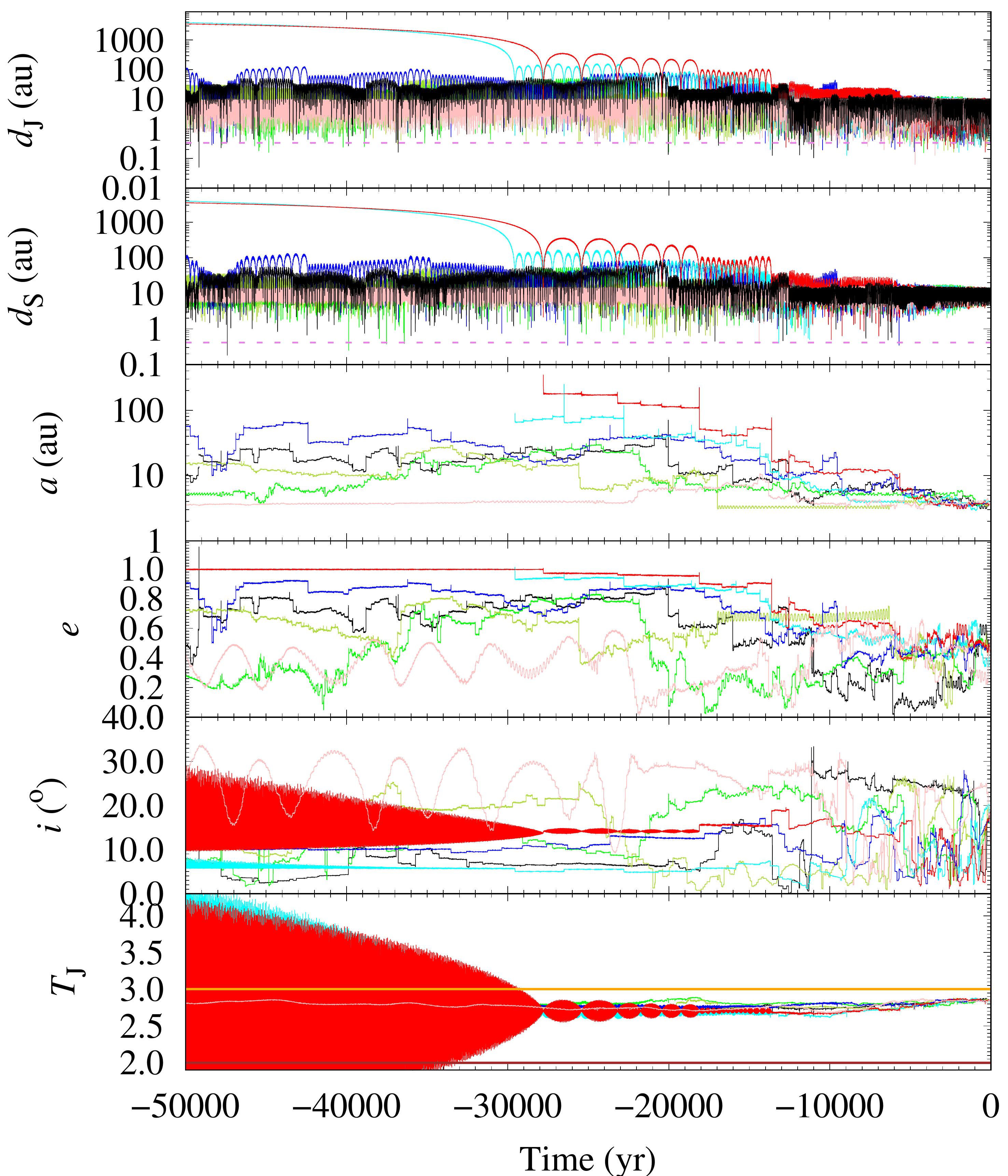}
   \caption{Short-term past evolution of relevant parameters of 2003~BM$_{1}$. We show the evolution of the distance to
            Jupiter (top panel) and Saturn (second to top) of the nominal orbit (in black) as described by the 
            corresponding orbit determination in Table~\ref{elementsBM1} of Appendix~\ref{app:1} and those of control 
            orbits or clones with Cartesian vectors separated $+$3$\sigma$ (in green), $-$3$\sigma$ (in lime), 
            $+$6$\sigma$ (in blue), $-$6$\sigma$ (in cyan), $+$9$\sigma$ (in red), and $-$9$\sigma$ (in pink) from the 
            nominal values in Table~\ref{vectorBM1} of Appendix~\ref{app:2}. The Hill radii of Jupiter, 0.338~au, and 
            Saturn, 0.412~au, are shown in red. The third to top panel shows the evolution of the semimajor axis, $a$. 
            The third to bottom panel shows the evolution of the eccentricity, $e$. The second to bottom panel displays 
            the inclination, $i$. The bottom panel shows the variations of the Tisserand's parameter, $T_{\rm J}$ 
            \citep{1999ssd..book.....M}, and includes the boundary references 2 (in brown) and 3 (in orange). The output 
            time-step size is 1~yr, the origin of time is epoch 2459600.5 TDB. The source of the input data is JPL's 
            {\tt Horizons}.
           }
   \label{fig:evolutionBM1}
\end{figure}

Figure~\ref{fig:evolutionBM1} shows the results of $N$-body integrations backward in time for the nominal orbit and 
representative control or clone orbits of 2003~BM$_{1}$ with Cartesian vectors separated $+$3$\sigma$ (in green), 
$-$3$\sigma$ (in lime), $+$6$\sigma$ (in blue), $-$6$\sigma$ (in cyan), $+$9$\sigma$ (in red), and $-$9$\sigma$ (in 
pink) from the nominal values in Table~\ref{vectorBM1} of Appendix~\ref{app:2}. These calculations have been carried out 
as described in Sect.~\ref{sec:2}. Our results are indicative of a very chaotic dynamical past driven by very frequent 
encounters with Jupiter but also with Saturn, inside the Hill radii of both planets. The evolution is so unstable that 
some control orbits in Fig.~\ref{fig:evolutionBM1} led to ejections from the Solar system ($-$6$\sigma$ control orbit in 
cyan and $+$9$\sigma$ in red); in other words, 2003~BM$_{1}$ has a small probability of having an origin in the Oort 
cloud and perhaps even in interstellar space. However, our calculations suggest that its most probable source is in the 
region between the orbits of Jupiter and Neptune, the Centaur orbital domain. The $-$9$\sigma$ control orbit (in pink) 
shows a capture in a von Zeipel-Lidov-Kozai secular resonance with anti-correlated eccentricity-inclination 
oscillations. 
%
%
%
%

\subsection{2020~UO$_{43}$}
\label{sec:4.2}
This object was first observed on October 20, 2020, by the Pan-STARRS~1 \citep{2004SPIE.5489...11K} telescope system at 
Haleakala. Its orbit determination is based on 13 observations with a data-arc span of 548~d (see 
Table~\ref{elementsUO43} in Appendix~\ref{app:1}). Asteroid 2020~UO$_{43}$ follows a quite eccentric orbit that crosses 
that of Mars, reaching aphelion well beyond Jupiter. The value of its semimajor axis, 4.14~au places it within the 
orbital realm of the Hilda asteroids, but it is not one of them. Figure~\ref{fig:best2nom}, right-hand side panels, 
shows that the nominal orbit brought 2020~UO$_{43}$ from interstellar space into the outer belt. 

\begin{figure}
  \centering
  \includegraphics[width=\linewidth]{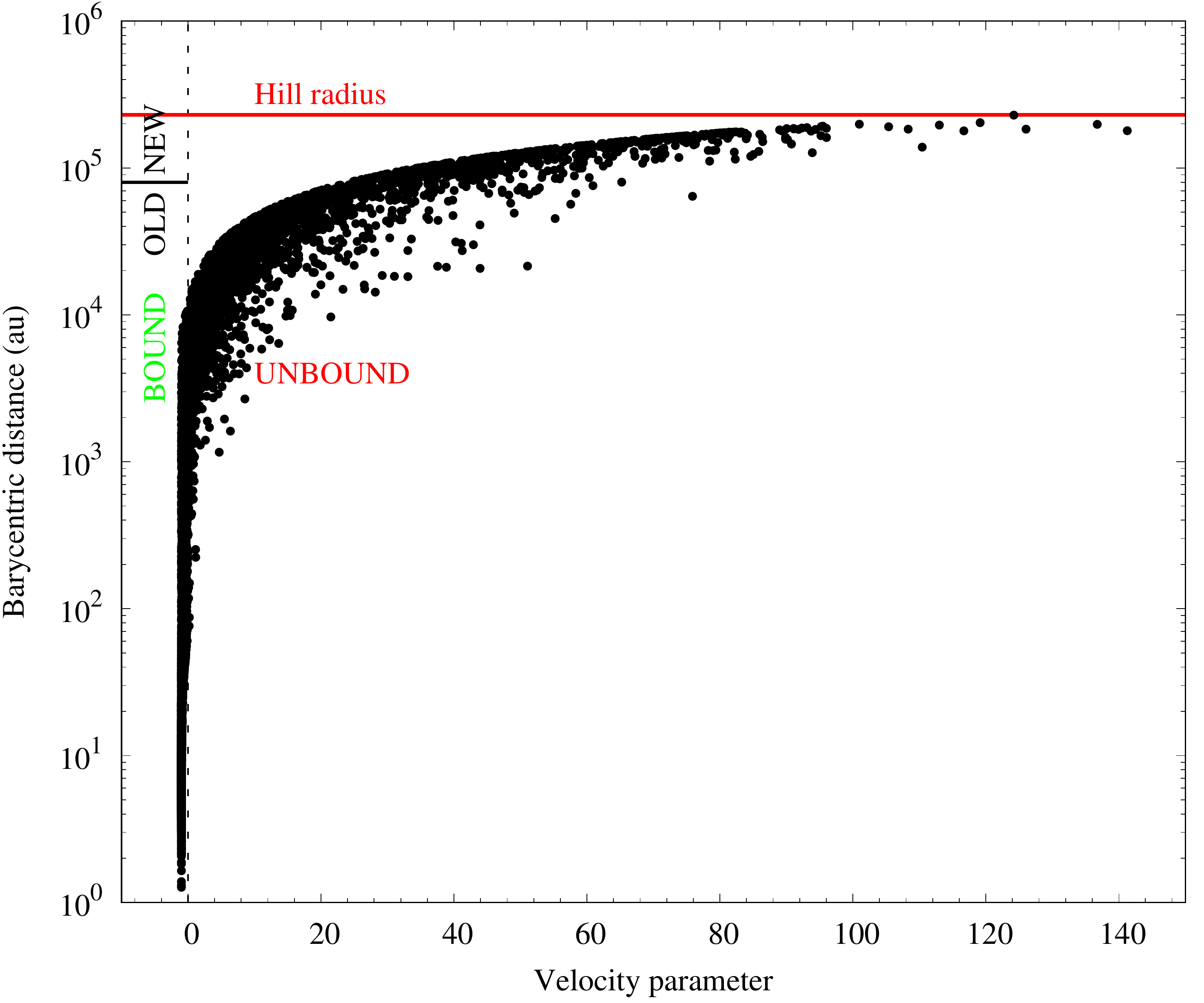}
  \caption{Values of the barycentric distance as a function of the velocity parameter 10$^{5}$~yr into the past for 
           8$\times$10$^{3}$ control orbits of 2020~UO$_{43}$ generated using the MCCM approach. The velocity parameter 
           is the difference between the barycentric and escape velocities at the computed barycentric distance in units 
           of the escape velocity. Positive values of the velocity parameter are associated with control orbits that 
           could be the result of capture. The thick black line corresponds to the aphelion distance ---$a \ (1 + e)$, 
           limiting case $e=1$--- that defines the domain of dynamically old comets with 
           {$a^{-1}>2.5\times10^{-5}$~au$^{-1}$} (see \citealt{2017MNRAS.472.4634K}); the thick red line signals the 
           radius of the Hill sphere of the Solar system (see for example \citealt{1965SvA.....8..787C}).
          }
  \label{originUO43}
\end{figure}

To understand its possible origin better, we performed integrations backward in time using MCCM to generate control 
orbits and found that the probability of having this object captured from interstellar space during the last 10$^{5}$~yr 
is 0.28$\pm$0.05 (average and standard deviation of 8$\times$10$^{3}$ experiments). Figure~\ref{originUO43} shows the 
results of these simulations. Most control orbits led to barycentric distances with values below the aphelion distance 
that defines the domain of dynamically old Oort cloud comets (see \citealt{2017MNRAS.472.4634K}). The most 
straightforward interpretation of these results is that 2020~UO$_{43}$ may have not arrived from interstellar space: It 
could be a dynamically old object instead, with a likely origin in the Solar system. However, an origin outside the 
Solar system cannot be rejected with the current orbit determination. In fact, this object might not be real but the 
result of bad linkage by the MPC.\footnote{The 2019 Catalina Sky Survey observations are of 108761 (2001~OK$_{46}$); the 
2020 observations do not correspond to any known object, but are vague enough to have a meaningless orbit and make it 
effectively unrecoverable (Deen 2022, private communication).}
%
%
%
%

\subsection{210718 (2000 ST$_{252}$)}
\label{sec:4.3}
\begin{figure}
  \centering
  \includegraphics[width=\linewidth]{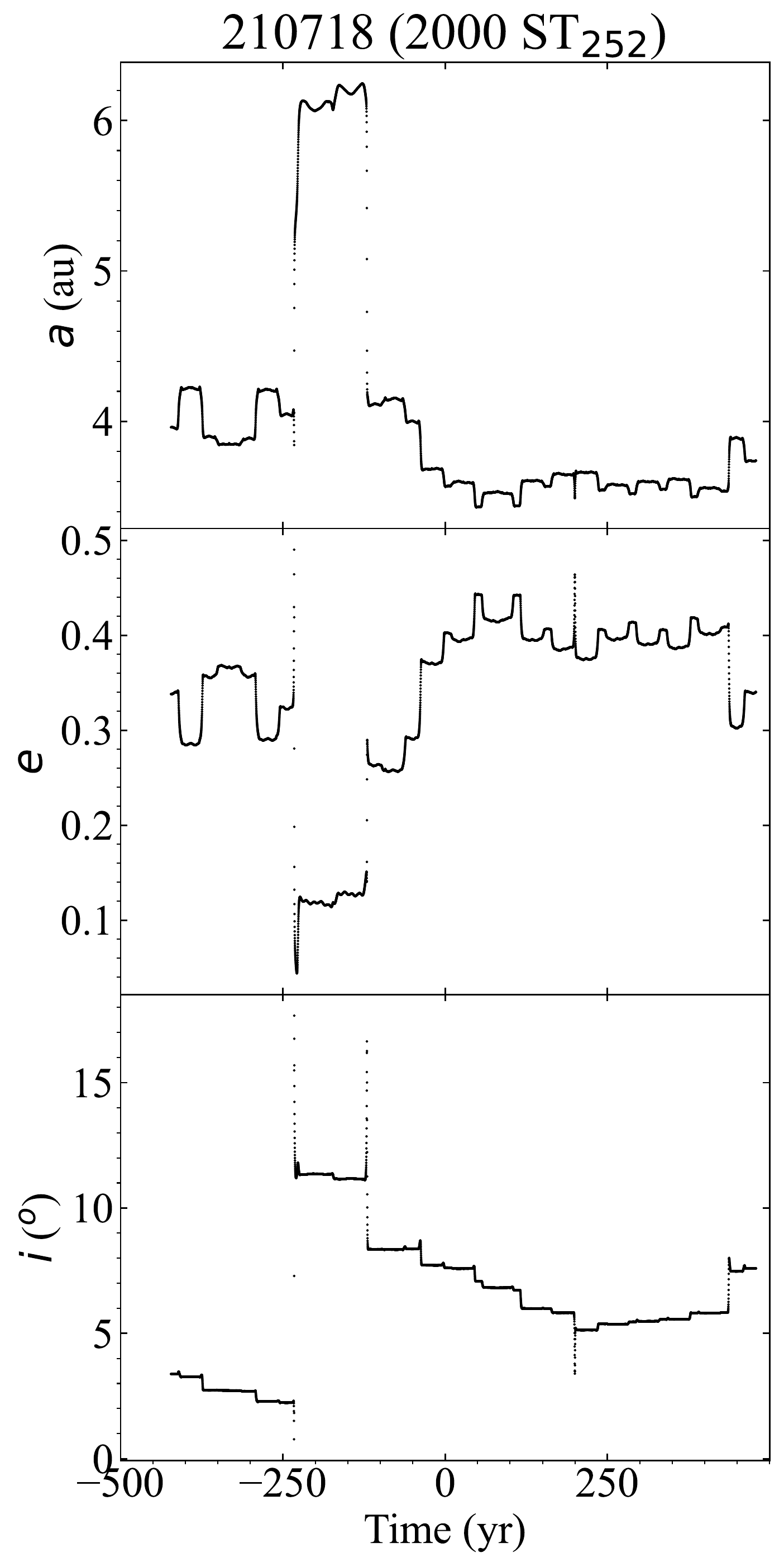}
  \caption{Evolution of the orbital elements semimajor axis (top panels), eccentricity (middle panels), and inclination 
           (bottom panels) for the nominal orbit of 210718 (2000 ST$_{252}$) that experiences a short-term excursion 
           well outside the outer asteroid belt within the explored time frame. The origin of time is the epoch 
           2459600.5 JD Barycentric Dynamical Time (2022-Jan-21.0 00:00:00.0 TDB) and the output cadence is 30~d. The 
           source of the data is JPL's {\tt Horizons}. 
          }
  \label{fig:best1nom}
\end{figure}

Asteroid 210718 (2000 ST$_{252}$) is not in Table~\ref{tab:1}, but it is included here because it experienced a 
short-term excursion well outside the outer asteroid belt within the explored time frame. This object was first observed 
on September 24, 2000, by the Lincoln Near-Earth Asteroid Research (LINEAR) project \citep{2000Icar..148...21S} from 
Socorro, New Mexico. Its orbit determination is based on 584 observations with a data-arc span of 7822~d (see 
Table~\ref{elementsST252} in Appendix~\ref{app:1}). Asteroid 210718 follows a moderately eccentric orbit that never gets 
close to Mars, reaching aphelion inside Jupiter's orbit. The value of its semimajor axis, 3.61~au places it in the 
orbital realm of the Cybele asteroids, but it is not one of them. Figure~\ref{fig:best1nom}, shows that the nominal 
orbit led 210718 into the outer belt from beyond Jupiter, but 400~yr ago it was part of the outer asteroid belt.
%
%
%
%

\begin{figure}
   \centering
   \includegraphics[width=\linewidth]{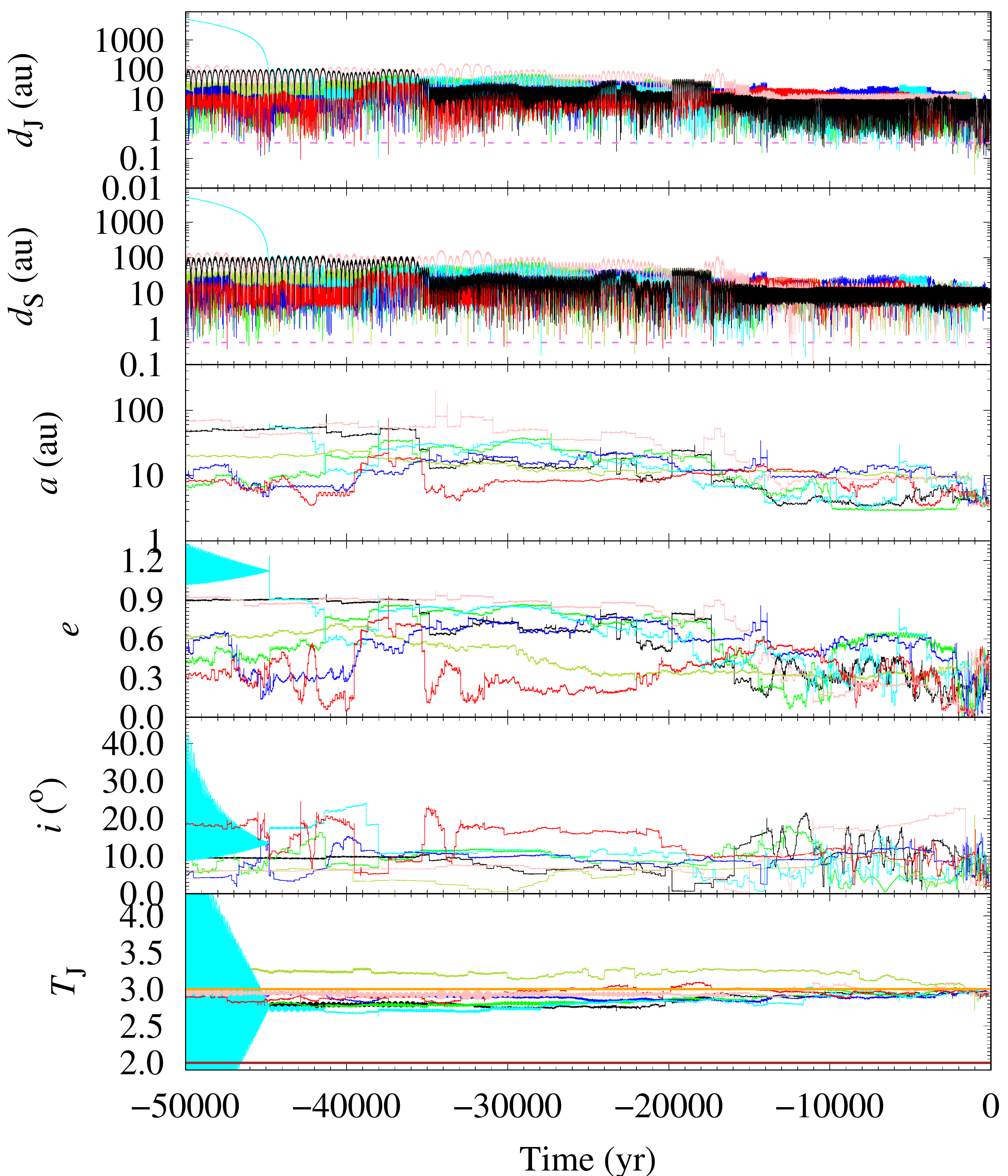}
   \caption{Short-term past evolution of relevant parameters of 210718 (2000~ST$_{252}$). We show the evolution of the 
            distance to Jupiter (top panel) and Saturn (second to top) of the nominal orbit (in black) as described by 
            the corresponding orbit determination in Table~\ref{elementsST252} of Appendix~\ref{app:1} and those of 
            control orbits or clones with Cartesian vectors separated $+$3$\sigma$ (in green), $-$3$\sigma$ (in lime), 
            $+$6$\sigma$ (in blue), $-$6$\sigma$ (in cyan), $+$9$\sigma$ (in red), and $-$9$\sigma$ (in pink) from the 
            nominal values in Table~\ref{vectorST252} of Appendix~\ref{app:2}. The Hill radii of Jupiter, 0.338~au, and 
            Saturn, 0.412~au, are shown in red. The third to top panel shows the evolution of the semimajor axis, $a$. 
            The third to bottom panel shows the evolution of the eccentricity, $e$. The second to bottom panel displays 
            the inclination, $i$. The bottom panel shows the variations of the Tisserand's parameter, $T_{\rm J}$ 
            \citep{1999ssd..book.....M}, and includes the boundary references 2 (in brown) and 3 (in orange). The output 
            time-step size is 1~yr, the origin of time is epoch 2459600.5 TDB. The source of the input data is JPL's 
            {\tt Horizons}.
           }
   \label{fig:evolutionST252}
\end{figure}

Figure~\ref{fig:evolutionST252} shows the result of $N$-body integrations backward in time for the nominal orbit and
representative control orbits of 210718. As in the case of 2003~BM$_{1}$, our results show a very chaotic dynamical past 
driven by frequent encounters with Jupiter and Saturn, inside the Hill radii of both planets. The evolution is quite 
unstable and one control orbit in Fig.~\ref{fig:evolutionST252} led to an ejection from the Solar system ($-$6$\sigma$ 
control orbit in cyan); therefore, 210718 has a very small probability of having an origin in the Oort cloud and
perhaps even in interstellar space. However, our calculations suggest that its most probable source, like in the case of 
2003~BM$_{1}$, is in the region between the orbits of Jupiter and Neptune, the Centaur orbital domain. 
%
%
%
%

\subsection{2011 QQ$_{99}$}
\label{sec:4.4}
This object was first observed on September 8, 1996, at the Steward Observatory in Kitt Peak and assigned the 
provisional designation 1996~RR$_{10}$. It was rediscovered on August 23, 2011, by the Pan-STARRS~1 
\citep{2004SPIE.5489...11K} telescope system at Haleakala and assigned the provisional designation 2011~QQ$_{99}$ to 
being later recognized as the former 1996~RR$_{10}$. Its orbit determination is based on 48 observations with a 
data-arc span of 8918~d (see Table~\ref{elementsQQ99} in Appendix~\ref{app:1}). Asteroid 2011~QQ$_{99}$ follows a 
moderately eccentric orbit that never approaches Mars, reaching aphelion beyond Jupiter that is the only planet that can 
directly perturb its path. The value of its semimajor axis, 3.80~au places it within the orbital parameter space of the 
Hilda asteroids, but it is not one of them. Figure~\ref{fig:best3nom}, second-to-right-hand side panels, shows that the 
nominal orbit led 2011 QQ$_{99}$ in and out of the outer belt. 

\begin{figure}
   \centering
   \includegraphics[width=\linewidth]{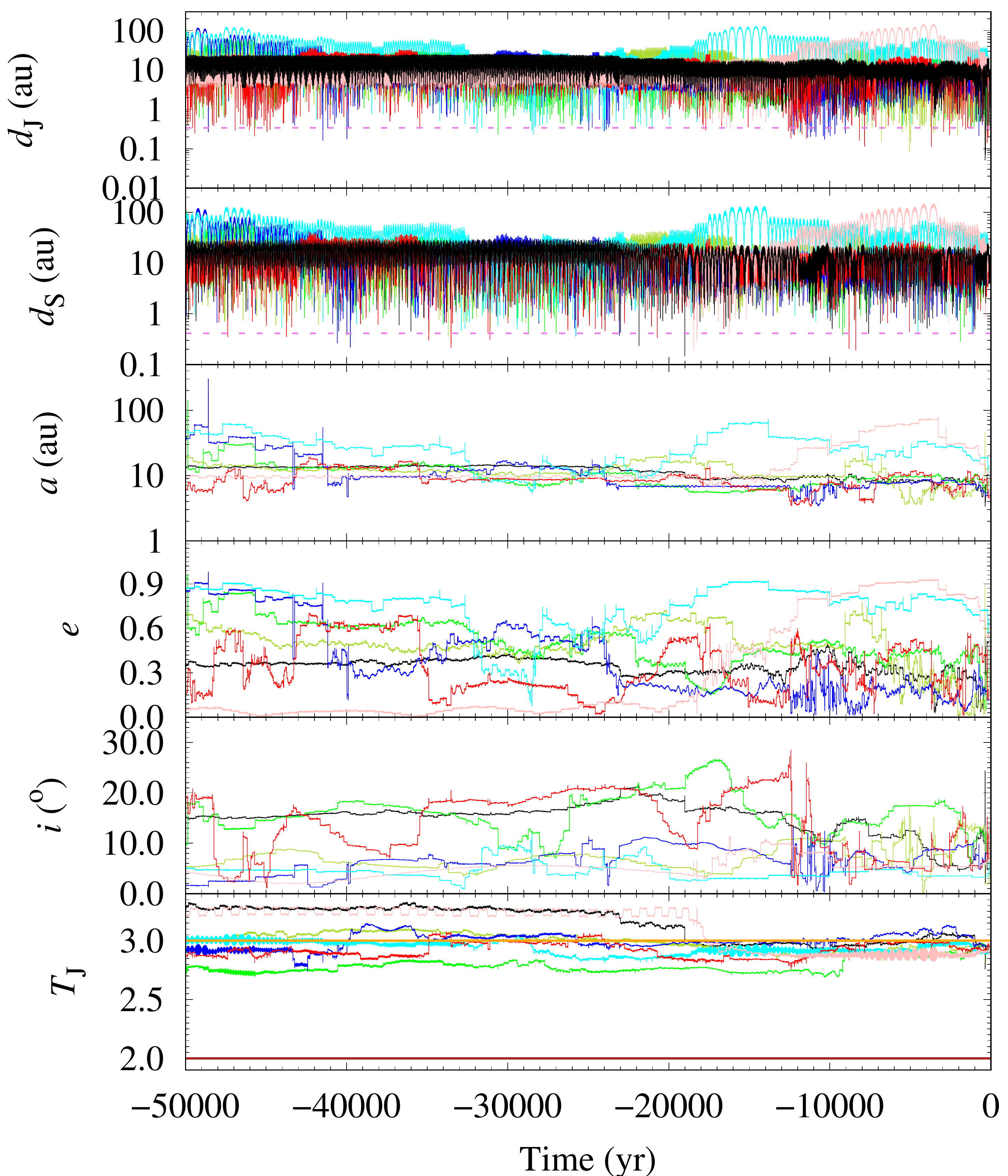}
   \caption{Short-term past evolution of relevant parameters of 2011~QQ$_{99}$. We show the evolution of the distance to 
            Jupiter (top panel) and Saturn (second to top) of the nominal orbit (in black) as described by the 
            corresponding orbit determination in Table~\ref{elementsQQ99} of Appendix~\ref{app:1} and those of control 
            orbits or clones with Cartesian vectors separated $+$3$\sigma$ (in green), $-$3$\sigma$ (in lime), 
            $+$6$\sigma$ (in blue), $-$6$\sigma$ (in cyan), $+$9$\sigma$ (in red), and $-$9$\sigma$ (in pink) from the 
            nominal values in Table~\ref{vectorQQ99} of Appendix~\ref{app:2}. The Hill radii of Jupiter, 0.338~au, and 
            Saturn, 0.412~au, are shown in red. The third to top panel shows the evolution of the semimajor axis, $a$. 
            The third to bottom panel shows the evolution of the eccentricity, $e$. The second to bottom panel displays 
            the inclination, $i$. The bottom panel shows the variations of the Tisserand's parameter, $T_{\rm J}$ 
            \citep{1999ssd..book.....M}, and includes the boundary references 2 (in brown) and 3 (in orange). The output 
            time-step size is 1~yr, the origin of time is epoch 2459600.5 TDB. The source of the input data is JPL's 
            {\tt Horizons}.
           }
   \label{fig:evolutionQQ99}
\end{figure}

Figure~\ref{fig:evolutionQQ99} shows the result of $N$-body integrations backward in time for the nominal orbit and
representative control orbits of 2011~QQ$_{99}$ (input Cartesian vectors from data in Table~\ref{vectorQQ99} of 
Appendix~\ref{app:2}). Our results show a very chaotic dynamical past driven again by frequent encounters with Jupiter 
and Saturn, inside the Hill radii of both planets. The evolution is however somewhat different from those of 210718 or 
2003~BM$_{1}$. Asteroid 2011~QQ$_{99}$ may remain confined within a relatively narrow volume in orbital parameter space, 
never venturing too far from the region of the giant planets. Its source region is probably between the orbits of 
Jupiter and Neptune, the Centaur orbital domain.
%
%
%
%

\subsection{2021~UJ$_{5}$}
\label{sec:4.5}
This object was first observed on October 28, 2021, by the Pan-STARRS~2 \citep{2004SPIE.5489...11K} telescope system at
Haleakala. Its orbit determination is based on 51 observations with a data-arc span of 68~d (see Table~\ref{elementsUJ5} 
in Appendix~\ref{app:1}). It has a semimajor axis of 3.40~au, eccentricity of 0.51, and orbital inclination of 
9.39$^{\circ}$. Its current orbit places it between those of Mars and Jupiter. It may have been a member of the 
Jupiter-family of comets \citep{2022ApJ...928...43G} or related populations such as the comet 29P/Schwassmann-Wachmann~1 
or 2020~MK$_{4}$ \citep{2021A&A...649A..85D}.

\begin{figure}
   \centering
   \includegraphics[width=\linewidth]{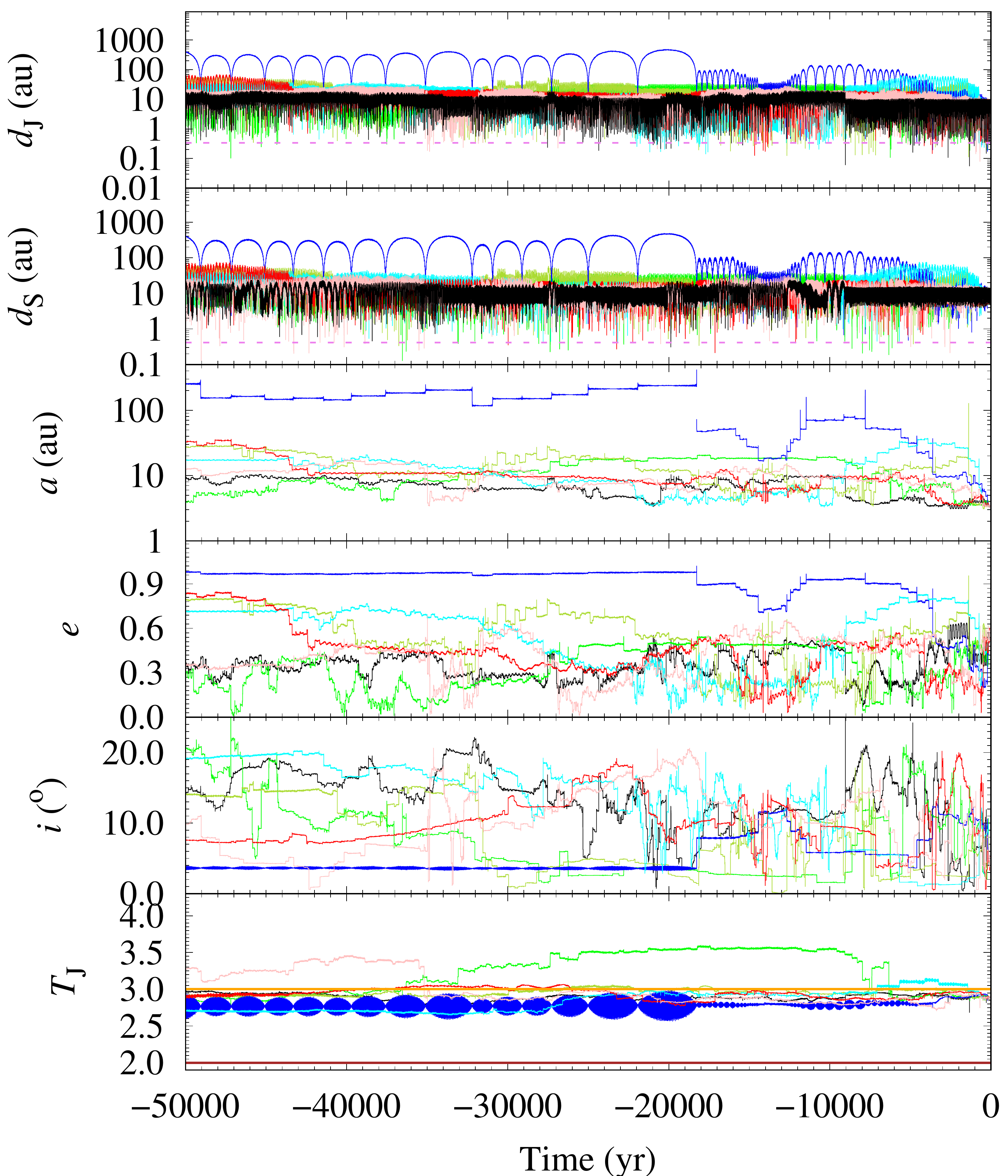}
   \caption{Short-term past evolution of relevant parameters of 2021~UJ$_{5}$. We show the evolution of the distance to 
            Jupiter (top panel) and Saturn (second to top) of the nominal orbit (in black) as described by the 
            corresponding orbit determination in Table~\ref{elementsUJ5} of Appendix~\ref{app:1} and those of control 
            orbits or clones with Cartesian vectors separated $+$3$\sigma$ (in green), $-$3$\sigma$ (in lime), 
            $+$6$\sigma$ (in blue), $-$6$\sigma$ (in cyan), $+$9$\sigma$ (in red), and $-$9$\sigma$ (in pink) from the 
            nominal values in Table~\ref{vectorUJ5} of Appendix~\ref{app:2}. The Hill radii of Jupiter, 0.338~au, and 
            Saturn, 0.412~au, are shown in red. The third to top panel shows the evolution of the semimajor axis, $a$. 
            The third to bottom panel shows the evolution of the eccentricity, $e$. The second to bottom panel displays 
            the inclination, $i$. The bottom panel shows the variations of the Tisserand's parameter, $T_{\rm J}$ 
            \citep{1999ssd..book.....M}, and includes the boundary references 2 (in brown) and 3 (in orange). The output 
            time-step size is 1~yr, the origin of time is epoch 2459600.5 TDB. The source of the input data is JPL's 
            {\tt Horizons}.
           }
   \label{fig:evolutionUJ5}
\end{figure}

Figure~\ref{fig:evolutionUJ5} shows the result of $N$-body integrations backward in time for the nominal orbit and
representative control orbits of 2021~UJ$_{5}$. Our results show a very chaotic dynamical past driven by frequent 
encounters with Jupiter and Saturn, inside the Hill radii of both planets. Our calculations suggest that its most 
probable source is in the Centaur population or, less likely, the trans-Neptunian region.
%
%
%
%

\section{Discussion}
\label{sec:5}
Our statistical analyses and calculations indicate that the present-day main asteroid belt may not contain a sizeable 
population of minor bodies recently scattered outward by the terrestrial planets. However, it does host a non-negligible 
population of newcomers scattered inward by the giant planets and perhaps even captured from the Oort cloud or 
interstellar space. These conclusions suggest that the dynamical pathways that may have populated the early main belt 
may still be active today, but only in the case of material scattered inward. While asteroids from the main belt are 
currently being scattered inward, toward the terrestrial planets \citep{2017A&A...598A..52G}, the flux in the opposite 
direction may have ceased altogether. On the other hand, \citet{2016Ap&SS.361..371G} predicted that former Centaurs and 
TNOs could be found in the outer asteroid belt. Our results confirm this prediction but all the objects in 
Table~\ref{tab:1} have $a$ in the range 3.5--4.2~au. 

The presence of 457175 (2008~GO$_{98}$) among the newcomers hints at a connection between them and the small group of
active objects present in the main asteroid belt. Although the first active member of the main asteroid belt was found 
in 1979 (originally discovered as 1979~OW$_{7}$, rediscovered as 1996~N$_{2}$, and now known as comet 133P/Elst-Pizarro, 
see for example \citealt{2014AJ....147..117J}), the nature of this population remained ambiguous. It was not until some 
time later that it became clear that a population of comets was present in the main asteroid belt 
\citep{2006Sci...312..561H,2022arXiv220301397J}. Only a very small fraction of the known members of the asteroid belt 
has been found to display comet-like features. Although the dominant process behind the observed cometary activity in 
the main belt remains unclear, possible mechanisms at work include sublimation of icy materials, impacts, rotational 
breakups, and electrostatic effects (see for example \citealt{2012AJ....143...66J,2016A&A...589A.111G,
2022ApJ...928...43G}). The population of main-belt comets includes several bizarre objects. The first known binary 
comet, 288P/(300163) 2006~VW$_{139}$ \citep{2017Natur.549..357A,2020A&A...643A.152A}, which is perhaps triple 
\citep{2020DPS....5221701K}, belongs to this population that also includes a multi-tailed comet, P/2013~P5 
\citep{2013ApJ...778L..21J}, and objects involved in impacts and disruption events as in the cases of (596) Scheila 
\citep{2011ApJ...733L...4J}, (6478) Gault \citep{2019A&A...624L..14M}, P/2015~X6 \citep{2016ApJ...826..137M}, P/2016~G1 
\citep{2016ApJ...826L..22M}, P/2019~A4 (PANSTARRS) and P/2021~A5 (PANSTARRS) \citep{2021MNRAS.506.1733M} or 248370 
(2005~QN$_{173}$) \citep{2021ApJ...922L...8C,2021ApJ...922L...9H}. The $N$-body calculations presented in the previous 
section suggest that most newcomers, including 457175, 2011~QQ$_{99}$, and 2021~UJ$_{5}$, might have had an origin as 
debris from the 29P/Schwassmann-Wachmann~1--P/2008~CL94 (Lemmon)--P/2010~TO20 (LINEAR-Grauer) cometary complex in 
Centaur orbital parameter space. 

The outer section of the main asteroid belt has been previously studied within the context of the chaotic motion and the 
Lyapunov time, $T_{L}$ (the inverse of the maximum Lyapunov exponent), which could be very short for certain objects in
this region (see for example \citealt{1994AJ....108.2323M,1996AJ....112.1278H,2010A&A...523A..67W}). The Lyapunov time 
is the characteristic timescale for the exponential divergence of initially close orbits. Focusing on the objects in 
Table~\ref{tab:1} with the most reliable orbit determinations, we obtained $T_{L}$=1500~yr (1100~yr for integrations 
into the future) for 2003~BM$_{1}$, 500~yr (500~yr) for 2011~QQ$_{99}$, 500~yr (600~yr) for 2020~UO$_{43}$, and 450~yr 
(800~yr) for 2021~UJ$_{5}$. Therefore, our relevant results for real objects confirm the conclusions of the analysis in 
\citet{2010A&A...523A..67W} and lend credence to the notion that some small bodies may indeed have reached the main belt 
within the last few hundred years. However, they may not last long in their current orbits as their $T_{L}$ are also 
very short for integrations into the future.

The four asteroids mentioned above have Lyapunov times as short as their chaotic transport times or diffusion times. In 
general, this is not the case. The statistical relationship between these two timescales is not straightforward;
it has a large scatter and its fitting is approximately quadratic (see for example \citealt{1997AJ....114.1246M,
1998PhLA..241...53S}). A dramatic example of this fact is in the asteroid 522 Helga, which is both chaotic and stable, 
with a short Lyapunov time but infinite diffusion time \citep{1993CeMDA..56..323M}. This topic has been recently 
revisited by \citet{2022PhyD..43033101C}.

\begin{figure}
  \centering
  \includegraphics[width=0.495\linewidth]{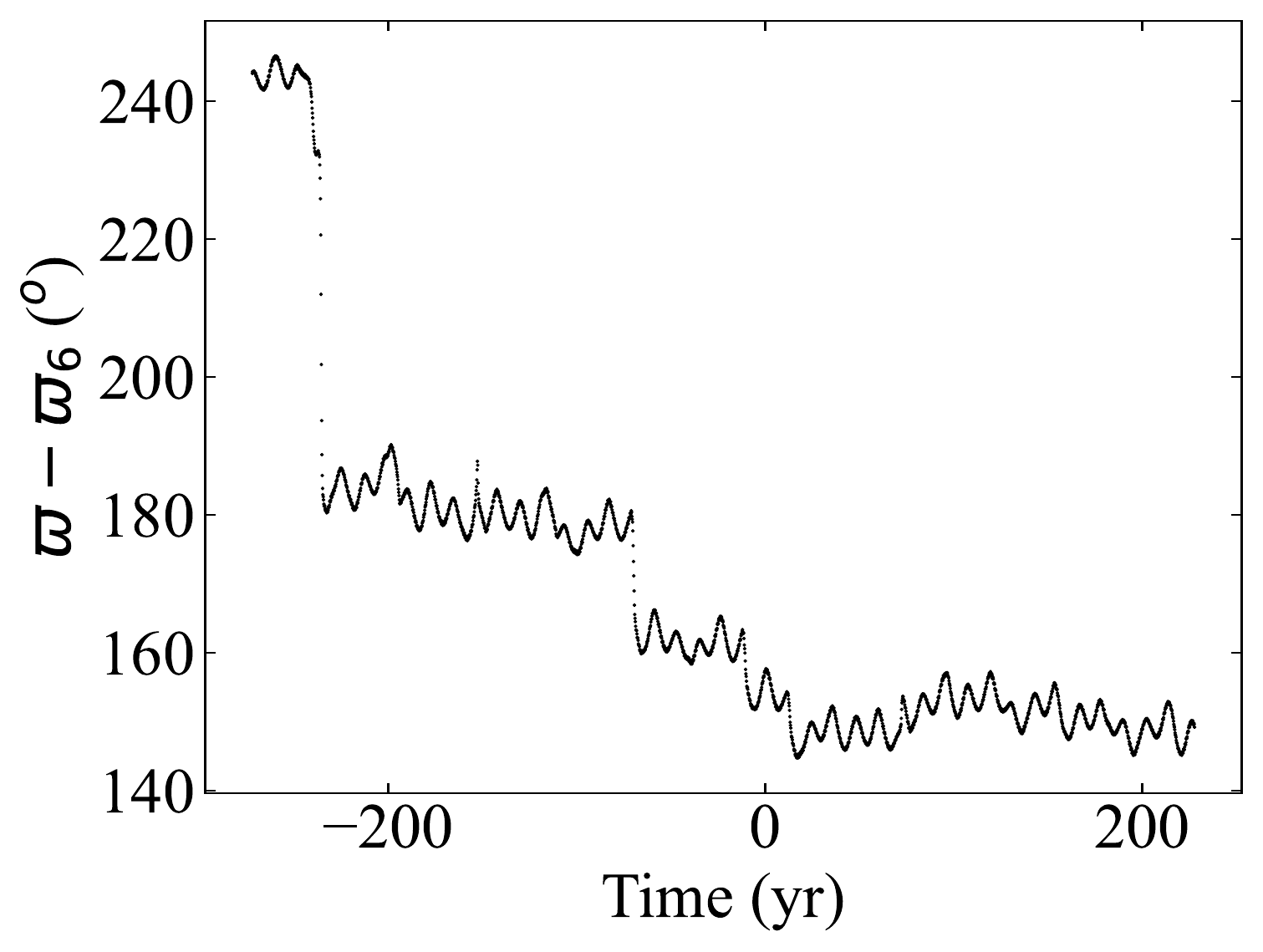}
  \includegraphics[width=0.495\linewidth]{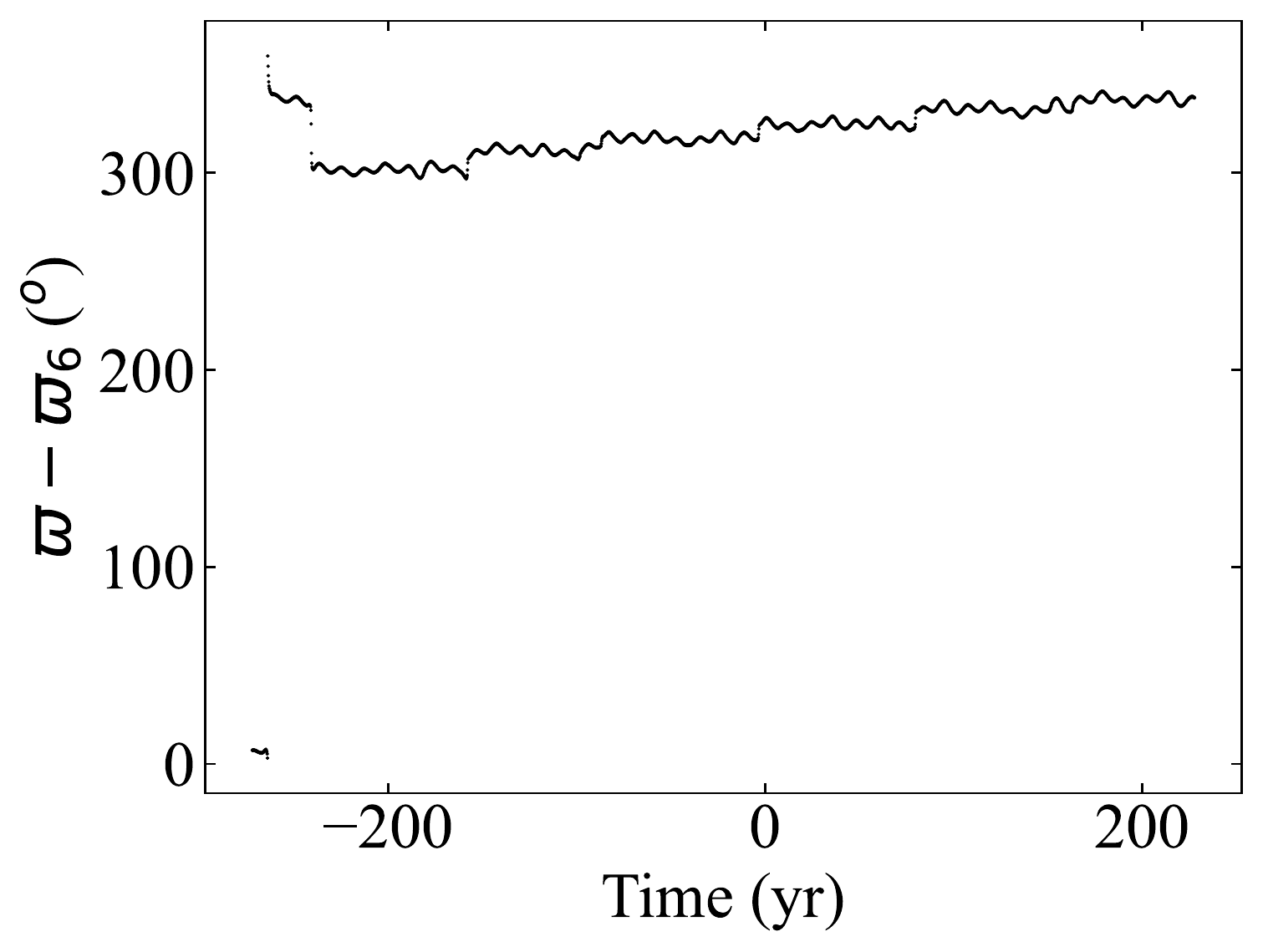}
  \caption{Evolution of the resonant argument $\sigma_6$=$\varpi - \varpi_6$ for 457175 (2008~GO$_{98}$) and 
           2021~UJ$_{5}$, left- and right-hand side panels, respectively. The origin of time is the epoch 2459600.5~JD 
           Barycentric Dynamical Time (2022-Jan-21.0 00:00:00.0 TDB) and the output cadence is 60~d. The source of the 
           data is JPL's {\tt Horizons}. 
          }
  \label{fig:nu6_a}
\end{figure}

\begin{figure}
  \centering
  \includegraphics[width=0.325\linewidth]{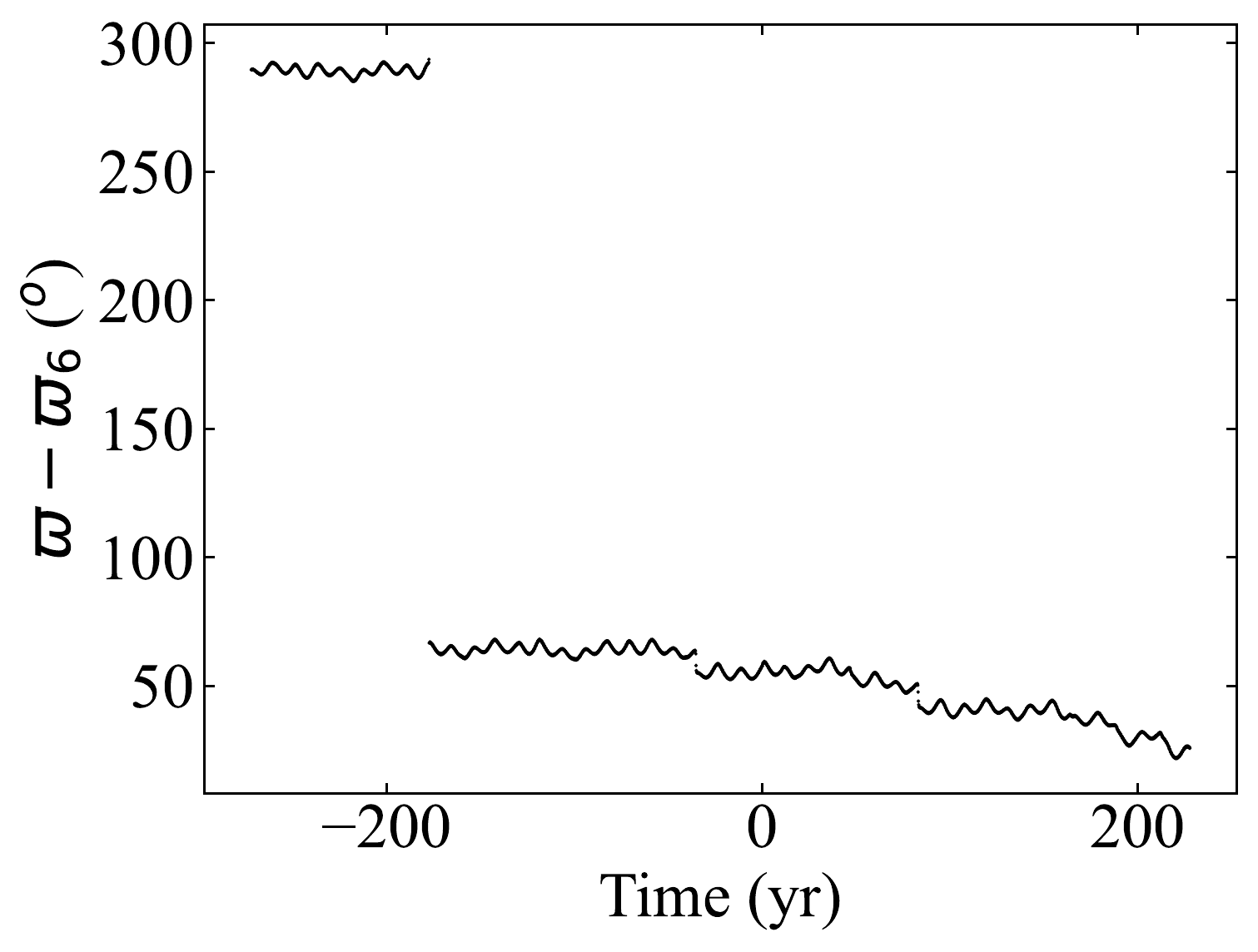}
  \includegraphics[width=0.325\linewidth]{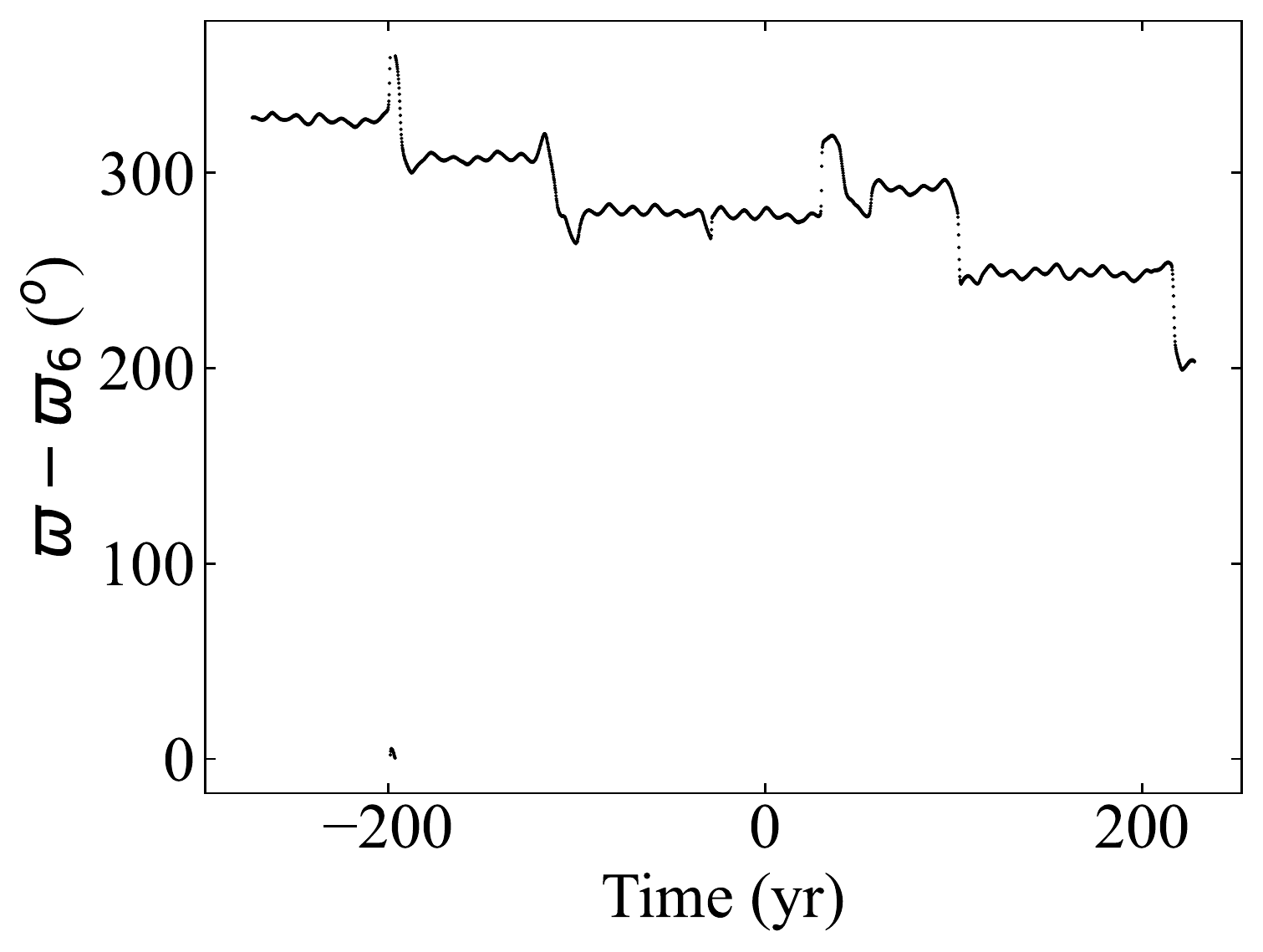}
  \includegraphics[width=0.325\linewidth]{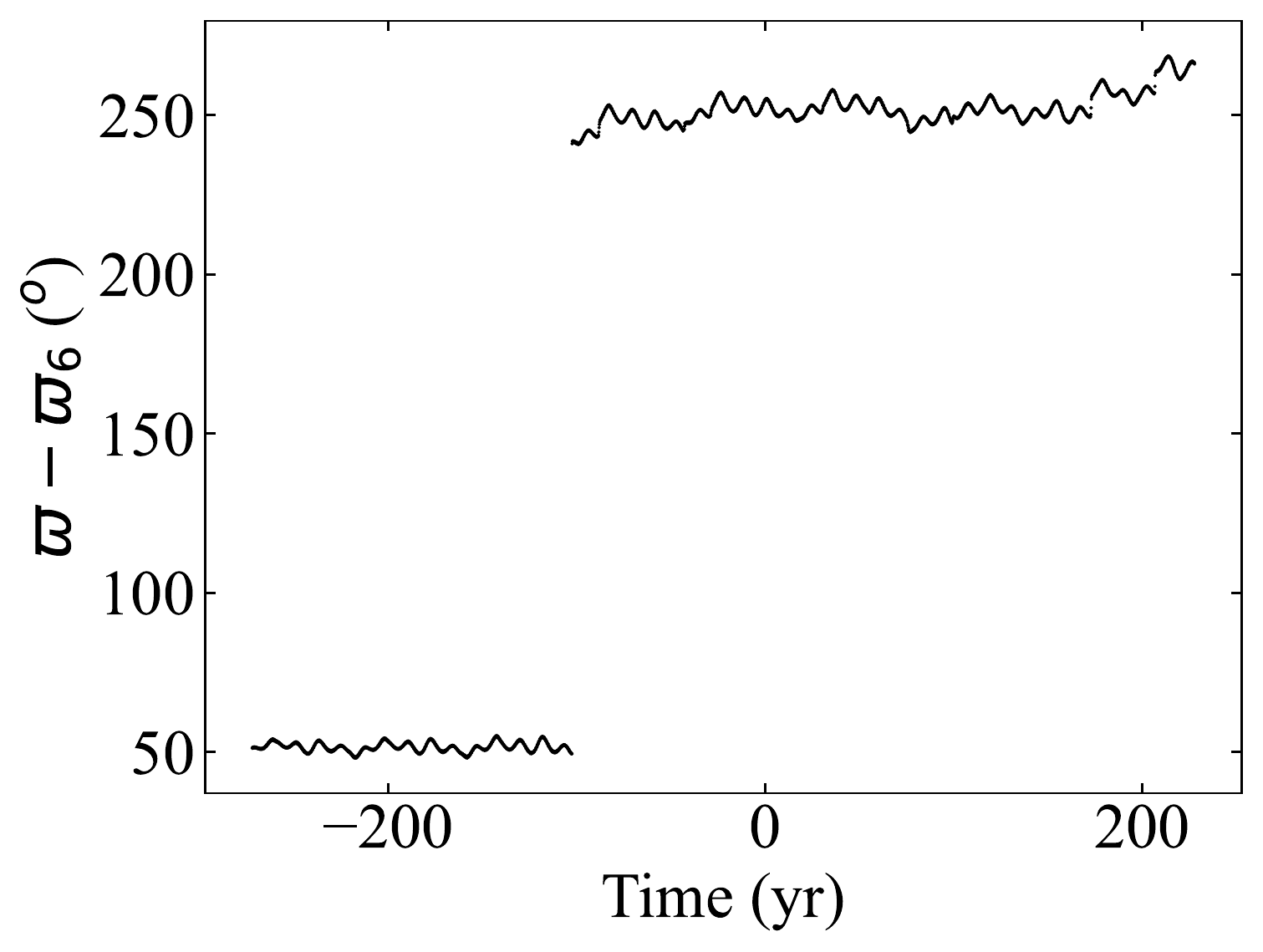}
  \caption{Evolution of the resonant argument $\sigma_6$=$\varpi - \varpi_6$ for (from left to right) 2003~BM$_{1}$, 
           2011~QQ$_{99}$, and 2020~UO$_{43}$. The origin of time is the epoch 2459600.5~JD Barycentric Dynamical Time 
           (2022-Jan-21.0 00:00:00.0~TDB) and the output cadence is 60~d. The source of the data is JPL's 
           {\tt Horizons}. 
          }
  \label{fig:nu6_b}
\end{figure}

We have pointed out above that secular resonances (other than von Zeipel-Lidov-Kozai's) may bring some of these objects 
back and forth into the belt from beyond Jupiter. For instance, asteroids could be perturbed by secular resonances 
involving the values of the longitude of perihelion and the ascending node relative to those of Saturn (see for example
\citealt{1969PhDT.........2W,1986A&A...166..326F,1987A&A...179..294F,1988CeMec..43..113F,1989CeMDA..46..231F,
1987CeMec..40..233Y,1991CeMDA..51..131M,1991CeMDA..51..169M}). The role of the so-called $\nu_6$ secular 
resonance can be explored by studying the evolution of the resonant argument $\sigma_6$=$\varpi - \varpi_6$, where 
$\varpi$=$\Omega + \omega$ is the longitude of perihelion of the asteroid and the one of Saturn is $\varpi_6$. According 
to \citet{1969PhDT.........2W}, two resonant states are possible with libration of $\sigma_6$ around 0\degr (when the 
perihelia of the asteroid and Saturn are approximately aligned) or 180\degr (when the perihelia are somewhat 
anti-aligned), for additional details see for example \citet{2011MNRAS.412.2040C} and \citet{2018MNRAS.481.1707H}. The 
secular nodal resonance with Saturn can be assessed by studying the resonant argument $\Omega - \Omega_6$. Using JPL's 
{\tt Horizons} data, we have explored these secular resonances for 457175, 2003~BM$_{1}$, 2011~QQ$_{99}$, 
2020~UO$_{43}$, and 2021~UJ$_{5}$. The secular nodal resonance does not seem to be currently active for any of these 
objects, but Fig.~\ref{fig:nu6_a} shows that at least two objects are, have been, or will be subjected to the $\nu_6$ 
secular resonance. Comet 362P or 457175 may have engaged in anti-aligned $\nu_6$ secular resonant behavior in the past 
and 2021~UJ$_{5}$ seems to be currently trapped in the $\nu_6$ secular resonance with $\sigma_6$ oscillating about 
0\degr. Figure~\ref{fig:nu6_b} shows that librations of $\sigma_6$ are also observed for other objects, but not about 
0\degr or 180\degr. However and given the rather short values of $T_{L}$ pointed out above, any resonant engagements are 
expected to be relatively brief.
%
%
%
%
%
%
%
%

Our analyses have also uncovered a possible gravitational trap in the outer main belt for interstellar objects entering 
the Solar system at relatively slow speeds with respect to the barycenter; in other words, objects following hyperbolic 
orbits with low excess eccentricity, $\sim$1 to 1.3. Asteroid 2020~UO$_{43}$ might be one of such objects but its 
current orbit determination is not precise enough to confirm (or reject) an interstellar origin (see also the cautionary 
note above). 

As pointed out in Sect.~\ref{sec:2}, we have not considered the Yarkovsky and Yarkovsky--O'Keefe--Radzievskii--Paddack 
(YORP) effects (see for example \citealt{2006AREPS..34..157B}) in our calculations, although both effects have been 
confirmed to be detectable in both near-Earth and main-belt asteroid populations (see for example 
\citealt{2013Icar..224....1F,2016AJ....151..164C,2017MNRAS.469.4400C,2018A&A...617A..61D,2018A&A...609A..86D,
2019Icar..321..564G,2020AJ....159...92G}). However, ignoring these effects has no significant impact on the evaluation
of the short-term orbital evolution of the objects discussed here because for an average value for the Yarkovsky drift 
of 10$^{-9}$~au~yr$^{-1}$ (see for example \citealt{2012AJ....144...60N}), the timescale to enter the main belt is of
the order of Myr, which is several orders of magnitude longer than the time intervals discussed in this research. On 
the other hand, accurate modelling of the Yarkovsky force requires relatively precise knowledge of the physical 
properties ---such as rotation rate, albedo, bulk density, surface conductivity, or emissivity--- of the objects under 
study, which is not the case for the relevant small bodies studied in this work.

\section{Conclusions}
\label{sec:6}
In this paper, we have discussed the details of a data mining experiment aimed at identifying present-day members of the 
main asteroid belt that may have reached the belt during the last few hundred years. We have used publicly available 
data from JPL's SBDB and {\tt Horizons}. Our analyses show which regions of the belt are the most and the least stable, 
and how the various mean-motion resonances shape the orbital architecture of the main belt. Our conclusions can be 
summarized as follows.
\begin{enumerate}
   \item The inner belt and the core of the main belt do not include any known object that could be regarded as a recent 
         arrival to the belt.
   \item The outer main belt hosts a small but very interesting population of orbital newcomers. A few dozen objects 
         may have been inserted into this region within the last few hundred years. The list of interlopers with 
         reliable orbit determinations includes 457175 (2008~GO$_{98}$), 2003~BM$_{1}$, 2011~QQ$_{99}$, 2020~UO$_{43}$,
         and 2021~UJ$_{5}$. At least one object has been found to exhibit cometary activity, 457175 or comet 362P, which 
         opens an interesting connection between the interlopers singled out here and the small group of active objects 
         present in the main asteroid belt. 
   \item Asteroid 2020~UO$_{43}$ has a non-negligible probability of having an origin in the Oort cloud or even 
         interstellar space, but its orbit determination is not sufficiently robust to reach a conclusive answer to
         the question of its origin, if the object is confirmed as real.
   \item Most interlopers found here may come from the Centaur orbital realm. 
\end{enumerate}
Our results suggest that the dynamical pathways that inserted material from beyond Jupiter into the main asteroid belt 
early in the history of the Solar system may continue to be active today but those sending debris from the orbital realm 
of the terrestrial planets into the belt may be currently inactive or at least their present-day strength could be 
significantly weaker than that of those scattering material inward.

\begin{acknowledgements}
   We thank the referee, V. Carruba, for his constructive, thorough, and helpful reports, a second referee for pointing 
   out the non-straightforward statistical relationship between Lyapunov time and diffusion time, S.~J. Aarseth for 
   providing one of the codes used in this research, A.~B. Chamberlin for helping with the new JPL's Solar System 
   Dynamics website, S. Deen for finding precovery images of some of the objects discussed here and for additional 
   comments including the issue of the dubious linkage of 2020~UO$_{43}$, and A.~I. G\'omez de Castro for providing 
   access to computing facilities. Part of the calculations and the data analysis were completed on the Brigit HPC 
   server of the `Universidad Complutense de Madrid' (UCM), and we thank S. Cano Als\'ua for his help during this stage. 
   This work was partially supported by the Spanish `Ministerio de Econom\'{\i}a y Competitividad' (MINECO) under grant 
   ESP2017-87813-R and by the `Agencia Estatal de Investigaci\'on (Ministerio de Ciencia e Innovaci\'on)' under grant
   PID2020-116726RB-I00 /AEI/10.13039/501100011033. In preparation for this paper, we made use of the NASA Astrophysics 
   Data System, the ASTRO-PH e-print server, and the MPC data server. 
\end{acknowledgements}

%
%

\section*{Statements and Declarations}
The data underlying this paper were accessed from JPL's SBDB (\url{https://ssd.jpl.nasa.gov/tools/sbdb_lookup.html\#/})
and {\tt Horizons} on-line solar system data and ephemeris computation service (\url{https://ssd.jpl.nasa.gov/horizons/}),
both provided by the Solar System Dynamics Group (\url{https://ssd.jpl.nasa.gov/}).
The derived data generated in this research will be shared on reasonable request to the corresponding author.

\bibliographystyle{spbasic}      
\bibliography{MBarrivals}   

\appendix

\section{Orbit determinations}
\label{app:1}
The orbit determinations of the objects discussed in detail in the sections are shown here (Tables \ref{elementsBM1} to
\ref{elementsUJ5}).
%
%
\begin{table}
   \centering
   \fontsize{8}{12pt}\selectfont
   \tabcolsep 0.14truecm
   \caption{\label{elementsBM1}Values of the Heliocentric Keplerian orbital elements and their respective 1$\sigma$ 
            uncertainties of 2003~BM$_{1}$. The orbit determination is referred to epoch JD 2459600.5 (2022-Jan-21.0) 
            TDB (Barycentric Dynamical Time, J2000.0 ecliptic and equinox), and it is based on 90 observations 
            with a data-arc span of 5140~d (solution date, 2021-Apr-14 21:40:32 PDT). Source: JPL's SBDB.
           }
   \begin{tabular}{lcc}
      \hline
        Orbital parameter                                 &   &                           \\ 
      \hline
        Semimajor axis, $a$ (au)                          & = &   3.640089$\pm$0.000002   \\
        Eccentricity, $e$                                 & = &   0.5143570$\pm$0.0000010 \\ 
        Inclination, $i$ (\degr)                          & = &  11.34667$\pm$0.00003     \\
        Longitude of the ascending node, $\Omega$ (\degr) & = & 350.5776$\pm$0.0002       \\
        Argument of perihelion, $\omega$ (\degr)          & = & 156.1809$\pm$0.0009       \\
        Mean anomaly, $M$ (\degr)                         & = & 247.5705$\pm$0.0002       \\
        Perihelion distance, $q$ (au)                     & = &   1.767783$\pm$0.000004   \\
        Aphelion distance, $Q$ (au)                       & = &   5.512394$\pm$0.000003   \\
        Absolute magnitude, $H$ (mag)                     & = &  18.2                     \\
      \hline
   \end{tabular}
\end{table}
%
%
%
%
\begin{table}
   \centering
   \fontsize{8}{12pt}\selectfont
   \tabcolsep 0.14truecm
   \caption{\label{elementsUO43}Values of the Heliocentric Keplerian orbital elements and their respective 1$\sigma$
            uncertainties of 2020~UO$_{43}$. The orbit determination is referred to epoch JD 2459396.5 (2021-Jul-01.0)
            TDB (Barycentric Dynamical Time, J2000.0 ecliptic and equinox), and it is based on 13 observations
            with a data-arc span of 548~d (solution date, 2021-Apr-15 23:16:16 PDT). Source: JPL's SBDB.
           }
   \begin{tabular}{lcc}
      \hline
        Orbital parameter                                 &   &                       \\
      \hline
        Semimajor axis, $a$ (au)                          & = &   4.1360$\pm$0.0011   \\
        Eccentricity, $e$                                 & = &   0.64513$\pm$0.00011 \\
        Inclination, $i$ (\degr)                          & = &   1.7584$\pm$0.0003   \\
        Longitude of the ascending node, $\Omega$ (\degr) & = &  83.552$\pm$0.009     \\
        Argument of perihelion, $\omega$ (\degr)          & = & 260.125$\pm$0.010     \\
        Mean anomaly, $M$ (\degr)                         & = &  40.38$\pm$0.02       \\
        Perihelion distance, $q$ (au)                     & = &   1.46775$\pm$0.00008 \\
        Aphelion distance, $Q$ (au)                       & = &   6.804$\pm$0.002     \\
        Absolute magnitude, $H$ (mag)                     & = &  19.50$\pm$0.15       \\
      \hline
   \end{tabular}
\end{table}
%
%
%
%
\begin{table}
   \centering
   \fontsize{8}{12pt}\selectfont
   \tabcolsep 0.14truecm
   \caption{\label{elementsST252}Values of the Heliocentric Keplerian orbital elements and their respective 1$\sigma$ 
            uncertainties of 210718 (2000~ST$_{252}$). The orbit determination is referred to epoch JD 2459600.5 
            (2022-Jan-21.0) TDB (Barycentric Dynamical Time, J2000.0 ecliptic and equinox), and it is based on 584 
            observations with a data-arc span of 7822~d (solution date, 2022-Apr-13 22:34:29 PDT). Source: JPL's 
            SBDB.
           }
   \begin{tabular}{lcc}
      \hline
        Orbital parameter                                 &   &                             \\ 
      \hline
        Semimajor axis, $a$ (au)                          & = &   3.60668553$\pm$0.00000004 \\
        Eccentricity, $e$                                 & = &   0.39219767$\pm$0.00000003 \\ 
        Inclination, $i$ (\degr)                          & = &   7.477859$\pm$0.000005     \\
        Longitude of the ascending node, $\Omega$ (\degr) & = &  74.36999$\pm$0.00005       \\
        Argument of perihelion, $\omega$ (\degr)          & = &  15.40597$\pm$0.00005       \\
        Mean anomaly, $M$ (\degr)                         & = & 342.689186$\pm$0.000010     \\
        Perihelion distance, $q$ (au)                     & = &   2.19215188$\pm$0.00000014 \\
        Aphelion distance, $Q$ (au)                       & = &   5.02121918$\pm$0.00000005 \\
        Absolute magnitude, $H$ (mag)                     & = &  15.18                      \\
      \hline
   \end{tabular}
\end{table}
%
%
%
%
\begin{table}
   \centering
   \fontsize{8}{12pt}\selectfont
   \tabcolsep 0.14truecm
   \caption{\label{elementsQQ99}Values of the Heliocentric Keplerian orbital elements and their respective 1$\sigma$ 
            uncertainties of 2011~QQ$_{99}$. The orbit determination is referred to epoch JD 2459600.5 (2022-Jan-21.0) 
            TDB (Barycentric Dynamical Time, J2000.0 ecliptic and equinox), and it is based on 48 observations 
            with a data-arc span of 8918~d (solution date, 2021-Apr-15 05:16:22 PDT). Source: JPL's SBDB.
           }
   \begin{tabular}{lcc}
      \hline
        Orbital parameter                                 &   &                             \\ 
      \hline
        Semimajor axis, $a$ (au)                          & = &   3.80163390$\pm$0.00000008 \\
        Eccentricity, $e$                                 & = &   0.4261700$\pm$0.0000003   \\ 
        Inclination, $i$ (\degr)                          & = &   3.21247$\pm$0.00003       \\
        Longitude of the ascending node, $\Omega$ (\degr) & = &   1.6633$\pm$0.0003         \\
        Argument of perihelion, $\omega$ (\degr)          & = &   8.9432$\pm$0.0003         \\
        Mean anomaly, $M$ (\degr)                         & = & 134.16675$\pm$0.00004       \\
        Perihelion distance, $q$ (au)                     & = &   2.1814916$\pm$0.0000013   \\
        Aphelion distance, $Q$ (au)                       & = &   5.42177625$\pm$0.00000012 \\
        Absolute magnitude, $H$ (mag)                     & = &  16.5                       \\
      \hline
   \end{tabular}
\end{table}
%
%
%
%
\begin{table}
   \centering
   \fontsize{8}{12pt}\selectfont
   \tabcolsep 0.14truecm
   \caption{\label{elementsUJ5}Values of the Heliocentric Keplerian orbital elements and their respective 1$\sigma$ 
            uncertainties of 2021~UJ$_{5}$. The orbit determination is referred to epoch JD 2459600.5 (2022-Jan-21.0) 
            TDB (Barycentric Dynamical Time, J2000.0 ecliptic and equinox), and it is based on 51 observations 
            with a data-arc span of 68~d (solution date, 2022-Jan-07 22:17:00 PST). Source: JPL's SBDB.
           }
   \begin{tabular}{lcc}
      \hline
        Orbital parameter                                 &   &                       \\ 
      \hline
        Semimajor axis, $a$ (au)                          & = &   3.3960$\pm$0.0002   \\
        Eccentricity, $e$                                 & = &   0.51475$\pm$0.00002 \\ 
        Inclination, $i$ (\degr)                          & = &   9.3875$\pm$0.0003   \\
        Longitude of the ascending node, $\Omega$ (\degr) & = &  65.33424$\pm$0.00014 \\
        Argument of perihelion, $\omega$ (\degr)          & = & 351.1743$\pm$0.0012   \\
        Mean anomaly, $M$ (\degr)                         & = &  11.9668$\pm$0.0010   \\
        Perihelion distance, $q$ (au)                     & = &   1.64789$\pm$0.00002 \\
        Aphelion distance, $Q$ (au)                       & = &   5.1440$\pm$0.0003   \\
        Absolute magnitude, $H$ (mag)                     & = &  19.72                \\
      \hline
   \end{tabular}
\end{table}
%
%

\section{Cartesian vectors}
\label{app:2}
The barycentric Cartesian state vectors of the objects discussed in detail in the sections are shown here (Tables 
\ref{vectorBM1} to \ref{vectorUJ5}).
%
%
\begin{table}
   \centering
   \fontsize{8}{12pt}\selectfont
   \tabcolsep 0.15truecm
   \caption{\label{vectorBM1}Barycentric Cartesian state vector of 2003~BM$_{1}$: components and associated 1$\sigma$
            uncertainties. Data are referred to epoch 2459600.5, 21-January-2022 00:00:00.0 TDB (J2000.0 ecliptic and 
            equinox). Source: JPL's {\tt Horizons}.
           }
   \begin{tabular}{ccc}
      \hline
        Component                         &   &    value$\pm$1$\sigma$ uncertainty                                \\
      \hline
        $X$ (au)                          & = &    4.897320775769619$\times10^{+0}$$\pm$1.34839974$\times10^{-5}$ \\
        $Y$ (au)                          & = & $-$5.212872303509024$\times10^{-1}$$\pm$5.05348133$\times10^{-5}$ \\
        $Z$ (au)                          & = &    5.752081909237879$\times10^{-2}$$\pm$1.23618711$\times10^{-5}$ \\
        $V_X$ (au/d)                      & = & $-$1.879730452630099$\times10^{-3}$$\pm$5.60658767$\times10^{-8}$ \\
        $V_Y$ (au/d)                      & = &    5.817006393756842$\times10^{-3}$$\pm$3.02313450$\times10^{-8}$ \\
        $V_Z$ (au/d)                      & = &    1.091705108513680$\times10^{-3}$$\pm$8.31948849$\times10^{-9}$ \\
       \hline
   \end{tabular}
\end{table}
%
%
%
%
\begin{table}
   \centering
   \fontsize{8}{12pt}\selectfont
   \tabcolsep 0.15truecm
   \caption{\label{vectorST252}Barycentric Cartesian state vector of 210718 (2000~ST$_{252}$): components and associated 
            1$\sigma$ uncertainties. Data are referred to epoch 2459600.5, 21-January-2022 00:00:00.0 TDB (J2000.0 
            ecliptic and equinox). Source: JPL's {\tt Horizons}.
           }
   \begin{tabular}{ccc}
      \hline
        Component                         &   &    value$\pm$1$\sigma$ uncertainty                                 \\
      \hline
        $X$ (au)                          & = &    1.536584439342031$\times10^{+0}$$\pm$4.74764379$\times10^{-7}$  \\
        $Y$ (au)                          & = &    1.775723062487080$\times10^{+0}$$\pm$1.73163239$\times10^{-7}$  \\
        $Z$ (au)                          & = & $-$1.324647636935953$\times10^{-1}$$\pm$2.62870805$\times10^{-7}$  \\
        $V_X$ (au/d)                      & = & $-$1.116266081601603$\times10^{-2}$$\pm$1.48210777$\times10^{-9}$  \\
        $V_Y$ (au/d)                      & = &    6.472903689610230$\times10^{-3}$$\pm$9.95711599$\times10^{-10}$ \\
        $V_Z$ (au/d)                      & = &    1.639978110112024$\times10^{-3}$$\pm$1.30296247$\times10^{-9}$  \\
       \hline
   \end{tabular}
\end{table}
%
%
%
%
\begin{table}
   \centering
   \fontsize{8}{12pt}\selectfont
   \tabcolsep 0.15truecm
   \caption{\label{vectorQQ99}Barycentric Cartesian state vector of 2011~QQ$_{99}$: components and associated 1$\sigma$
            uncertainties. Data are referred to epoch 2459600.5, 21-January-2022 00:00:00.0 TDB (J2000.0 ecliptic and 
            equinox). Source: JPL's {\tt Horizons}.
           }
   \begin{tabular}{ccc}
      \hline
        Component                         &   &    value$\pm$1$\sigma$ uncertainty                                \\
      \hline
        $X$ (au)                          & = & $-$5.088605123745555$\times10^{+0}$$\pm$1.06554528$\times10^{-6}$ \\
        $Y$ (au)                          & = &    9.382269468486346$\times10^{-1}$$\pm$3.14616410$\times10^{-6}$ \\
        $Z$ (au)                          & = &    6.091157848659857$\times10^{-2}$$\pm$9.27918730$\times10^{-7}$ \\
        $V_X$ (au/d)                      & = & $-$2.536954036137189$\times10^{-3}$$\pm$2.13416890$\times10^{-9}$ \\
        $V_Y$ (au/d)                      & = & $-$5.505587343920672$\times10^{-3}$$\pm$3.72951351$\times10^{-9}$ \\
        $V_Z$ (au/d)                      & = & $-$3.041369097484084$\times10^{-4}$$\pm$3.48439024$\times10^{-9}$ \\
       \hline
   \end{tabular}
\end{table}
%
%
%
%
\begin{table}
   \centering
   \fontsize{8}{12pt}\selectfont
   \tabcolsep 0.15truecm
   \caption{\label{vectorUJ5}Barycentric Cartesian state vector of 2021~UJ$_{5}$: components and associated 1$\sigma$
            uncertainties. Data are referred to epoch 2459600.5, 21-January-2022 00:00:00.0 TDB (J2000.0 ecliptic and 
            equinox). Source: JPL's {\tt Horizons}.
           }
   \begin{tabular}{ccc}
      \hline
        Component                         &   &    value$\pm$1$\sigma$ uncertainty                                \\
      \hline
        $X$ (au)                          & = & $-$2.335857504258083$\times10^{-1}$$\pm$9.68577354$\times10^{-6}$ \\
        $Y$ (au)                          & = &    1.780329374916908$\times10^{+0}$$\pm$3.05919872$\times10^{-5}$ \\
        $Z$ (au)                          & = &    1.565833140371389$\times10^{-1}$$\pm$5.84106645$\times10^{-6}$ \\
        $V_X$ (au/d)                      & = & $-$1.529097469644606$\times10^{-2}$$\pm$4.78188519$\times10^{-8}$ \\
        $V_Y$ (au/d)                      & = &    1.569062550517957$\times10^{-3}$$\pm$2.26363719$\times10^{-7}$ \\
        $V_Z$ (au/d)                      & = &    2.405822479905948$\times10^{-3}$$\pm$9.80029822$\times10^{-8}$ \\
       \hline
   \end{tabular}
\end{table}
%
%

\section{Additional new arrivals}
\label{app:3}
\begin{figure}
  \centering
  \includegraphics[width=0.193\linewidth]{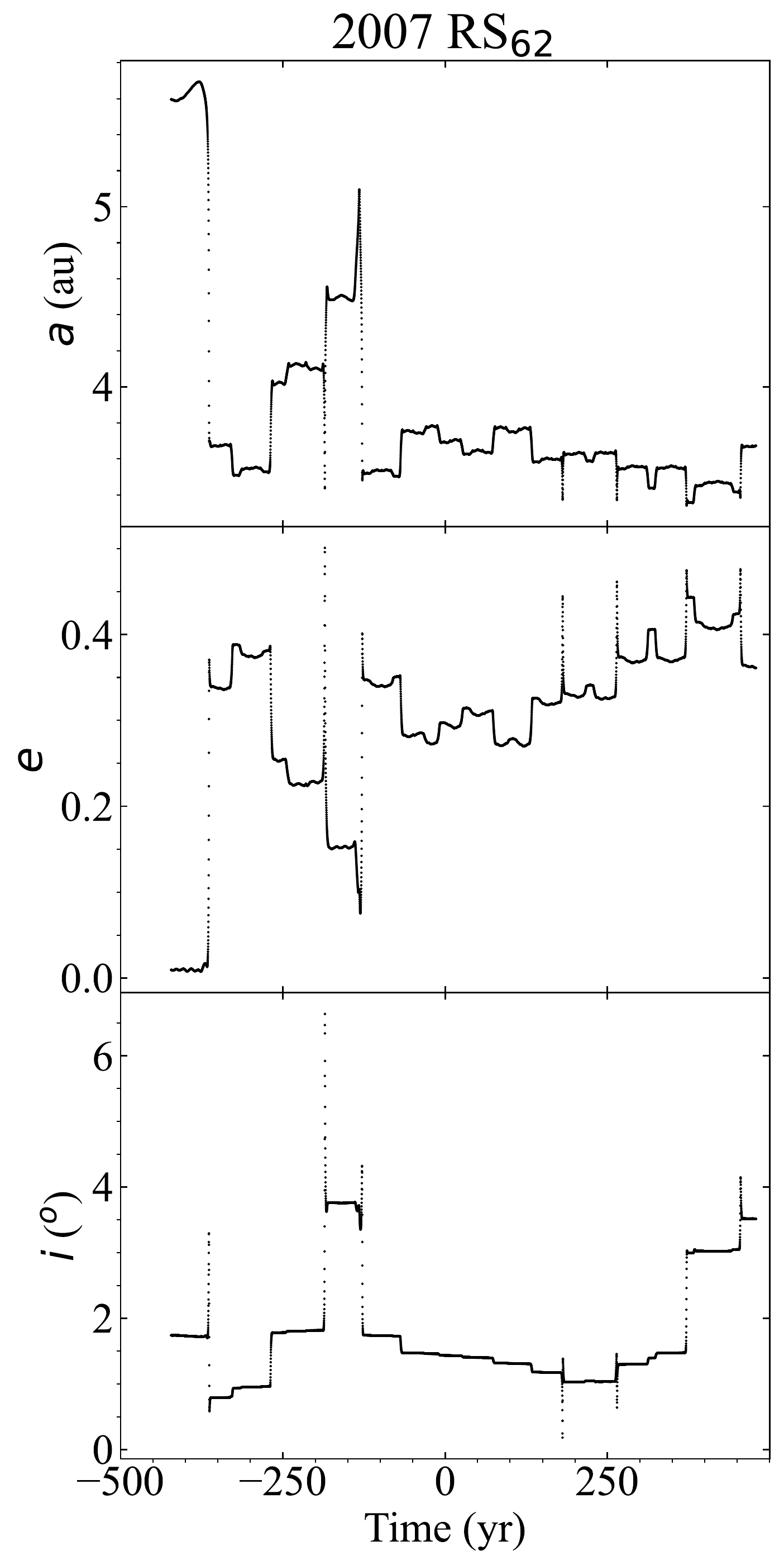}
  \includegraphics[width=0.193\linewidth]{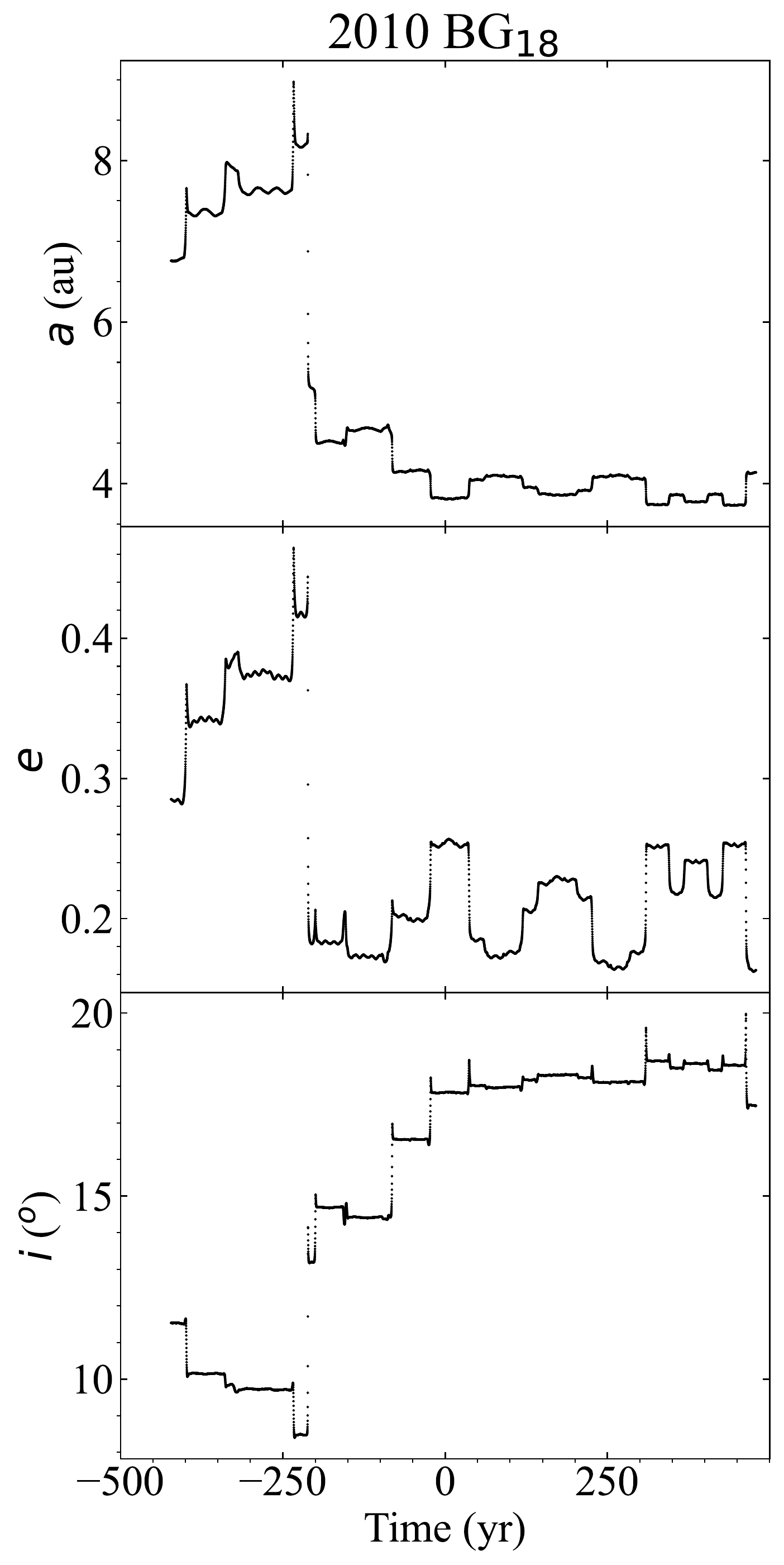}
  \includegraphics[width=0.193\linewidth]{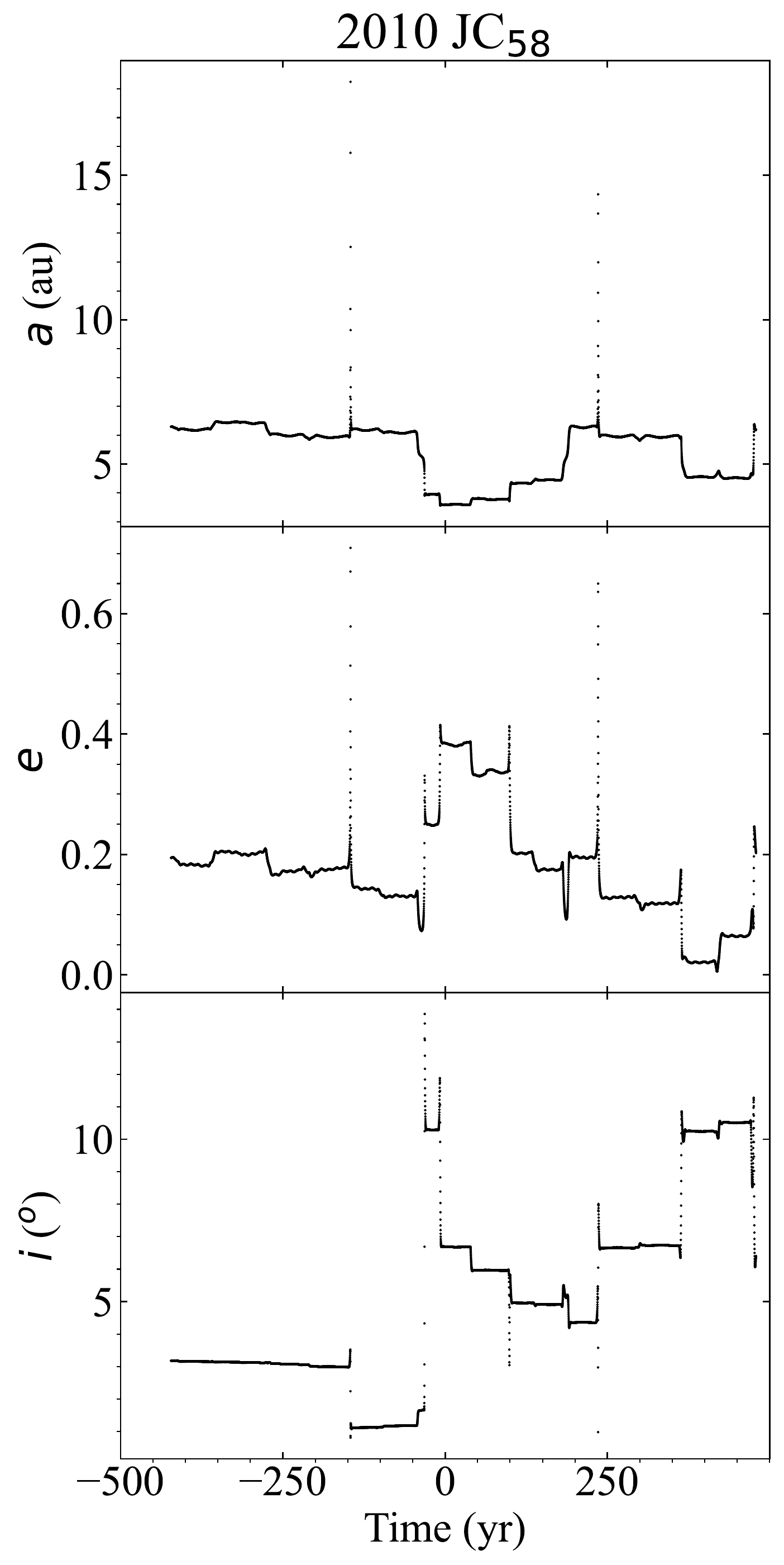}
  \includegraphics[width=0.199\linewidth]{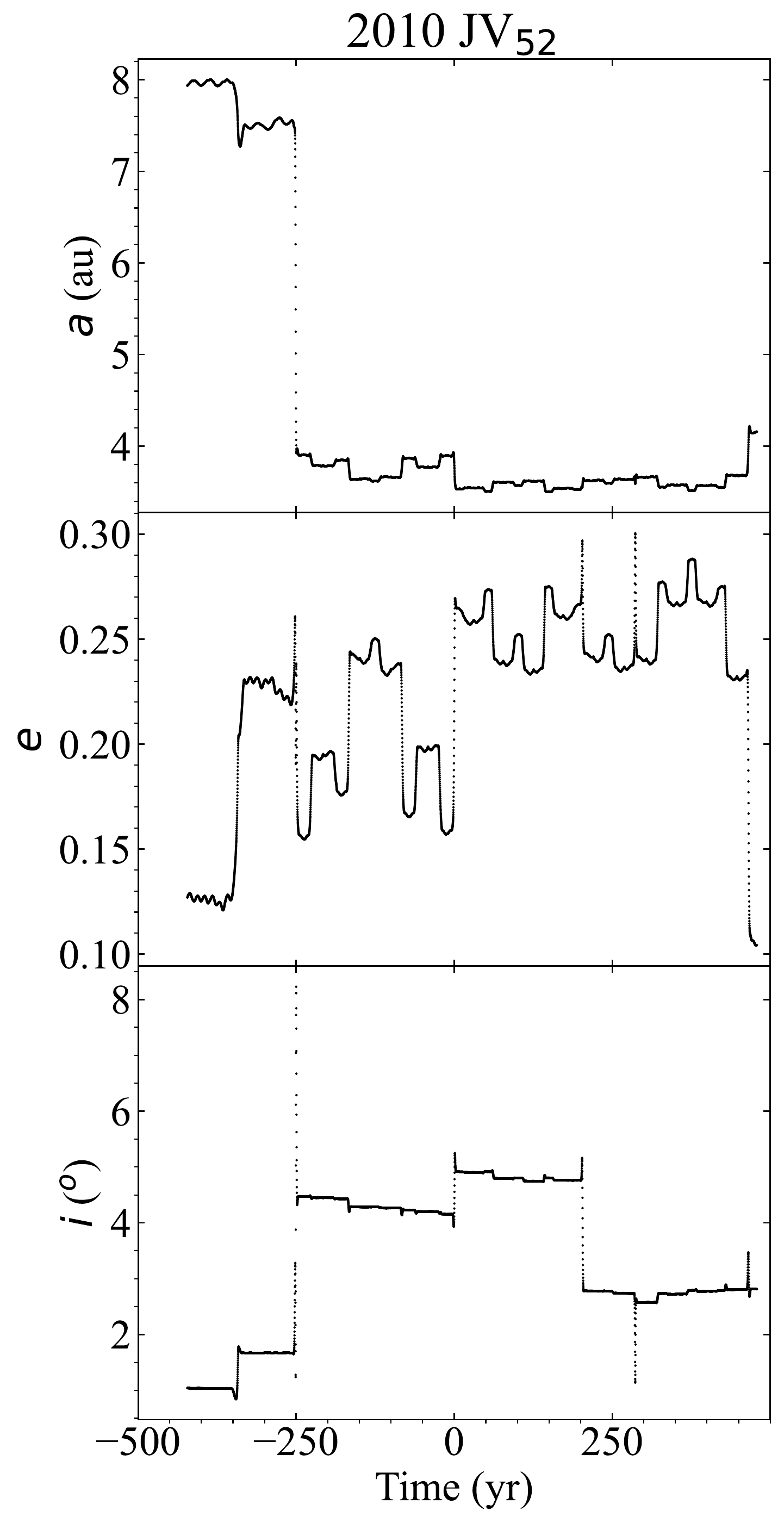}
  \includegraphics[width=0.199\linewidth]{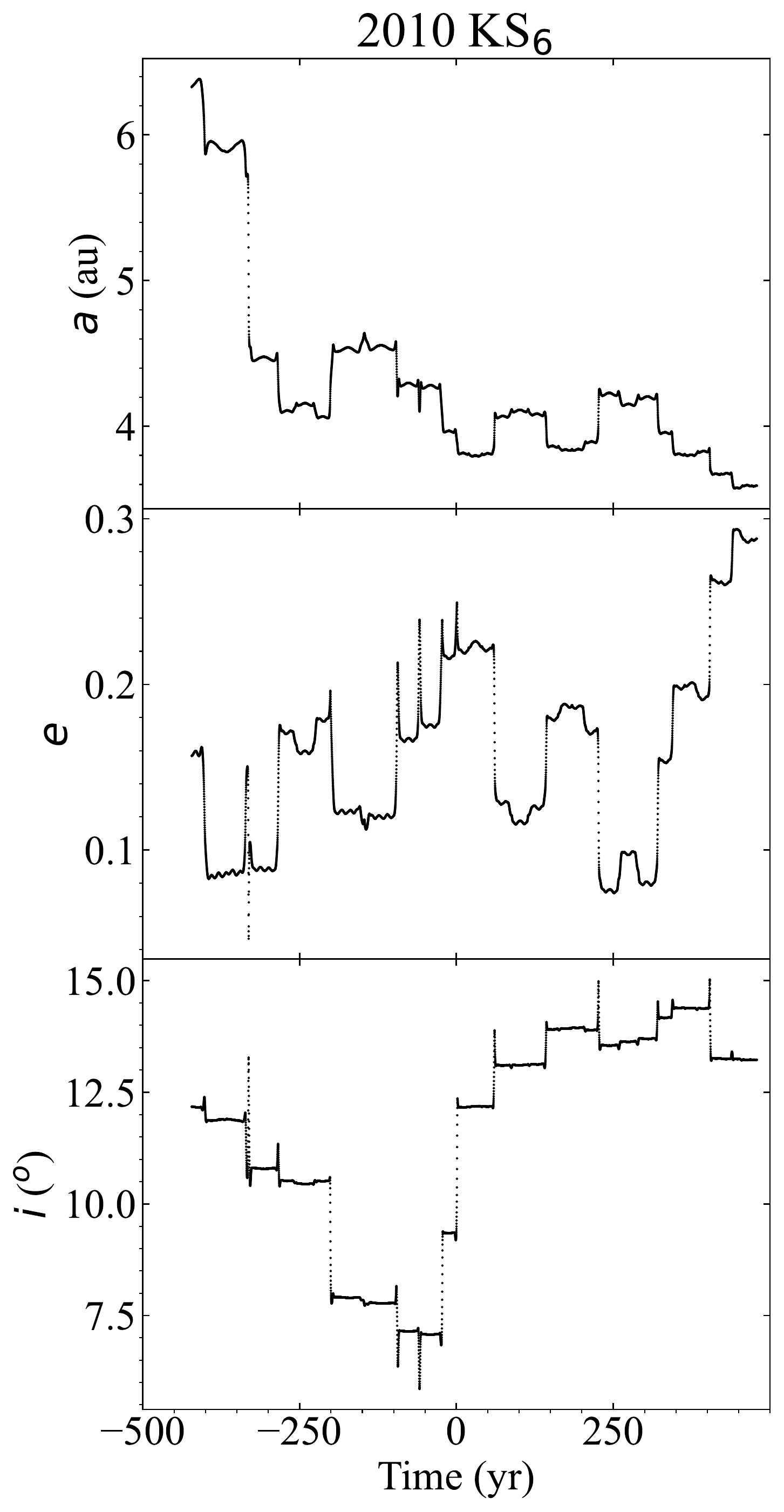}
  \includegraphics[width=0.195\linewidth]{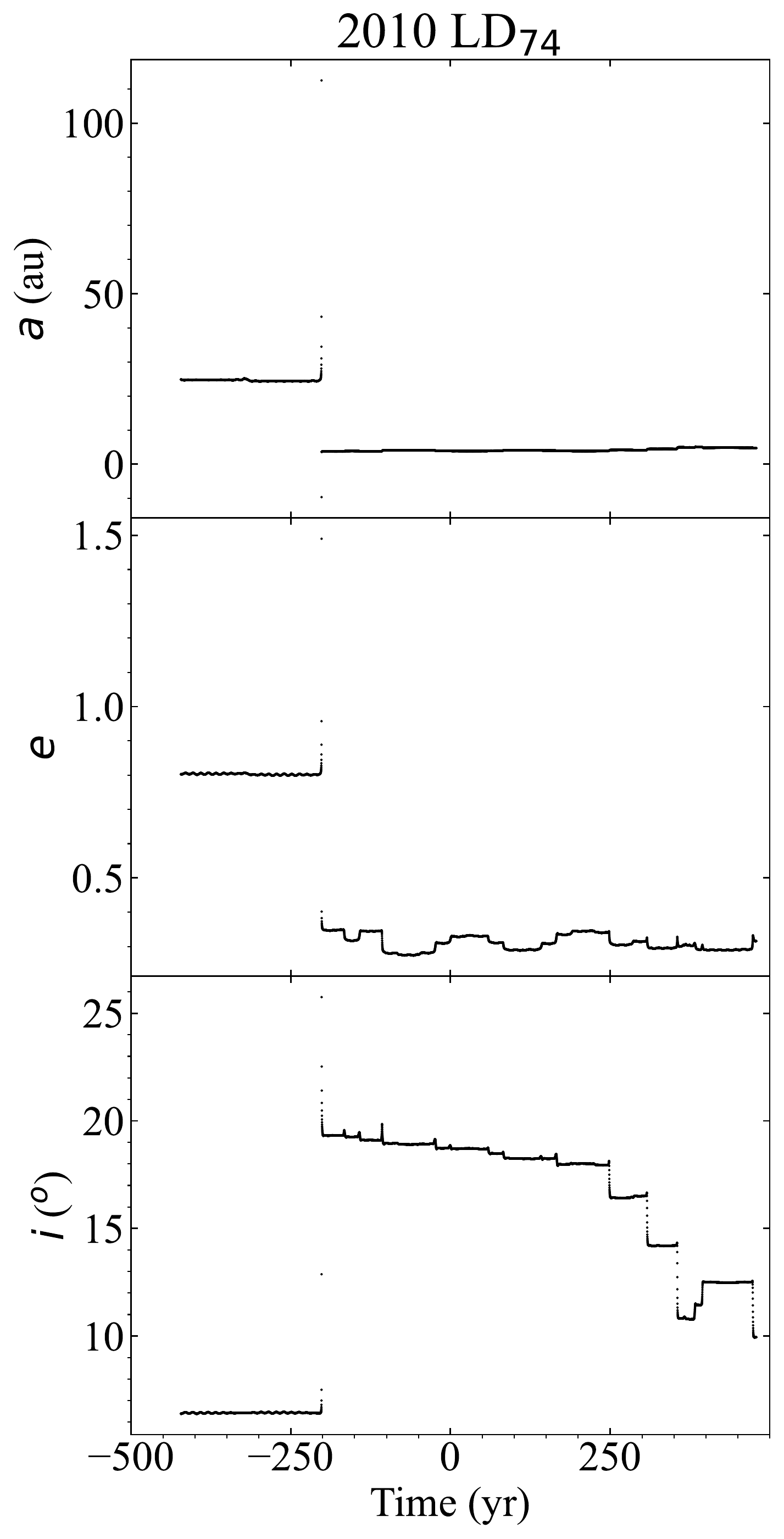}
  \includegraphics[width=0.195\linewidth]{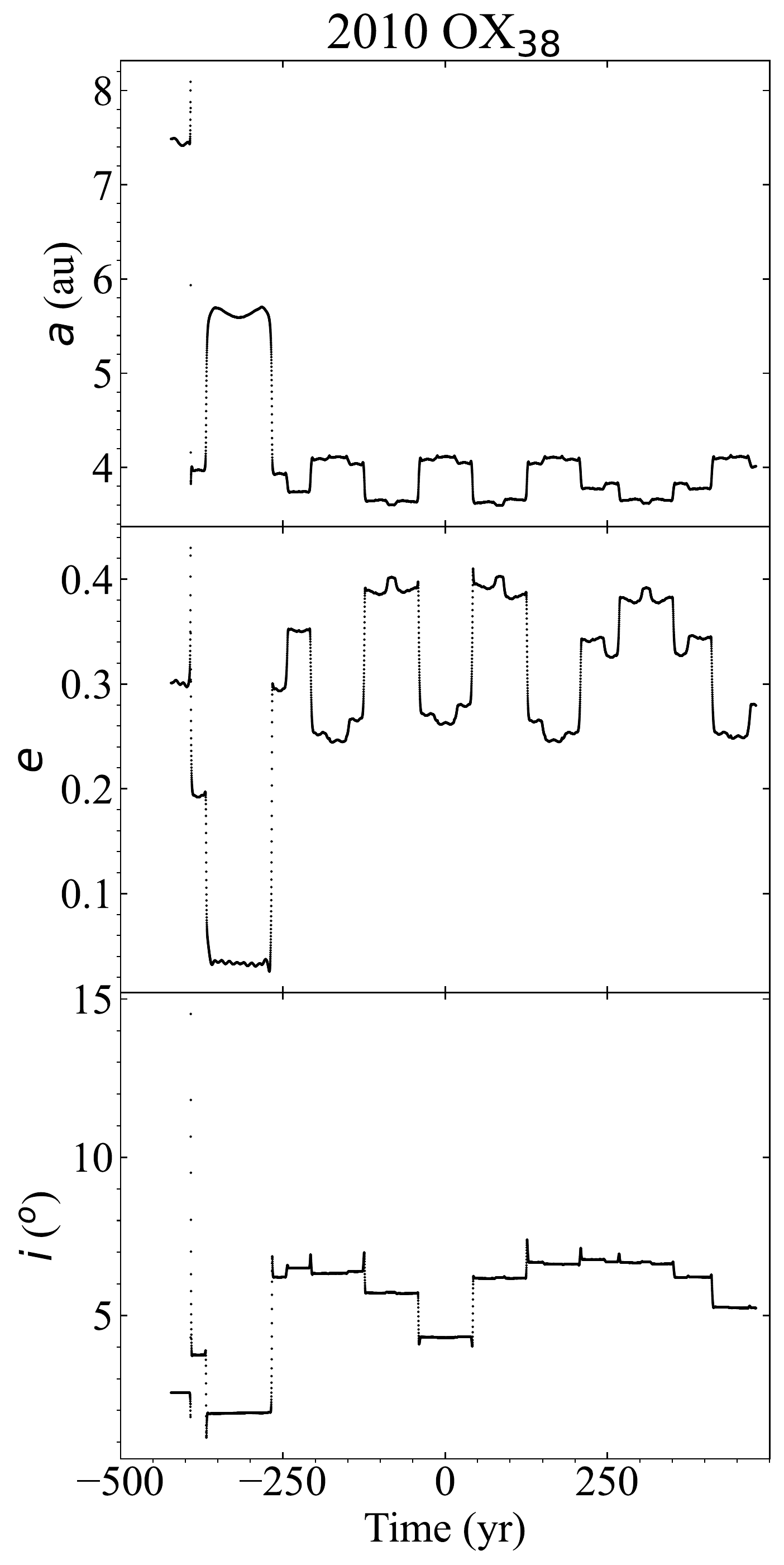}
  \includegraphics[width=0.195\linewidth]{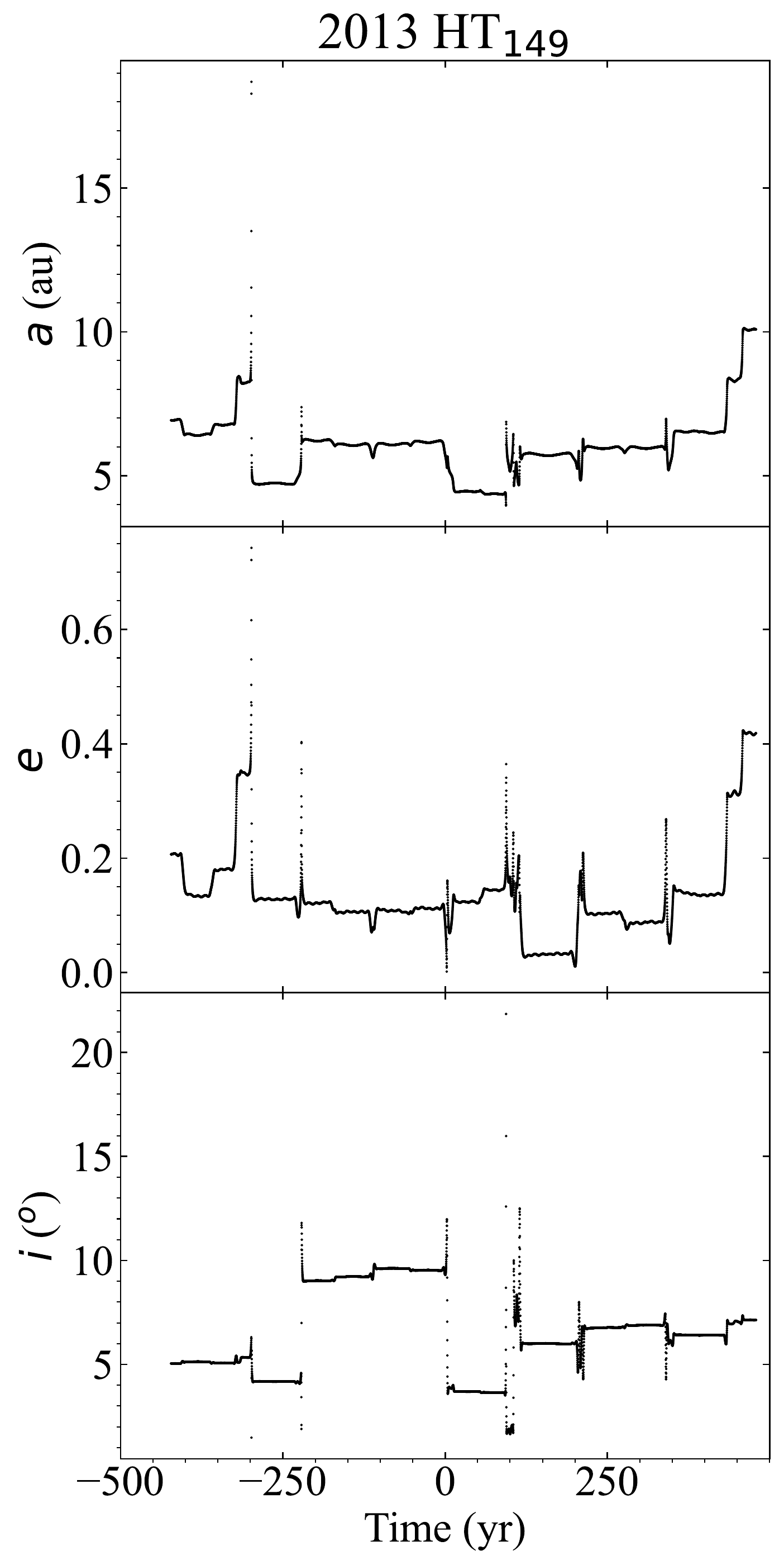}
  \includegraphics[width=0.195\linewidth]{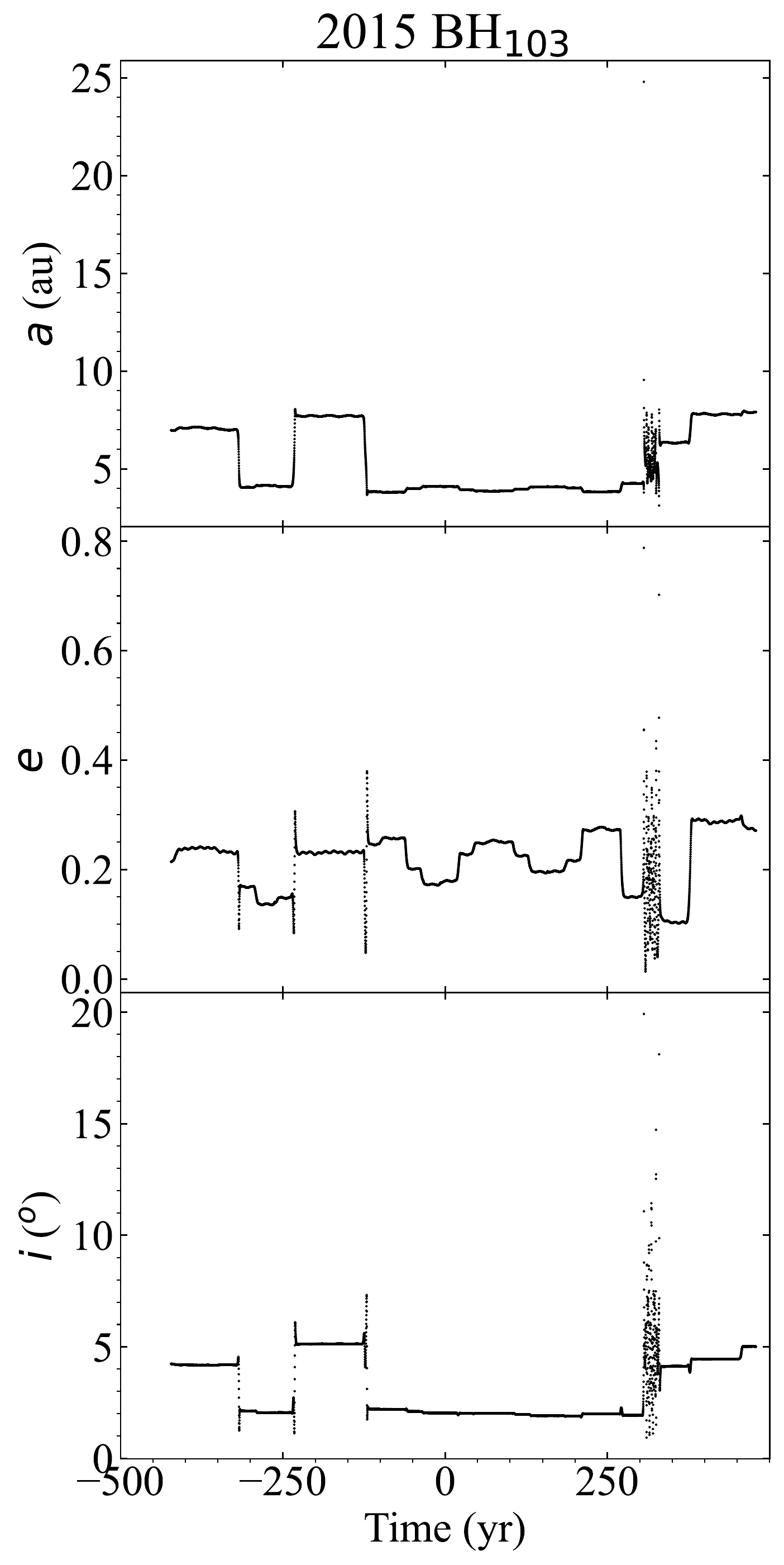}
  \includegraphics[width=0.195\linewidth]{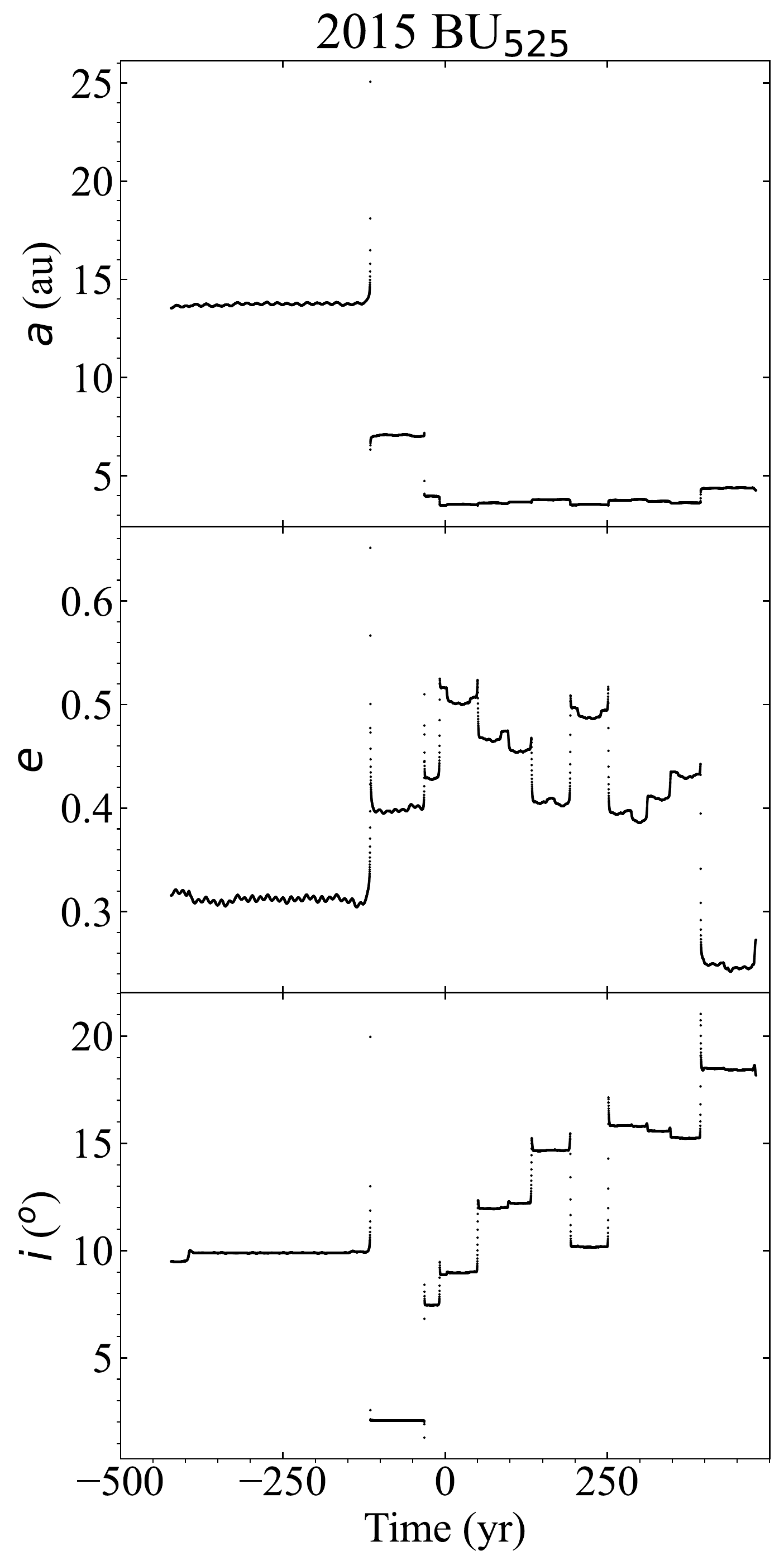}
  \includegraphics[width=0.198\linewidth]{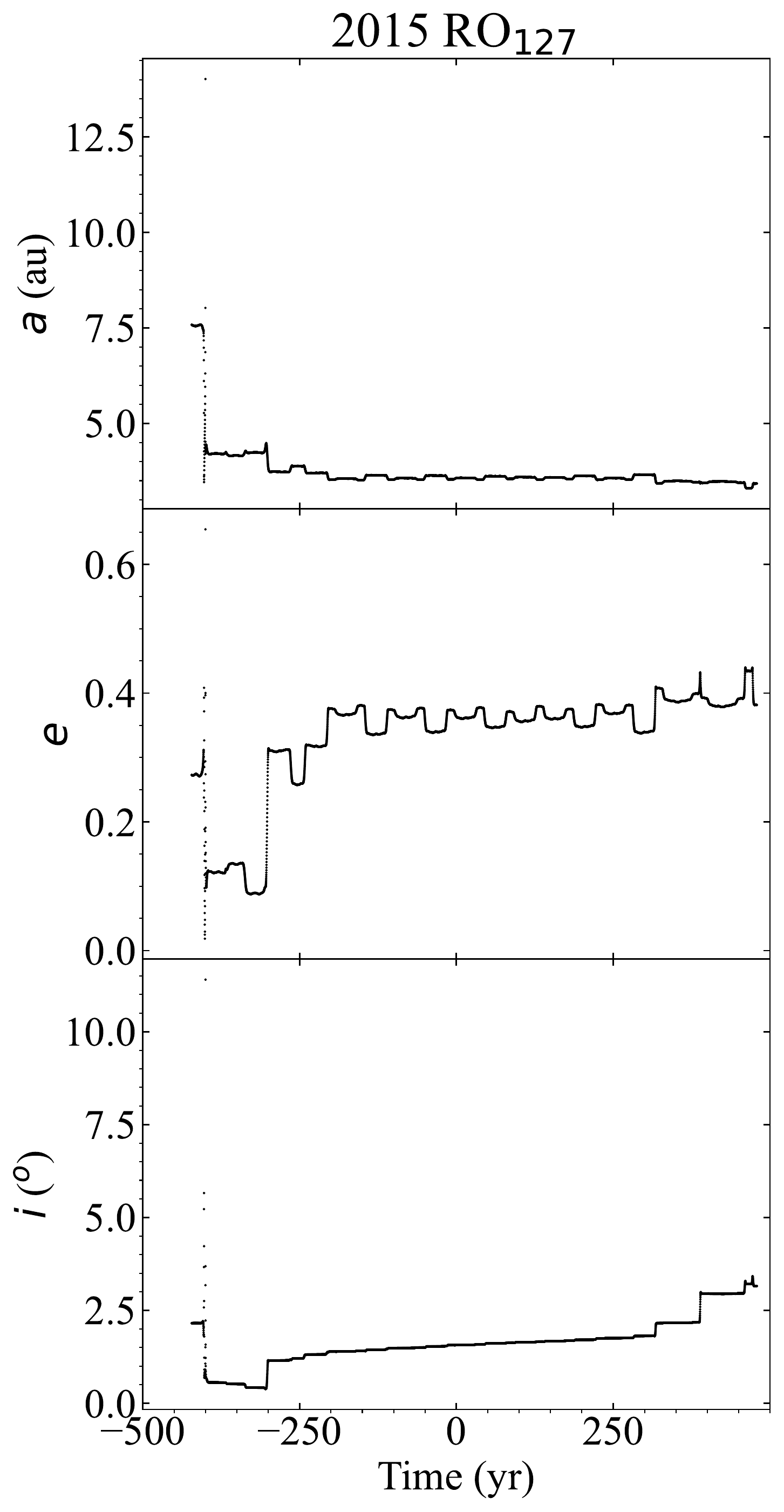}
  \includegraphics[width=0.193\linewidth]{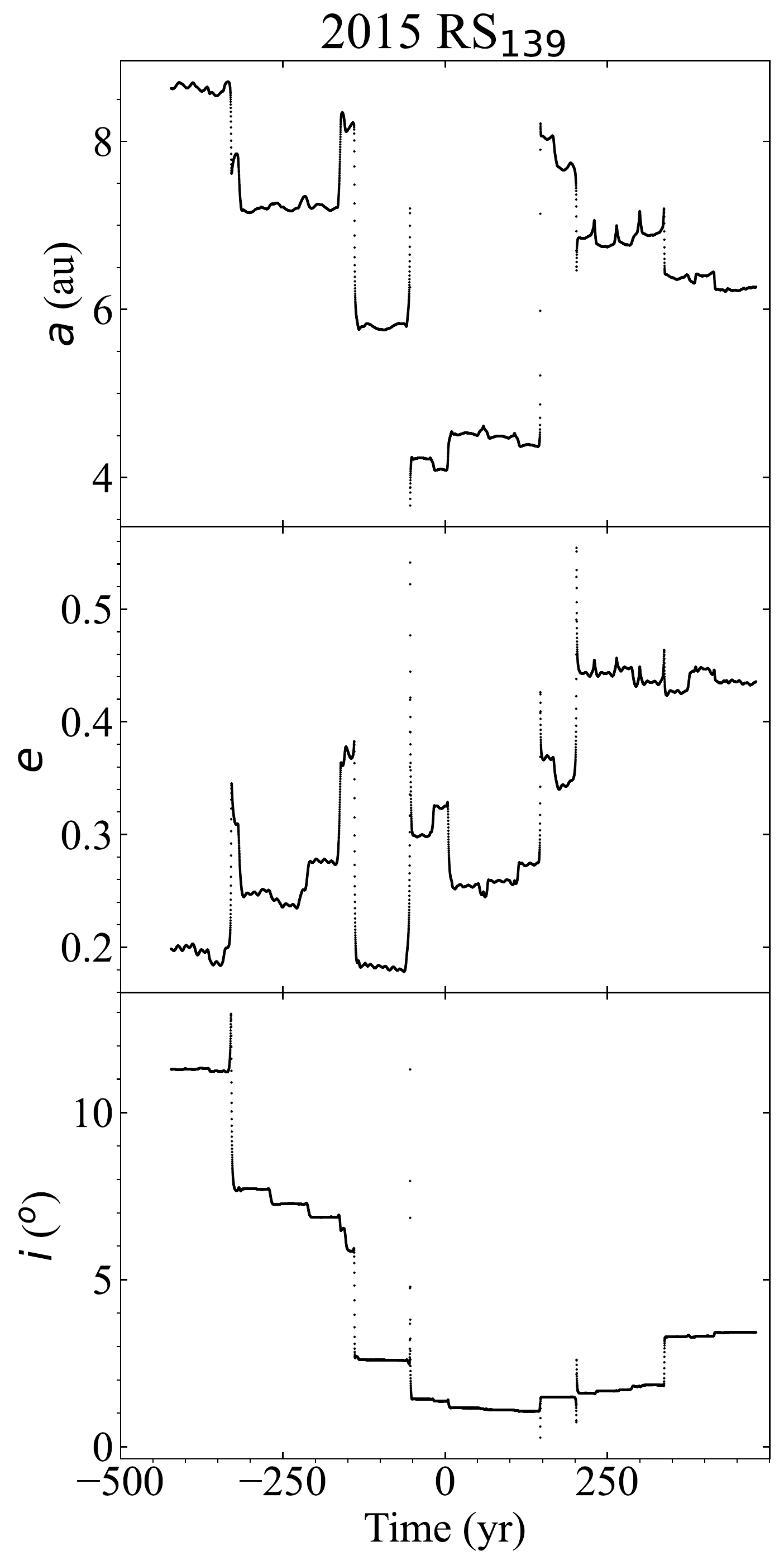}
  \includegraphics[width=0.199\linewidth]{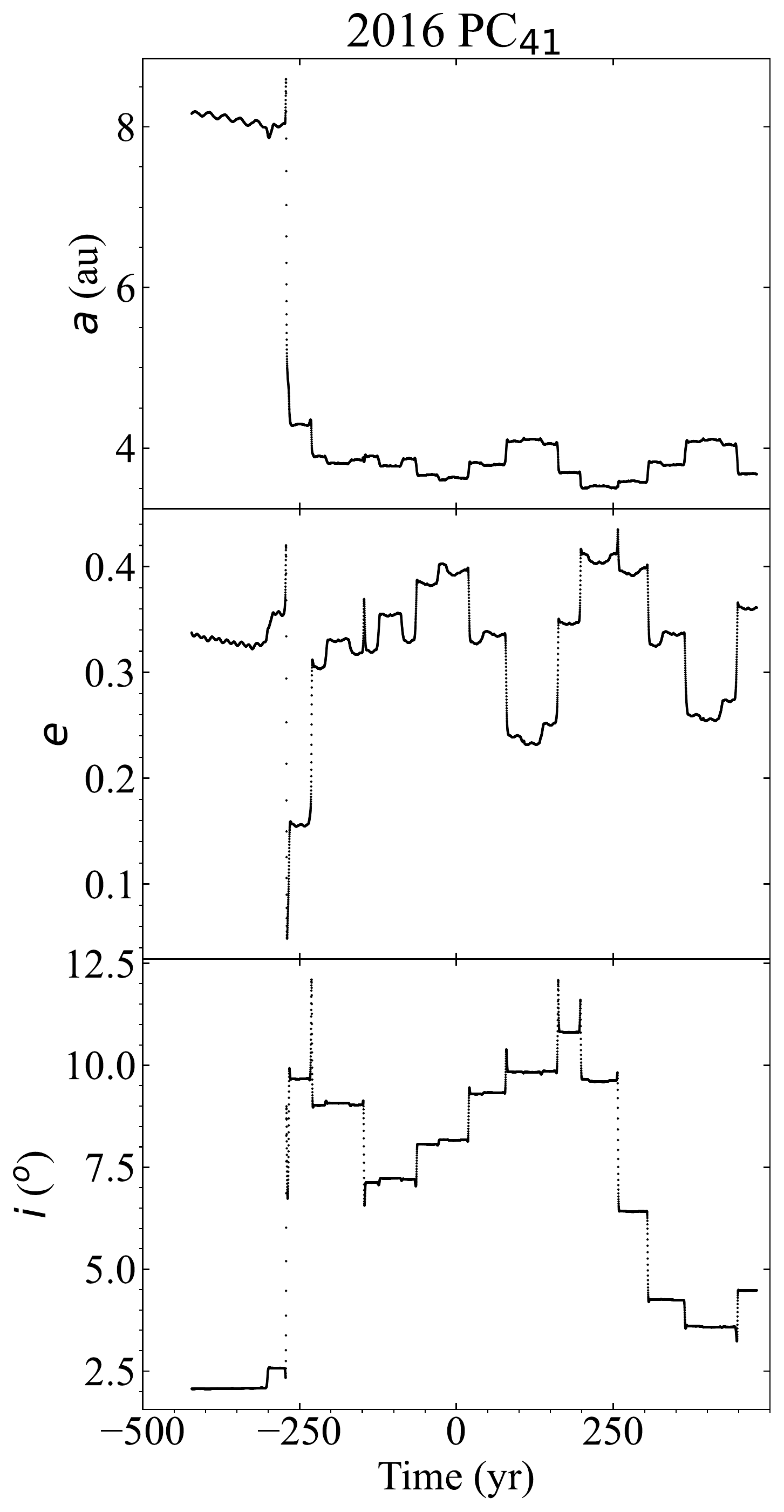}
  \includegraphics[width=0.193\linewidth]{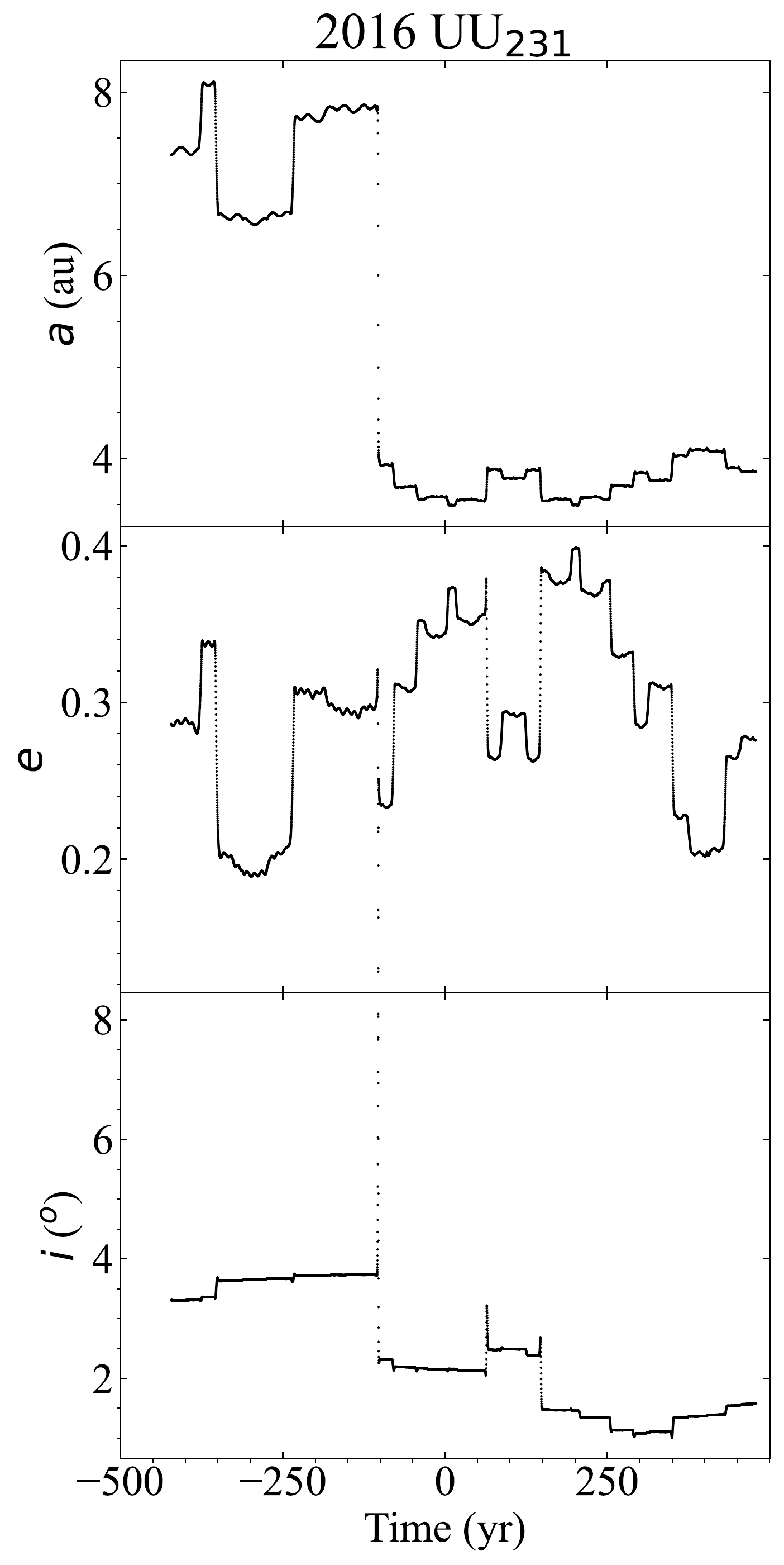}
  \includegraphics[width=0.193\linewidth]{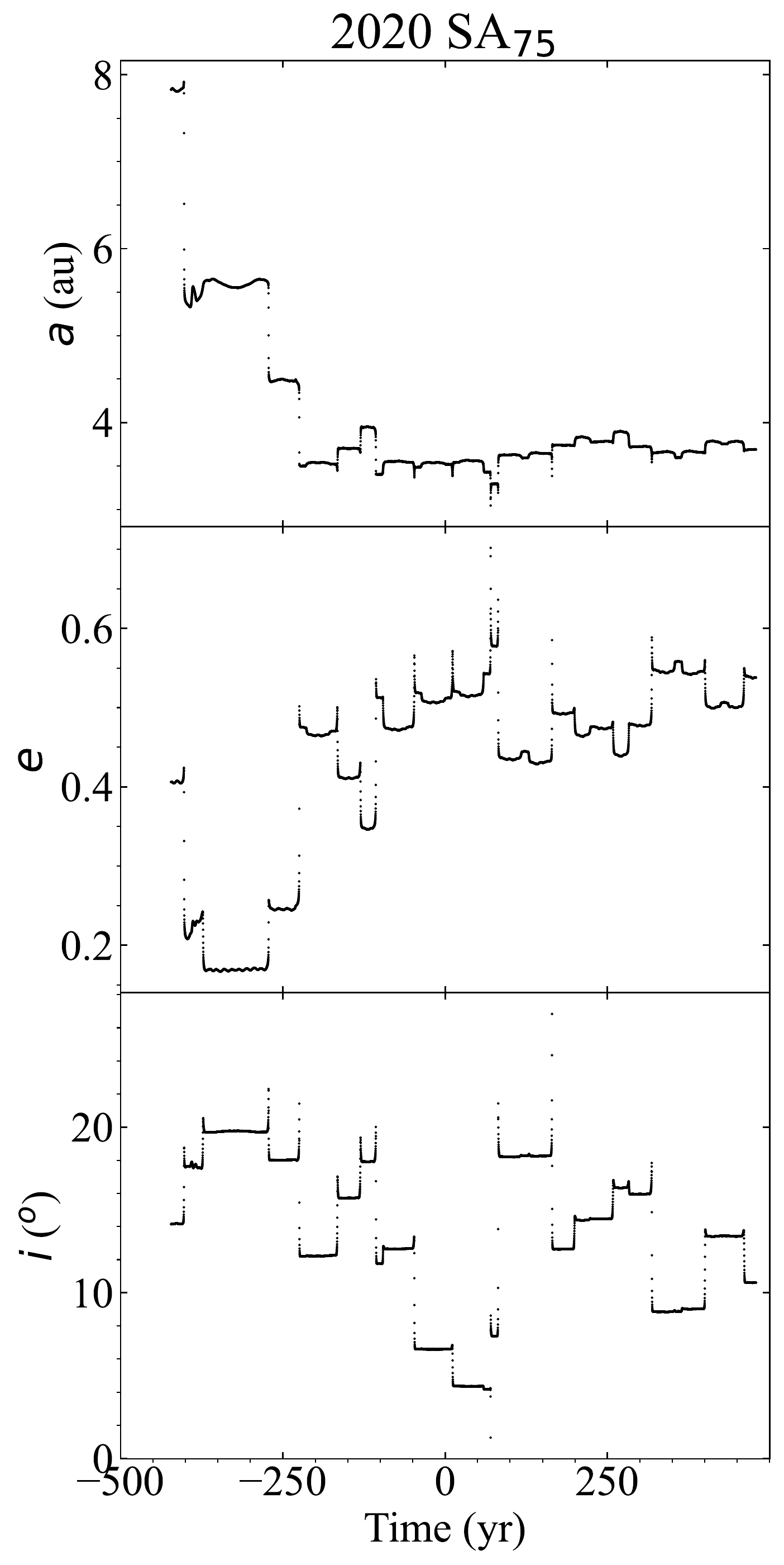}
  \caption{Evolution of the orbital elements semimajor axis (top panels), eccentricity (middle panels), and inclination 
           (bottom panels) for the nominal orbits of 2007~RS$_{62}$, 2010~BG$_{18}$, 2010~JC$_{58}$, 2010~JV$_{52}$, 
           2010~KS$_{6}$, 2010~LD$_{74}$, 2010~OX$_{38}$, 2013~HT$_{149}$, 2015~BH$_{103}$, 2015~BU$_{525}$, 
           2015~RO$_{127}$, 2015~RS$_{139}$, 2016~PC$_{41}$, 2016~UU$_{231}$, and 2020~SA$_{75}$. The origin of time is 
           the epoch 2459600.5 JD Barycentric Dynamical Time (2022-Jan-21.0 00:00:00.0 TDB) and the output cadence is 
           30~d. The source of the data is JPL's {\tt Horizons}.
          }
  \label{fig:others}
\end{figure}

Figure~\ref{fig:others} shows the evolution of relevant orbital elements for the objects in Table~\ref{tab:1} with the
most uncertain orbit determinations. All of them appear to have a short-term origin in Centaur orbital parameter space 
and a few might be debris from the 29P/Schwassmann-Wachmann~1--P/2008~CL94 (Lemmon)--P/2010~TO20 (LINEAR-Grauer) 
cometary complex.
%
%
%
%

\end{document}